\documentclass[a4paper, 12pt]{article}

\usepackage[sort&compress]{natbib}
\bibpunct{(}{)}{;}{a}{}{,} 

\usepackage{amsthm, amsmath, amssymb, mathrsfs, multirow, url, subfigure}
\usepackage{graphicx} 
\usepackage{ifthen} 
\usepackage{amsfonts}
\usepackage[usenames]{color}
\usepackage{fullpage}
\usepackage{tikz}
\usetikzlibrary{shapes.symbols, hobby, positioning}
\usepackage[ruled,vlined]{algorithm2e}

\theoremstyle{plain} 
\newtheorem{thm}{Theorem}

\theoremstyle{definition}

\newtheorem*{asmp}{Assumptions}

\theoremstyle{remark}

\newtheorem{ques}{Question}

\newcommand{\prob}{\mathsf{P}}

\newcommand{\unif}{{\sf Unif}}
\newcommand{\nm}{{\sf N}}

\newcommand{\gam}{{\sf Gamma}}

\newcommand{\chisq}{{\sf ChiSq}}

\newcommand{\RR}{\mathbb{R}}

\newcommand{\XX}{\mathbb{X}}

\newcommand{\UU}{\mathbb{U}}

\newcommand{\TT}{\mathbb{T}}

\newcommand{\G}{\mathscr{G}}
\newcommand{\Gbar}{\overline{\mathscr{G}}}

\newcommand{\eps}{\varepsilon}

\newcommand{\prior}{\mathsf{Q}}

\newcommand{\kernel}{\mathsf{K}}
\newcommand{\marg}{\mathsf{M}}

\newcommand{\Gauss}{\mathsf{G}}

\newcommand{\cred}{\mathscr{C}}

\newcommand{\uPi}{\mathsf{\Pi}}

\newcommand{\uOmega}{\mathsf{\Omega}}
\newcommand{\lOmega}{\underline{\Omega}}
\newcommand{\uGamma}{\mathsf{\Gamma}}

\title{No-prior Bayes reIMagined: probabilistic approximations of possibilistic inferential models}
\author{Ryan Martin\footnote{Department of Statistics, North Carolina State University, {\tt rgmarti3@ncsu.edu}}
}
\date{\today}

\begin{document}

\maketitle 

\begin{abstract}
When prior information is lacking, the go-to strategy for probabilistic inference is to combine a ``default prior'' and the likelihood via Bayes's theorem.  Objective Bayes, (generalized) fiducial inference, etc.~fall under this umbrella.  This construction is natural, but the corresponding posterior distributions generally only offer limited, approximately valid uncertainty quantification.  The present paper takes a reimagined approach that yields posterior distributions with stronger reliability properties.  The proposed construction starts with an inferential model (IM), one that takes the mathematical form of a data-driven possibility measure and features exactly valid uncertainty quantification, and then returns a so-called inner probabilistic approximation thereof.  This inner probabilistic approximation inherits many of the original IM's desirable properties, including credible sets with exact coverage and asymptotic efficiency.  The approximation also agrees with the familiar Bayes/fiducial solution in applications where the model has a group invariance structure.  A Monte Carlo method for evaluating the probabilistic approximation is presented, along with numerical illustrations.   

\smallskip

\end{abstract}

\section{Introduction}
\label{S:intro}

The program in which science advances by formulating what's currently known as prior beliefs and then applying a normative procedure to update those prior beliefs in light of new data is deeply rooted in all of us who've studied probability and statistics.  Despite its appeal, this program is challenging to implement for several reasons, not the least of which is that genuine prior information is rarely available, hence no particular formulation of prior beliefs is warranted.  Brad Efron, in his presentation at the 2016 Joint Statistical Meetings in Chicago, said that ``scientists like to work on new problems,'' meaning that there's often no history or experience to draw from in order to build a meaningful prior distribution.  For similar reasons, \citet{efron.cd.discuss} stated more formally that ``...the most important unresolved problem in statistical inference is the use of Bayes theorem in the absence of prior information.''
The present paper offers new insights, theory, and methodology with the ambitious goal of resolving this most important unresolved problem.  

\citet{fisher1933, fisher1935a, fisher1935b} and his fiducial argument was the first proposed solution to the above problem, ``a bold attempt to make the Bayesian omelet without breaking the Bayesian egg'' \citep{savage1961}; see \citet{zabell1992} and \citet{dawid2020} for excellent surveys.  The consensus is that Fisher's solution failed, but even his ``biggest blunder'' \citep{efron1998} had a major impact, inspiring fundamental advances such as confidence limits \citep{neyman1941}, imprecise probability \citep{dempster1966}, and new proposed solutions, including generalized fiducial inference \citep{hannig.review, murph.etal.fiducial}, confidence distributions \citep{xie.singh.2012, thornton.xie.cd, cox2006, schweder.hjort.book}, and objective/non-informative Bayes \citep{jeffreys1946, berger2006, berger.objective.book}.  In what follows, to keep terminology simple, I'll refer to all of these approaches as {\em no-prior Bayes solutions}.  

A difficulty with Bayesian-like probabilistic inference in this no-prior setting concerns the interpretation of the posterior probabilities themselves.  When genuine prior information is available, then the Bayesian posterior probabilities are the unique coherent updates of prior beliefs given the observed data.  But when there's no prior beliefs to update, and a default prior is used in its place, of course the aforementioned ``updating'' interpretation disappears and, consequently, it's unclear if the corresponding posterior distribution is meaningful at all.  
Indeed, ``[Bayes's formula] does not create real probabilities from hypothetical probabilities'' \citep{fraser.copss} and, more forcefully, 
\begin{quote}
...any serious mathematician would surely ask how you could use [Bayes's theorem] with one premise missing by making up an ingredient and thinking that the conclusions of the [theorem] were still available \citep
{fraser2011.rejoinder}.
\end{quote}
Fortunately, coherent updating of prior beliefs isn't necessary for a posterior to have belief-forming inferential weight.  But making this case requires, first, defining what it takes for a framework to have belief-forming inferential weight and, second, demonstrating that a proposed framework has this property.  To me, the only way forward is to demonstrate that the posterior probabilities are {\em reliable} for a sufficiently wide class of hypotheses, i.e., it's a demonstrably rare event that the posterior assigns high (resp.~low) probabilities to hypotheses about the unknown that are false (resp.~true).  The advantage of such a requirement is that the belief-forming inferential weight is inherited from Fisher's inductive logic: if, e.g., assigning low probability to a true assertion is a rare event and, in the application at hand, the assertion is assigned low probability, then inferring that the assertion is false is warranted because rare events effectively don't happen.  No-prior Bayes solutions often have some form of reliability, but the {\em false confidence theorem} \citep{balch.martin.ferson.2017, martin.nonadditive} says that none have the strong reliability property just described.  Since no-prior Bayes solutions can't offer reliable belief assignments, there's motivation to look beyond these existing approaches.  

{\em Inferential models} (IMs) \citep[e.g.,][]{imbasics, imbook, imreview} are an alternative to the aforementioned probabilistic approaches, whose output is a data-dependent possibility measure that offers provably reliable uncertainty quantification about unknown parameters.  Further details about possibility measures and a specific IM construction are given in Section~\ref{S:background}.  Importantly, my proposed shift from probabilistic to possibilistic uncertainty quantification facilitates strong reliability and, among other things, this strong reliability implies the usual frequentist error rate control without giving up on fully conditional Bayesian-like reasoning.  

Despite the benefits of possibilistic IMs compared to probabilistic no-prior Bayes solutions, I'm under no delusion that statisticians will abandon probabilism in favor of possibilism in the near future. But that doesn't mean IMs have to wait for the distant future to make a contribution.  The ``reIMagined'' approach I'm proposing here starts with the provably reliable, likelihood-based possibilistic IM and then reads off and returns an {\em inner probabilistic approximation} as a novel no-prior Bayes solution.  As I'll demonstrate below, the proposed inner probabilistic approximation inherits some---but necessarily not all---of the original possibilistic IM's strong reliability properties.  The motivation for this new approach is that there's only so much that can be achieved by constructing default priors and checking if the corresponding posterior distributions are reliable.  My suggestion here is to prioritize the reliability properties and directly construct a data-dependent probability that has those desired properties.   

The remainder of the paper is organized as follows.  Some background on possibility theory and IMs is presented in Section~\ref{S:background}.  A characterization of the IM's credal set is presented in Section~\ref{S:inner}, which is then used to define the corresponding inner probabilistic approximation.  Various properties of the proposed inner probabilistic approximation are developed in Section~\ref{S:properties}, including an agreement with the familiar no-prior Bayes solutions in invariant models and a version of the celebrated Bernstein--von Mises theorem that establishes its asymptotic efficiency.  A strategy for (approximately) computing the IM's inner probabilistic approximation is presented in Section~\ref{S:computation} and a new, valid, and efficient solution to the technically challenging and practically relevant Behrens--Fisher problem is presented in Section~\ref{S:bf}.  Some concluding remarks are given in Section~\ref{S:discuss} and the supplementary material 
contains some further technical details and illustrations.

\section{Background}
\label{S:background}

\subsection{Possibility theory}
\label{SS:possibility}

Possibility measures \citep[e.g.,][]{dubois.prade.book} are among the simplest imprecise probability models, closely linked to fuzzy set theory \citep[e.g.,][]{zadeh1978} and Dempster--Shafer theory \citep[e.g.,][]{shafer1976, shafer1987}.  Applications in statistics are described in \citet{dubois2006} and \citet{dubois.denoeux.2010}; see, also, Section~\ref{SS:ims}. 

The mathematical differences between probability and possibility theory can be succinctly summarized as follows: optimization is to possibility theory what integration is to probability theory.  That is, a possibility measure $\uPi$ defined on a space $\TT$ is determined by a function $\pi: \TT \to [0,1]$ with the property that $\sup_{t \in \TT} \pi(t) = 1$.  This function is called the {\em possibility contour} and the supremum-equals-1 property is a normalization condition analogous to the familiar integral-equals-1 property of probability densities.  Then the possibility measure is determined by optimizing its contour, i.e., $\uPi(A) = \sup_{t \in A} \pi(t)$, for any $A \subseteq \TT$, just like a probability measure is determined by integrating its density.  

The different calculus has a number of implications.  Of particular relevance for the developments in this paper is that the aforementioned supremum-equals-1 normalization condition ensures that $\uPi$ is a coherent upper probability \citep[e.g.,][]{cooman.poss1, cooman.aeyels.1999} in the spirit of \citet{walley1991} and others.  Among other things, this means that $\uPi$ determines a non-empty (closed and convex) set of probabilities it dominates:
\begin{equation}
\label{eq:credal}
\cred(\uPi) = \{ \prior \in \mathcal{P}(\TT): \prior \preceq \uPi \}, 
\end{equation}
where $\mathcal{P}(\TT)$ is the set of all probabilities supported on the Borel $\sigma$-algebra of measurable subsets of $\TT$ and ``$\prior \preceq \uPi$'' means ``$\prior(H) \leq \uPi(H)$ for all measurable $H$.'' The set $\cred(\uPi)$ is called the {\em credal set} corresponding to $\uPi$.  Aside from being relatively simple, a key advantage to possibilistic uncertainty quantification is that the associated credal set has a statistically oriented characterization, which is important for the developments below. 

Note that the members $\prior$ of the credal set $\cred(\uPi)$ are not constrained to be absolutely continuous with respect to any given dominating measure.  That is, some members are discrete with a mass function, some are continuous with a smooth density function with respect to Lebesgue measure, among other things.  The only constraint on the members $\prior$ of $\cred(\uPi)$ is dominance: $\prior \preceq \uPi$.

Critical to the developments below is a notion of approximating a probability by a possibility and vice versa.  I'll focus here on the former, and leave the latter for Section~\ref{S:inner}.  Suppose that $\prior$ is a probability on $\TT$ with density function $q$ relative to a given dominating measure.  The {\em probability-to-possibility transform} \citep[e.g.,][]{dubois.etal.2004, hose2022thesis} of $\prior$, with respect to the order determined by $q$, yields a possibility measure $\uPi$ with contour 
\[ \pi(t) = \prior(\{ \tau \in \TT: q(\tau) \leq q(t) \}), \quad t \in \TT. \]
This $\uPi$ has two key properties: first, $\cred(\uPi)$ contains $\prior$, so that $\uPi \succeq \prior$; second, $\uPi$ is the ``least imprecise'' possibilistic approximation of $\prior$ in the sense that it has the smallest credal set among other suitable possibilistic approximations.  More formally, $\cred(\uPi) = \bigcap \cred(\uOmega)$, where the intersection is over all possibility measures $\uOmega$ with $\uOmega \succeq \prior$ and whose associated contour $\omega$ is comonotone with $q$, i.e., 
\[ \{ \omega(t) - \omega(\tau) \} \times \{ q(t) - q(\tau) \} \geq 0, \quad t,\tau \in \TT. \]
Comonotonicity is motivated by a fundamental principle in this possibility-theory context: ``what is probable must be plausible'' \citep[p.~121]{dubois.prade.book}.  Since $\uPi$ is the dominating possibility with the smallest credal set, it's often called an {\em outer possibilistic approximation} of $\prior$ \citep[e.g.,][]{dubois.prade.1990}.

\subsection{Possibilistic IMs}
\label{SS:ims}

An inferential model (IM) is a mapping from data, model, etc.~to an imprecise-probabilistic quantification of uncertainty about relevant unknowns.  Critical to this approach and perspective is that the IM's uncertainty quantification be reliable, as a function of data, in a sense to be described below.  It's this insistence on reliability that implies the IM output must take on the form of an imprecise probability.  The first IM developments were built up using random sets and described using belief function terminology.  More recent developments directly appeal to possibility-theoretic tools and reasoning in the IM construction and interpretation, and I refer to these as {\em possibilistic IMs}; see \citet{imreview} for more details.  This paper focuses on this latter brand of IMs. 

Consider a model $\{ \prob_\theta: \theta \in \TT \}$ consisting of probability distributions supported on a sample space $\XX$, with parameter space $\TT$.  Suppose that the observable data $X$, taking values in $\XX$, is a sample from the distribution $\prob_\Theta$, where $\Theta \in \TT$ is the unknown/uncertain ``true value.''  Prior information about $\Theta$ is assumed to be vacuous, but see \citet{martin.partial2} for generalizations.  The model and observed data $X=x$ determine a relative likelihood function
\begin{equation}
\label{eq:rellik}
R(x,\theta) = \frac{p_\theta(x)}{\sup_{\vartheta \in \TT} p_\vartheta(x)}, \quad \theta \in \TT,
\end{equation}
where $p_\theta$ is the density of $\prob_\theta$. I will assume throughout that the supremum in  \eqref{eq:rellik} is finite for almost all $x$, and that this value is attained at $\hat\theta_x \in \arg\max_\vartheta p_\vartheta(x)$, a maximum likelihood estimator of $\Theta$, given $X=x$. 



The relative likelihood defines a possibility contour, i.e., a non-negative function such that $\sup_\theta R(x,\theta) = 1$ for almost all $x$, which itself can be used for data-driven uncertainty quantification about $\Theta$.  This has been extensively studied in \citep[e.g.,][]{shafer1982, wasserman1990b, denoeux2006, denoeux2014}, and it has a number of desirable properties.  What it lacks, however, is a formal calibration property justifying that the ``possibilities'' assigned to hypotheses about $\Theta$ have belief-forming inferential weight.  

\begin{figure}[t]
\begin{center}
\scalebox{0.5}{\includegraphics{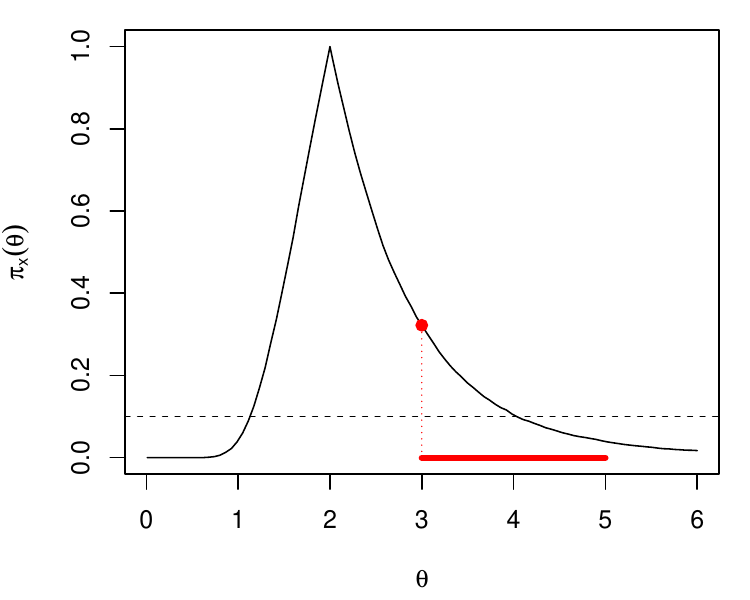}}
\end{center}
\caption{Illustration of a typical possibility contour $\theta \mapsto \pi_x(\theta)$ and how the possibility measure $\uPi_x$ is determined from it.  Here, the hypothesis $H$ is the interval $[3,5]$ and the maximum value $\uPi_x(H)$ on contour is highlighted. Horizontal dashed line at $\alpha=0.1$ determines the 90\% confidence interval $C_{0.1}(x)$ in \eqref{eq:region}.}
\label{fig:pl.toy}
\end{figure}

Fortunately, this reliability-focused calibration is easy to achieve, by ``validifying'' \citep{martin.partial, martin.partial2} the relative likelihood.  That is, this possibilistic IM construction proceeds by applying a version\footnote{The sense in which \eqref{eq:contour} is a ``version'' of the probability-to-possibility transform is too lengthy to be described in detail here.  Roughly, the likelihood function can be interpreted as an ``upper density'' for $(X,\Theta)$ under vacuous prior information for $\Theta$; then $R(x,\theta)$ is a normalized version of that ``upper density'' and \eqref{eq:contour} is basically the corresponding (imprecise) probability-to-possibility transform. The interested reader can consult \citet[][Sec.~5--6]{martin.partial2} for details.} of the probability-to-possibility transform to the relative likelihood, returning contour function:
\begin{equation}
\label{eq:contour}
\pi_{x}(\theta) = \prob_\theta\bigl\{ R(X,\theta) \leq R(x, \theta) \bigr\}, \quad \theta \in \TT.
\end{equation}
Note that $\pi_x(\hat\theta_x) = 1$, as required in Section~\ref{SS:possibility}, so $\pi_x$ is a genuine possibility contour---this is why normalizing the likelihood with a supremum in \eqref{eq:rellik} is needed.  The possibility measure $\uPi_x$ corresponding to $\pi_x$ in \eqref{eq:contour} is defined via optimization just like in Section~\ref{SS:possibility}: 
\begin{equation}
\label{eq:maxitive}
\uPi_{x}(H) = \sup_{\theta \in H} \pi_{x}(\theta), \quad H \subseteq \TT.
\end{equation}
An illustration of a typical contour $\pi_x$ is shown in Figure~\ref{fig:pl.toy}.  Some further details about this simple example are presented at the end of Section~\ref{S:inner} below. 

The relative likelihood $R(x,\theta)$ in \eqref{eq:rellik} is a very natural choice of ranking function when interest is in the full parameter $\Theta$.  Of course, other functions $R$ could be used in \eqref{eq:contour} instead of the relative likelihood, and this makes sense---and is even recommended---when there are other priorities involved: for example, if $\Theta$ contains nuisance parameters or if there are user-imposed shape constraints on the IM contour; see \citet[][Sec.~6.1]{imreview} and Sections~\ref{SS:caveats} and \ref{S:bf} below.  But I must emphasize that $R$ isn't arbitrary: care is needed to ensure that the transformation of $R$ in \eqref{eq:contour} yields a genuine possibility contour, i.e., $\sup_\theta \pi_x(\theta) = 1$.  Otherwise, $\uPi_x$ is incoherent \citep[][Prop.~7.14]{lower.previsions.book} and $\cred(\uPi_x)$ is empty.  This would obviously thwart any effort, like that below, to rigorously approximate the IM output by a probability in $\cred(\uPi_x)$.  More importantly, since a goal of the IM developments is to reconcile the performance-focused priorities of frequentism with the coherence-focused priorities of Bayesianism, the IM output must at least satisfy the imprecise-probabilistic version of coherence. 

As in Section~\ref{S:intro}, an essential feature of this possibilistic IM is its {\em validity} property:
\begin{equation}
\label{eq:valid}
\sup_{\theta \in \TT} \prob_\theta\bigl\{ \pi_{X}(\theta) \leq \alpha \bigr\} \leq \alpha, \quad \text{$\alpha \in [0,1]$}. 
\end{equation}
In words, \eqref{eq:valid} means that it's a rare event for the IM to assign small possibility $\pi_X(\Theta)$ to the true $\Theta$, which, of course, was important to the inductive logic explained in Section~\ref{S:intro}.  This has a number of important practical consequences as well.  First, \eqref{eq:valid} immediately implies that the upper $\alpha$ level set of the possibility contour, 
\begin{equation}
\label{eq:region}
C_\alpha(x) = \{\theta \in \TT: \pi_{x}(\theta) \geq \alpha\}, \quad \alpha \in [0,1],
\end{equation}
is a $100(1-\alpha)$\% confidence set, i.e., 
\[ \sup_{\theta \in \TT} \prob_\theta\bigl\{ C_\alpha(X) \not\ni \theta \bigr\} \leq \alpha. \] Figure~\ref{fig:pl.toy} depicts $C_\alpha(x)$ with $\alpha=0.1$; in one dimension, $C_\alpha(x)$ is typically a bounded interval and, more generally, a bounded set. Second, from \eqref{eq:maxitive} and \eqref{eq:valid}, 
\begin{equation}
\label{eq:valid.alt}
\sup_{\theta \in H} \prob_\theta\bigl\{ \uPi_{X}(H) \leq \alpha \bigr\} \leq \alpha, \quad \text{$\alpha \in [0,1]$, $H \subseteq \TT$}. 
\end{equation}
In words, a valid IM assigns possibility $\leq \alpha$ to true hypotheses at rate $\leq \alpha$ as a function of data $X$.  This gives the IM its ``inferential weight''---\eqref{eq:valid.alt} implies that $\uPi_{x}(H)$ is not expected to be small when $H$ is true, so one is inclined to doubt the truthfulness of a hypothesis $H$ if $\uPi_{x}(H)$ is small, and the strength of this inclination is determined by the magnitude of $\uPi_{x}(H)$.  This of course can be turned into a formal test procedure with frequentist error rate guarantees: by \eqref{eq:valid.alt}, the test ``reject $H$ if $\uPi_x(H) \leq \alpha$'' controls the Type~I error probability at level $\alpha$.  Importantly, \eqref{eq:valid} implies more than fixed-$H$ error rate control, as the following {\em uniform-in-hypotheses} version of \eqref{eq:valid.alt} shows:
\[ \sup_\theta \prob_\theta\{ \uPi_X(H) \leq \alpha \text{ for some $H$ with $H \ni \theta$} \} \leq \alpha. \]
Third, these calibration properties ensure that the possibilistic IM is safe from false confidence \citep{balch.martin.ferson.2017, martin.nonadditive}, unlike other no-prior Bayes solutions.  Further IM details are presented in Appendix~A in the supplement;
see, also, \citet{imreview}. 

Since the IM output is a coherent imprecise probability, there is an associated (non-empty) credal set, $\cred(\uPi_x)$ as in \eqref{eq:credal}.  The members of $\cred(\uPi_x)$, which I'll denote by $\prior_x$, need not correspond to Bayesian posterior distributions under any prior.  Fortunately, an interpretation can be given to the members of $\cred(\uPi_x)$, and this is key to my developments in Section~\ref{S:inner}.  


I'll end this section with two technical remarks.  First, the reader  recognizes the contour function $\pi_x(\theta)$ in \eqref{eq:contour} as a {\em p-value function} corresponding to the likelihood ratio statistic.  Aside from p-value function, this has appeared in the literature under many other names, e.g., preference functions \citep{spjotvoll1983}, significance functions \citep{fraser1991}, and confidence curves \citep{blaker.spjotvoll.2000, birnbaum1961, schweder.hjort.2002, schweder.hjort.book, xie.singh.2012}.  What distinguishes the present approach from these others is that IMs embrace the inherent imprecision and leverages relevant imprecise-probabilistic results.  That is, an IM determines a coherent and fully conditional imprecise-probabilistic quantification of uncertainty for inference and prediction \citep{imreview}, decision-making \citep{imdec.ext}, etc.  Even the more Bayesian-leaning developments along these lines \citep[e.g.,][]{MR4426408, grunwald.epost, cortinovis.caron.2024} don't take full advantage of what imprecise probability has to offer. 

Second, with the big-picture assessment in the previous paragraph, one might compare IMs to other imprecise-probabilistic solutions like that in \citet{walley1991}.  A critical observation is that, when prior information about $\Theta$ is vacuous, as assumed here, Walley's generalized Bayes posterior is also vacuous, i.e., no learning is achieved.  Since Bayesian reasoning isn't fully satisfactory in such cases, new ideas, like those reviewed above, are needed to achieve inference that's both valid and efficient.   


\section{Inner probabilistic approximations}
\label{S:inner}

\subsection{Intuition}
\label{SS:intuition}

Section~\ref{SS:possibility} discussed approximating a given probability distribution by a possibility measure, the workhorse being the probability-to-possibility transform.  The goal of the present section is to reverse this process: to approximate the possibilistic IM by a suitable probability distribution.  Before giving a general description, I'll offer some intuition using a simple example where all the calculations can be done explicitly.  

Let $X \sim \prob_\Theta = \nm(\Theta, 1)$.  Given $X=x$, the standard no-prior Bayes solution, corresponding to the flat, Jeffreys prior, returns $\prior_x = \nm(x, 1)$ as the posterior distribution for $\Theta$.  For the IM construction, the relative likelihood is $R(x,\theta) = \exp\{-(x-\theta)^2/2\}$ and the contour function is 
\begin{align*}
\pi_x(\theta) & = \prob_\theta\{ R(X,\theta) \leq R(x,\theta) \} \\
& = \prob_\theta\{ e^{-(X-\theta)^2/2} \leq e^{-(x-\theta)^2/2} \} \\
& = \prob\{ \chisq(1) \geq (x-\theta)^2 \} \\
& = 1 - F_1\bigl( (x-\theta)^2 \bigr), 
\end{align*}
where $F_\nu$ is the $\chisq(\nu)$ distribution function.  How are $\prior_x$ and the corresponding $\uPi_x$ related?  One way to see it, following Section~\ref{SS:possibility}, is to apply the probability-to-possibility transform to $\prior_x$ and see what comes out.  In this case, with $q_x$ the $\nm(x,1)$ density function 
\begin{align*}
\text{prob-to-poss}_{\prior_x}(\theta) & := \prior_x\{ q_x(\Theta) \leq q_x(\theta)\} \\
& = \prior_x\{ e^{-(\Theta-x)^2/2} \leq e^{-(\theta-x)^2/2} \} \\
& = \prob\{ \chisq(1) \geq (\theta-x)^2 \} \\
& = \pi_x(\theta).
\end{align*}
That is, the outer possibilistic approximation of $\prior_x$, obtained via the probability-to-possibility transform, is precisely the possibilistic IM $\uPi_x$.  This is an {\em outer} approximation because, roughly, it's the ``least imprecise'' possibility measure among those that $\preceq$-dominate $\prior_x$.  It's only natural, then, to turn this relationship around and say that $\prior_x$ is an {\em inner probabilistic approximation} of $\uPi_x$ in the sense of being the ``maximally diffuse'' probability distribution that's $\preceq$-dominated by $\uPi_x$.  These details will be elaborated on in Section~\ref{SS:characterization} below. 

As a preview, there are at least two equivalent ways to extract a probability from a possibility: one based on bounding tail probabilities and another based on level set matching.  Here I'll focus on the latter approach, formalized in Section~\ref{SS:characterization}.  Since $\pi_x(\theta)$ is most naturally interpreted as the ``plausibility'' of $\Theta=\theta$ given $x$, and since posterior densities are (informally) interpreted in the same way, it makes sense that the density $q_x$ of $\prior_x$ would have equivalent level sets as $\pi_x$.  Akin to slice sampling \citep[e.g.,][]{neal2003}, a probability $\prior_x$ is determined by:
\begin{itemize}
\item first sampling the level set;
\item then sampling a point on the chosen level set.  
\end{itemize} 
Here, the level sets are the endpoints of the confidence intervals $C_\alpha(x) = \{ \theta: \pi_x(\theta) \geq \alpha \}$, indexed by $\alpha \in [0,1]$.  
Then sampling the level sets is equivalent to sampling the level itself, say $A$, from a distribution supported on $[0,1]$.  If $\prior_x$ represents the distribution of $\Theta$ under the two-stage sampling scheme above, then $\Theta \in C_\alpha(x)$ if and only if $A > \alpha$.  Then dominance $\prior_x \preceq \uPi_x$ implies 
\begin{align*}
\prior_x(A > \alpha) & \equiv \prior_x\{ C_\alpha(x) \} \\
& = 1 - \prior_x\{ C_\alpha(x)^c \} \\
& \geq 1 - \uPi_x\{ C_\alpha(x)^c \} \\
& = 1 - \alpha. 
\end{align*}
That is, dominance implies that $A$ is stochastically no smaller than $\unif(0,1)$.  At least intuitively, the ``maximally diffuse'' probability distribution $\prior_x^\star$ dominated by $\uPi_x$ corresponds to taking $A \sim \unif(0,1)$ and, given the level $A=\alpha$, sampling $\Theta$ on $C_\alpha(x)$ with maximal variance, i.e., uniformly on the endpoints of $C_\alpha(x)$.  Those endpoints are $x \pm z_{\alpha/2}$, where $z_\beta$ is the upper-$\beta$ quantile of $\nm(0,1)$, and then the distribution function of $\prior_x^\star$ is 
\begin{align*}
\prior_x^\star\bigl( (-\infty, \theta] \bigr) & = \int_0^1 \tfrac12 \, 1\{x - z_{\alpha/2} \leq \theta\} \, d\alpha + \int_0^1 \tfrac12 \, 1\{ x + z_{\alpha/2} \leq \theta\} \, d\alpha \\
& = \int_0^1 \tfrac12 \, 1\{ z_{\alpha/2} \geq x-\theta \} \, d\alpha + \int_0^1 \tfrac12 \, 1\{ z_{\alpha/2} \leq \theta-x \} \, d\alpha.
\end{align*}
As a function of $A \sim \unif(0,1)$, the random variable $z_{A/2}$ has the same distribution as $|Z|$ with $Z \sim \nm(0,1)$.  So,
\[ \prior_x^\star\bigl( (-\infty, \theta] \bigr) = \tfrac12 \, \prob(|Z| \geq x-\theta) + \tfrac12 \, \prob( |Z| \leq \theta-x). \]
Using some basic properties of the standard normal distribution, the right-hand side simplifies, leading to 
\[ \prior_x^\star\bigl( (-\infty, \theta] \bigr) = \prob(Z \leq \theta-x), \]
and, therefore, this ``maximally diffuse'' probability distribution $\prior_x^\star$ dominated by $\uPi_x$ is just the standard no-prior Bayes solution, i.e., $\prior_x^\star = \nm(x,1)$.

\subsection{Characterization}
\label{SS:characterization}

The goal here is to formalize and generalize the intuition developed above.  Towards this, I'll focus attention on continuous-data cases where $\prob_\theta$ has a density with respect to Lebesgue measure.  This is not necessary for the IM formulation, or for the theory that follows, but working with the characterization described below is difficult when $\pi_x$ is not a continuous function, which is typical for discrete-data models.  As $n \to \infty$, the  aforementioned discontinuities disappear, so all of what I have to say below applies at least approximately for discrete-data models with moderate-to-large sample sizes.  Details for the discrete-data case will be presented elsewhere. 

Start with a well-known characterization of a possibility measure's credal set \citep[e.g.,][]{cuoso.etal.2001, destercke.dubois.2014} which, in our case, says
\begin{equation}
\label{eq:char0}
\begin{split}
\prior_x & \in \cred(\uPi_x) \iff \prior_x\{C_\alpha(x)\} \geq 1-\alpha \;\; \text{for all $\alpha \in [0,1]$}, 
\end{split}
\end{equation}
where $C_\alpha(x)$ is as defined in \eqref{eq:region} with $\pi_x$ the contour corresponding to $\uPi_x$.  Again, the IM's validity property implies that $C_\alpha(x)$ is a $100(1-\alpha)$\% confidence set, so, since the elements $\prior_x$ of $\cred(\uPi_x)$ assign probability at least $1-\alpha$ to the $100(1-\alpha)$\% confidence set $C_\alpha(x)$, it's not inappropriate to call the elements of $\cred(\uPi_x)$ {\em confidence distributions}.  This definition aligns with that in recent work \citep[e.g.,][]{taraldsen.2021.cd, thornton.xie.cd} and generalizes those definitions \citep{xie.singh.2012, schweder.hjort.cd.discuss, prsa.conf} that focus mainly on scalar parameter cases; see Section~\ref{SS:caveats}.  

Refinements to the characterization in \eqref{eq:char0} can be given.  For example, \citet[][Theorem~2.1]{wasserman1990} characterizes the credal set associated with a belief function as a collection of suitable mixture distributions.  The result below gives a similar mixture characterization for possibility measures, including for the IM output $\uPi_x$.  

\begin{thm}[\citealt{immc}]
\label{thm:char}
Let $\pi_x(\theta)$ be the contour associated with the possibility measure $\uPi_x$, and set $C_\alpha(x) = \{\theta: \pi_x(\theta) > \alpha\}$ as in \eqref{eq:region}.  Then $\prior_x \in \cred(\uPi_x)$ if and only if it can be represented as 
\begin{equation}
\label{eq:char}
\prior_x(\cdot) = \int_0^1 \kernel_x^\alpha(\cdot) \, \marg_x(d\alpha), 
\end{equation}
for some kernel $\kernel_x^\alpha$, fully supported on $C_\alpha(x)$, so that $\kernel_x^\alpha\{\bar C_\alpha(x)\} = 1$ for each $\alpha \in [0,1]$, and some probability measure $\marg_x$ on $[0,1]$ whose distribution function is upper-bounded by that of $\unif(0,1)$.
\end{thm} 

Note that choosing the kernel $\kernel_x^\alpha$ to be $\unif\{ C_\alpha(x) \}$ is a common idea in the literature, e.g., in the {\em pignistic probability} \citep{smets.kennes.1994} or in the {\em Shapley value} \citep{shapley1953} of a game. Also, \citet{dubois.prade.smets.2008} show that the least informative plausibility function that dominates the pignistic probability is consonant, hence corresponds to a possibility measure. 

I referred to $\prior_x$ in $\cred(\uPi_x)$ as confidence distributions, but a more appropriate name is {\em over-confidence distributions}, since the credibility $\prior_x\{C_\alpha(x)\}$ generally exceeds the confidence level $1-\alpha$.  The key question is if there are any genuine confidence distributions among those over-confidence distributions, i.e., if there exists members $\prior_x^\star$ of $\cred(\uPi_x)$ such that equality on the right-hand side of \eqref{eq:char0} is achieved for each $\alpha$, i.e., 
\begin{equation}
\label{eq:inner}
\prior_x^\star\{ C_\alpha(x) \} = 1-\alpha, \quad \text{for all $\alpha \in [0,1]$}.
\end{equation}
This formalizes the notion of ``maximally diffuse'' in Section~\ref{SS:intuition}. As explained there, those $\prior_x^\star$ for which \eqref{eq:inner} holds are called {\em inner probabilistic approximations} of the possibilistic IM $\uPi_x$.  Alternatively, one can interpret $\prior_x^\star$ as a solution to the distributional equation ``$\pi_x(\Theta) \sim \unif(0,1)$ when $\Theta \sim \prior_x$.''  This perspective aligns with the familiar notion of pivoting, i.e., to matching the familiar sampling distribution property ``$\pi_X(\theta) \sim \unif(0,1)$ when $X \sim \prob_\theta$'' like that established in \eqref{eq:valid}. 

There are cases where attaining equality in \eqref{eq:inner} is out of reach.  For instance, suppose that $\theta \mapsto \pi_x(\theta)$ is bounded away from 0 for some $x$.  This is rare in applications, but not impossible.  For example, the data might not be particularly informative about a feature $\Phi=m(\Theta)$ of $\Theta$, and then the corresponding marginal IM (see Section~\ref{SS:caveats}) has a non-vanishing contour---that is, the data can't exclude any values of $\Phi$ with high degree of confidence.  

Some clues about what it takes to satisfy \eqref{eq:inner} were already presented in Section~\ref{SS:intuition}.  Indeed, to get a ``maximally diffuse'' member of the credal set, I suggested taking the marginal distribution $\marg_x$ to be $\unif(0,1)$ and taking the kernel $\kernel_x^\alpha$ to be fully supported on the boundary, $\partial C_\alpha(x)$, of the level set $C_\alpha(x)$.  To justify this, note that, for each $\alpha \in [0,1]$, a $\prior_x$ of the form \eqref{eq:char} satisfies 
\begin{align*}
\prior_x\{ C_\alpha(x) \} & = \int_0^1 \kernel_x^\beta\{ C_\alpha(x) \} \, \marg_x(d\beta) \\
& = \int_0^\alpha \underbrace{\kernel_x^\beta\{ C_\alpha(x) \}}_{\geq 0} \, \marg_x(d\beta) + \int_\alpha^1 \underbrace{\kernel_x^\beta\{ C_\alpha(x) \}}_{=1} \, \marg_x(d\beta) \\
& \geq \marg_x([\alpha,1]) \\
& \geq 1-\alpha.
\end{align*}
Then it's clear that equality in \eqref{eq:inner} is achieved by taking $\marg_x = \unif(0,1)$ and $\kernel_x^\alpha$ fully supported on $\partial C_\alpha(x)$, so that $\kernel_x^\beta\{ C_\alpha(x) \} = 0$ for all $\beta < \alpha$.  

How the kernel allocates its mass to $\partial C_\alpha(x)$ isn't determined by \eqref{eq:inner}, hence the inner probabilistic approximation of $\uPi_x$ isn't unique.  I'll say more about the role played by the kernel later and in Appendix~B.
The choice of kernel corresponds to a choice of a {\em disintegration} along the mapping $\theta \mapsto \pi_x(\theta)$, or a decomposition of a joint distribution into conditionals and a marginal \citep[e.g.,][]{chang.pollard.1997}.  This formalizes the existence of a two-step procedure like the one eluded to in Section~\ref{SS:intuition}.  Here I present my broad and generic recommendation.  Note that the sets $C_\alpha(x)$ are typically bounded; in fact, when the sample size is moderate to large, they're approximately ellipsoids (Section~\ref{SS:concentration}).  Then choosing the kernel $\kernel_x^\alpha$ to be a uniform distribution on $\partial C_\alpha(x)$, for each $\alpha \in (0,1)$, is quite natural, and consistent with both the maximum entropy \citep[e.g.,][]{cover.thomas.book, bernardo.smith.book} and indifference principles \citep[e.g.,][]{keynes.probability, jaynes2003}.  It'll also be seen in Section~\ref{SS:concentration} that uniformity is needed for asymptotic efficiency of the inner probabilistic approximation.  

To summarize, my proposed inner probabilistic approximation $\prior_x^\star$ of $\uPi_x$ is a uniform mixture of uniform distributions.  More specifically, a sample $(\Theta \mid x) \sim \prior_x^\star$ can be drawn via the following two-step procedure:
\begin{enumerate}
\item sample $(A \mid x) \sim \unif(0,1)$, and 
\item sample $(\Theta \mid x, A=\alpha) \sim \kernel_x^\alpha := \unif\{ \partial C_\alpha(x) \}$.
\end{enumerate}
Step~2 is far simpler to state than it is to carry out, and I'll discuss how this computational challenge can be addressed---at least approximately---in Section~\ref{S:computation}.  

Having some assurances about the existence of an inner probabilistic approximation and a suggested strategy for finding one, one might ask if this has any desirable properties.  A common practice among Bayesians and others  \citep[e.g.,][]{liu.liu.xie.2022} is constructing credible sets for $\Theta$ via their probabilistic uncertainty quantification.  This construction starts with the choice of a ranking function, say, $h_x$, and chooses a cutoff $\eta_{x,\alpha}$ such that 
\[ \prior_x^\star(\{ \theta: h_x(\theta) \leq \eta_{x,\alpha} \}) = \alpha. \]
Standard examples of $h_x$ include the distance $h_x(\theta) = \|\theta - \tilde\theta_x\|$ to an estimator $\tilde\theta_x$ and the density function $q_x^\star$ corresponding to $\prior_x^\star$.  Then the set $\{\theta: h_x(\theta) > \eta_{x,\alpha}\}$ is a $100(1-\alpha)$\% credible set; if $h_x=q_x^\star$, then it's called the highest posterior density credible set.  A desirable property is that the credible sets are ``probability matching'' \citep[e.g.,][]{datta.ghosh.1995} in the sense that the credibility and the coverage probabilities agree:
\[ \sup_\theta \prob_\theta\{ h_X(\theta) \leq \eta_{x,\alpha}\} \leq \alpha. \]
I'll refer to this property as $h$-{\em probability matching}.  The following theorem describes those $h$ for which an inner probabilistic approximation is $h$-probability matching. 

\begin{thm}
\label{thm:matching}
If $\prior_x^\star$ is an inner probabilistic approximation of $\uPi_x$, then it's $h$-probability matching for any $h$ such that $h_x$ is comonotone with $\pi_x$.
\end{thm}

\begin{proof}
Take $\theta_{x,\alpha}$ such that $\eta_{x,\alpha} = h_x(\theta_{x,\alpha})$.  Comonotonicity implies that
\[ h_x(\theta) \leq h_x(\theta_{x,\alpha}) \iff \pi_x(\theta) \leq \pi_x(\theta_{x,\alpha}). \]
According to \eqref{eq:inner}, the event on the right-hand side above has $\prior_x^\star$-probability $\alpha$ if $\pi_x(\theta_{x,\alpha}) = \alpha$.  In this case, the credible set with $h$ is equivalent to $C_\alpha(x)$, which is a confidence set, hence $\prior_x^\star$ is $h$-probability matching. 
\end{proof}

In some cases, including the directional data illustration in Section~\ref{SS:group}, especially Figure~\ref{fig:wheel.theta}, the density associated with an inner probabilistic approximation is comonotone with the contour $\pi_x$, hence the highest posterior density credible set is probability matching.  But there are other cases when this comonotonicity fails, including the gamma illustration in Section~\ref{SS:gamma} below.  As expected, like with Bayes and fiducial credible sets, these discrepancies often disappear as the sample size increases, but Theorem~\ref{thm:matching} provides some insights about when probability matching can be achieved in finite samples.  While the construction of credible sets using the underlying density is common, this isn't the only construction.  For example, often the relative likelihood and the possibility contour are comonotone, or at least approximately so.  This observation justifies the recommendation in \citet[][Sec.~3.4]{immc} to reconstruct the possibility contour from (samples from) $\prior_x^\star$ via $\theta \mapsto \prior_x^\star\{ R(x,\Theta) \leq R(x,\theta)\}$.

\subsection{Illustration}
\label{SS:gamma}

\begin{figure}[t]
\begin{center}
\subfigure[Distribution function of $\prior_x^\star$]{\scalebox{0.5}{\includegraphics{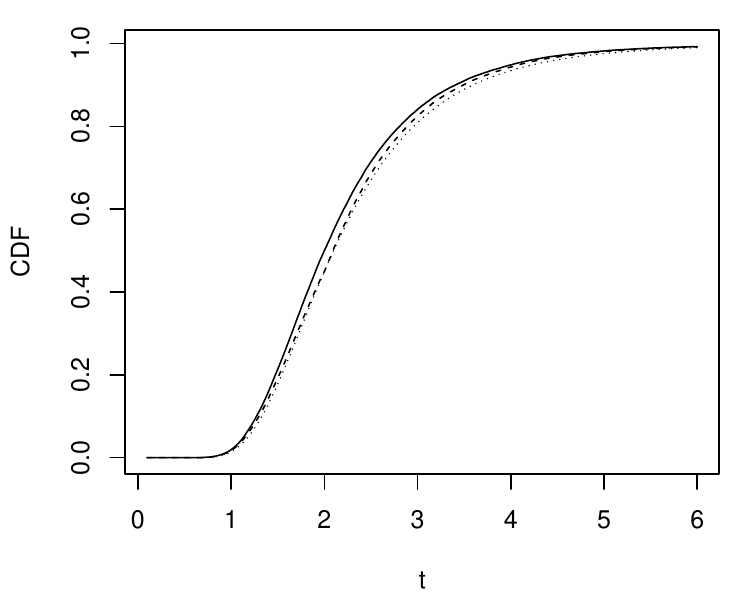}}}
\subfigure[Non-credibility of $\prior_x^\star$]{\scalebox{0.5}{\includegraphics{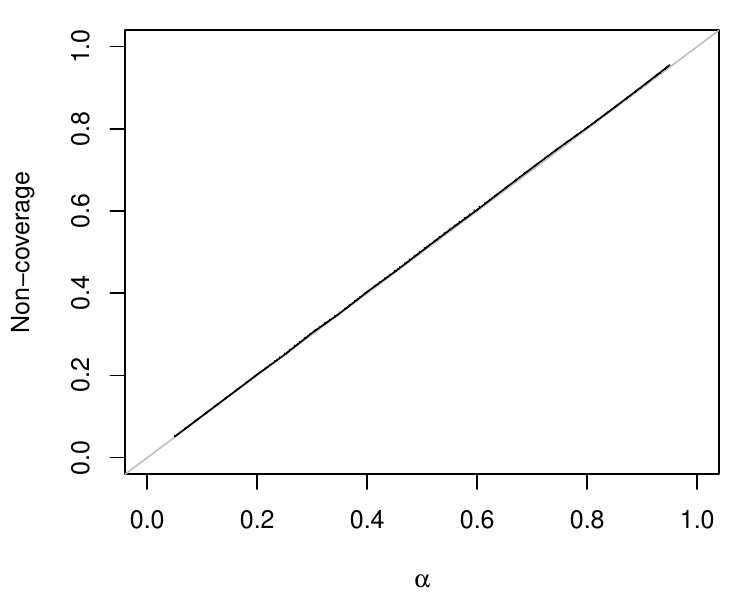}}}
\end{center}
\caption{Plots of three inner probabilistic approximations for the gamma illustration with $n=7$ and $x=14$: uniform kernel (solid), non-uniform kernel with $w \approx 0.45$ (dashed), and the Bayesian posterior distribution (dotted) based on the default prior $\theta \mapsto \theta^{-1}$.  Panel (a) shows the distribution function $t \mapsto \prior_x^\star(\Theta \leq t)$ and Panel~(b) shows the non-credibility function $\alpha \mapsto 1 - \prior_x^\star\{C_\alpha(x)\}$.}
\label{fig:cdf.toy}
\end{figure}

Consider a gamma model $\prob_\theta = \gam(n,\theta)$ for a single observation $X$, with $n$ a fixed constant; the more practical case of $n$ iid exponential samples where $X$ is the sum, the complete sufficient statistic, reduces to this one.  Then the likelihood-based IM has contour 
\[ \pi_x(\theta) = \prob_\theta\bigl\{ (X/\theta)^n e^{-X/\theta} \leq (x/\theta)^n e^{-x/\theta} \bigr\}, \quad \theta > 0. \]
The right-hand side has no simple closed-form expression but, since $X/\theta$ is a pivot under $\prob_\theta$, the computations are straightforward using Monte Carlo.  The contour shown in Figure~\ref{fig:pl.toy} is actually the one above, with $n=7$ and $x=14$.  Clearly, the level sets $C_\alpha(x)$ are intervals, say, $[a(x,\alpha), b(x,\alpha)]$, where $b \geq a$, but with $a(x,1) = b(x,1) = x/n$, the maximum likelihood estimator.  Then the kernel $\kernel_x^\alpha$ associated with an inner probabilistic approximation is a probability distribution supported on the endpoints $a(x,\alpha)$ and $b(x,\alpha)$, i.e., 
\[ \kernel_x^\alpha = w_{x,\alpha} \, \delta_{a(x,\alpha)} + (1-w_{x,\alpha}) \, \delta_{b(x,\alpha)}, \]
where $w_{x,\alpha} \in [0,1]$ and $\delta_z$ denotes a probability measure that puts all its mass at $z$.  The question I want to address here is: what do the inner probabilistic approximations $\prior_x^\star$ look like?  With constant weights, i.e., $w_{x,\alpha} \equiv w$, the argument used in the Gaussian example above can be generalized as follows. For a generic $\theta$, it follows that 
\[ \{ a(x,\alpha) \geq \theta \} \; \text{or} \; \{ b(x,\alpha) \leq \theta \} \iff \pi_x(\theta) \leq \alpha. \]
Using the mixture characterization in \eqref{eq:char}, the distribution function of $\Theta \sim \prior_x^\star$, at $\theta$ is  
\begin{align*}
\prior_x^\star\bigl( (-\infty, \theta] \bigr) & = \int_0^1 w \, 1\{ a(x,\alpha) \leq \theta \} \,d\alpha + \int_0^1 (1-w) \, 1\{ b(x,\alpha) \leq \theta\} \, d\alpha \\
& = \begin{cases} w \, \pi_x(\theta) & \text{if $\theta \leq x/n$} \\ 1 - (1-w) \, \pi_x(\theta) & \text{if $\theta > x/n$}. \end{cases} 
\end{align*}
Two such distribution functions are shown in Figure~\ref{fig:cdf.toy}(a), one based on my recommended choice $w=0.5$ (solid) and another based on $w \approx 0.45$ (dashed).  A third distribution function (dotted) is also shown, namely, that of the Bayesian posterior based on the right Haar prior with density $\theta \mapsto \theta^{-1}$, which is an inverse gamma distribution.  Unlike the flat (right Haar) prior Bayes posterior in the Gaussian example above, this Bayes posterior can't be expressed via a constant-weight kernel (Appendix~B);
the choice $w \approx 0.45$ was my best effort to do so.  But that this Bayes solution is also an inner probabilistic approximation is a general fact established in Section~\ref{SS:group} below.  The three distribution functions are very similar, but not the same.  To confirm that all three are, indeed, inner probabilistic approximations, Figure~\ref{fig:cdf.toy}(b) plots their respective non-credibility functions $\alpha \mapsto 1 - \prior_x^\star\{ C_\alpha(x) \}$.  According to \eqref{eq:inner}, the non-credibility should be equal to $\alpha$, which is confirmed by the plot.  

\section{Properties}
\label{S:properties}

\subsection{Agrees with Bayes in invariant models}
\label{SS:group}


Let $\G$ denote a group of transformations from $\XX$ to $\XX$, with function composition $\circ$ as the binary operation. 
As is customary, I'll write $gx$ for the image of $x \in \XX$ under transformation $g \in \G$.  Since $\G$ is a group, it's associative, i.e., $g_1 \circ (g_2 \circ g_3) = (g_1 \circ g_2) \circ g_3$ for all $g_1,g_2,g_3 \in \G$, it contains the identity transformation, and for every $g \in \G$, there exists an inverse $g^{-1} \in \G$ such that $g \circ g^{-1} = g^{-1} \circ g = \text{identity}$. Examples of transformations include translation, scaling, rotation, permutation, etc.  

The group $\G$ connects to the statistical model as follows.  Suppose that, for each $g \in \G$ and each $\theta \in \TT$, there exists a corresponding $\bar g \theta \in \TT$ such that 
\begin{equation}
\label{eq:invariant}
\prob_\theta(gX \in \cdot) = \prob_{\bar g \theta}(X \in \cdot), \quad (\theta,g) \in \TT \times \G. 
\end{equation}
For example, if the distribution of $X$ depends on a location parameter $\theta$, then the distribution of $X+\tau$ depends on parameter $\theta + \tau$.  When the statistical model $\{\prob_{\theta}: \theta \in \TT\}$ satisfies \eqref{eq:invariant}, it's called an {\em invariant statistical model}---that is, the model for $X$ is also the model for $gX$, with $g \in \G$.  Let $\Gbar$ denote the collection of transformations $\bar g$ on $\TT$, corresponding to the mappings $g \in \G$ and parameters $\theta \in \TT$; it's easy to check that $\Gbar$ is a group too.  More details on equivariant models can be found in \citet{fraser1968}, \citet{eaton1989}, and \citet[][Ch.~6]{schervish1995}, among others.  The necessary technical background is given in Appendix~C
of the supplementary material.

For invariant models, a standard and broadly agreed upon no-prior Bayes solution $\prior_x^\rho$ is that based on combining the likelihood with the so-called right Haar prior $\rho$ using Bayes's formula \citep[e.g.,][]{kass.wasserman.1996}.  This also agrees with the solution obtained via Fisher's fiducial argument, among others; see \citet{lindley1958}.  Under very mild conditions, the right Haar measure $\rho$ and the corresponding Bayes solution $\prior_x^\rho$ are unique.  But there are models, including the multivariate Gaussian models studied in \citet{sun.berger.2007}, that are invariant with respect to multiple groups; see Appendix~C.
The following result, which identifies the right Haar prior Bayes solution as an inner probabilistic approximation to the possibilistic IM, applies to each fixed group with respect to which the model is invariant.  If there's more than one such group, then all of the corresponding right Haar prior Bayes solutions are inner probabilistic approximations. 

\begin{thm}
\label{thm:group}
If the model is invariant with respect to a group $\G$, then, under the conditions laid out in Appendix~C,
the posterior $\prior_x^\rho$ relative to the right Haar prior $\rho$ is an inner probabilistic approximation of $\uPi_x$.  
\end{thm}

\begin{proof}
See Appendix~C
in the supplement.
\end{proof}

It's well-known that, for invariant models, the highest posterior density (with respect to right Haar measure) credible sets are confidence sets.  Theorem~\ref{thm:matching} provides a new proof of this fact, since invariance implies  $\pi_x(\theta) $ is comonotone with $R(x,\theta)$, which is proportional to the posterior density with respect to right Haar measure. 

Theorem~\ref{thm:group} justifies my previous claim that the Bayes posterior in the gamma illustration, in Section~\ref{SS:gamma}, is an inner probabilistic approximation.  


Remember that inner probabilistic approximations are not unique.  So, the right Haar prior Bayes solutions might not exactly match my recommended strategy that defines $\kernel_x^\beta$ to be uniform on $\partial C_\beta(x)$; the illustration in Figure~\ref{fig:cdf.toy} is such a case.  In fact, when there's a Bayesian solution that's an inner probabilistic approximation of the IM, then I might recommend the former since its computation is more familiar than the latter's.  But, like in Figure~\ref{fig:cdf.toy}, the difference between each inner probabilistic approximations will be negligible and will vanish as the information in data increases; see Section~\ref{SS:concentration}. 

\begin{figure}[t]
\begin{center}
\scalebox{0.45}{\includegraphics{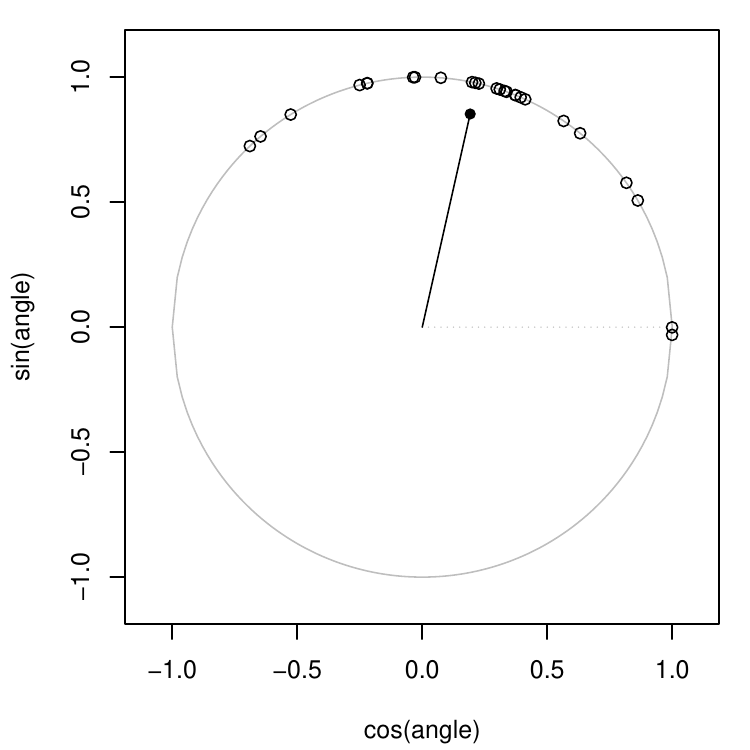}}
\end{center}
\caption{Plot of the directional data analyzed in Section~\ref{SS:group}.  Points on the circle determine the direction.  The small closed circle denotes the sample mean in Cartesian coordinates and the corresponding angle determined by that point, $\hat\theta_x=1.35$, is the maximum likelihood estimator of the mean angle.  The length of that vector, 0.87, is a to-be-conditioned-on ancillary statistic; see Appendix~C.2.
}
\label{fig:wheel}
\end{figure}

\begin{figure}[t]
\begin{center}
\subfigure[Possibilistic IM contour]{\scalebox{0.5}{\includegraphics{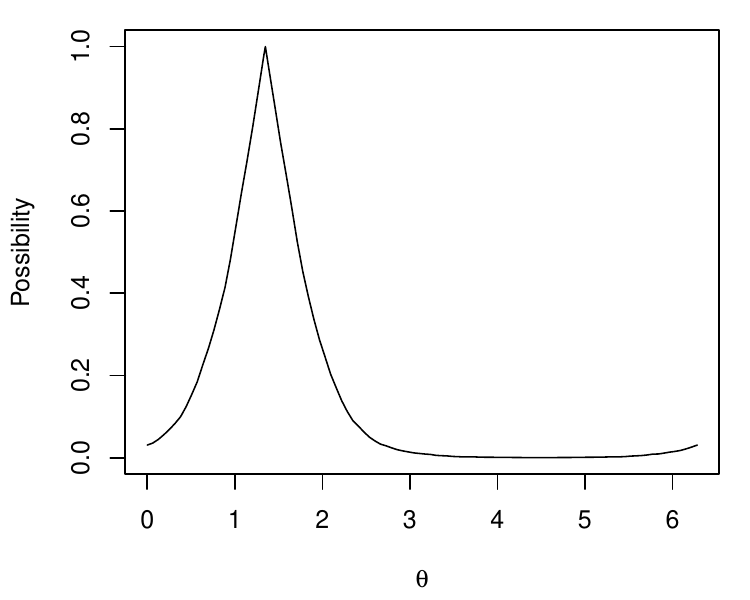}}}
\subfigure[Inner probabilistic approximation]{\scalebox{0.5}{\includegraphics{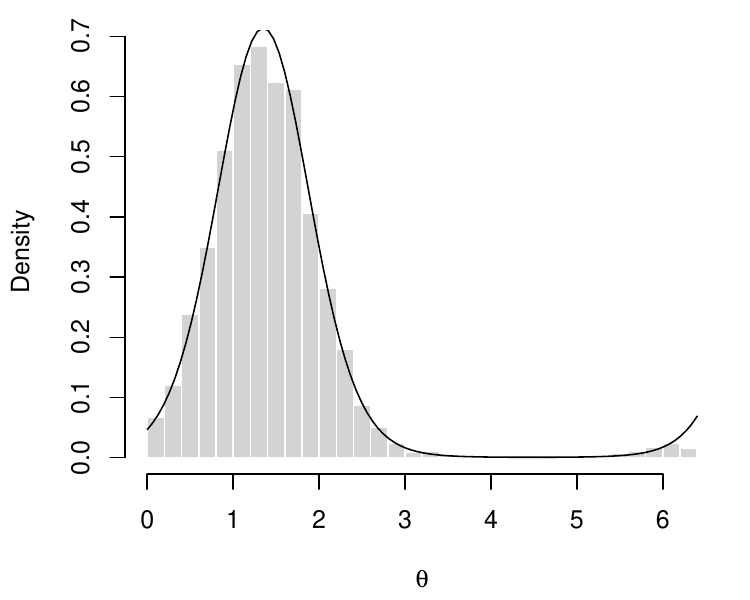}}}
\end{center}
\caption{Analysis of the directional data in Section~\ref{SS:group}.  Panel~(a) shows the possibilistic IM contour and Panel~(b) shows a histogram of samples from the corresponding inner probabilistic approximation with the flat-prior Bayes posterior density overlaid.}
\label{fig:wheel.theta}
\end{figure}

For illustration, suppose that observations correspond to directions---points on a unit circle in the plane---or simply angles with respect to a reference direction.  Real-world applications involving these kind of data include wind and animal movement studies; see \citet{mardia.jupp.book} for a comprehensive treatment.  More generally, direction measurements can be represented by points on the surface of a hypersphere.  This is common in astronomy, where the position of a planet or star can be described by a point on the celestial sphere.
The present example involves $n=25$ such planar directional observations, plotted in Figure~\ref{fig:wheel}.  These data are simulated from the so-called {\em von Mises distribution}, parametrized by a mean direction/angle $\theta$ and a concentration parameter $\kappa$.  Further details about this model and the corresponding solutions are given in Appendix~C.2
in the supplement.  Here I'll take the concentration parameter $\kappa=4$ to be fixed and known.  I'll also interpret the data as angles (radians) so that the parameter $\theta$ acts like a location paramter.  With this simplification, the $\theta$-only von Mises model is invariant with respect to addition modulo $2\pi$.  Therefore, Theorem~\ref{thm:group} implies that the Bayes solution is an inner probabilistic approximation of the possibilistic IM.  Figure~\ref{fig:wheel.theta} shows, first, a plot of the possibilistic IM's contour for these data, with its peak at the maximum likelihood estimator, $\hat\theta_x = 1.35$ and, second, a histogram of samples drawn from the inner probabilistic approximation (with $\kernel_x^\beta$ a uniform distribution) and the no-prior Bayesian posterior density overlaid.  The visual similarities provide confirmation of the claim in Theorem~\ref{thm:group}.  

\subsection{Asymptotic normality and efficiency}
\label{SS:concentration}


Next is a version of the celebrated {\em Bernstein--von Mises theorem} for the possibilistic IM, establishing its asymptotic normality and efficiency.  That is, for large $n$, the possibilistic IM's contour resembles a Gaussian possibility contour, defined below, with covariance matrix that matches the Cram\'er--Rao lower bound.  This implies a traditional Bernstein--von Mises theorem for the proposed inner probabilistic approximation and, therefore, it agrees with any other reasonable no-prior Bayesian solution asymptotically.  Moreover, this near-Gaussian form of the IM output offers some valuable insights and simplifications relevant to computing the inner probabilistic approximation; see Section~\ref{S:computation}.  The first order of business is to define Gaussian possibility.  

Let $g_{m,v}$ denote the $D$-dimensional Gaussian probability density function, parametrized by a mean vector $m \in \RR^D$ and a covariance matrix $v \in \RR_+^{D \times D}$.  Define the corresponding Gaussian possibility measure as the outer possibilistic approximation of $\nm_D(m, v)$, the probability-to-possibility transform applied to the aforementioned Gaussian distribution.  The corresponding possibility contour is 
\[ \gamma_{m, v}(y) = \prob\{ g_{m, v}(Y) \leq g_{m, v}(y) \}, \quad Y \sim \nm_D(m, v), \]
which can easily be rewritten as
\begin{align*}
\gamma_{m,v}(y) 
& = 1 - F_D\bigl( (y-m)^\top v^{-1} (y - m) \bigr), 
\end{align*}
where $F_D$ is the $\chisq(D)$ distribution function.  Write $\uGamma_{m,v}$ for the corresponding possibility measure obtained by maximizing the contour as described in Section~\ref{SS:possibility}.  Other authors have used a different form of Gaussian possibility \citep[e.g.,][]{denoeux.fuzzy.2022, denoeux.fuzzy.2023}, but I find the one above appealing and convenient because of the following duality between $\nm_D(m,v)$ and $\uGamma_{m,v}$: the former is an inner probabilistic approximation of the latter, and the latter is the outer possibilistic approximation of the former.  

For some intuition, consider a simple Gaussian location problem, $X \sim \prob_\Theta = \nm_D(\Theta, v)$, with known covariance matrix $v$.  Then the relative likelihood is $R(x,\theta) = \exp\{ -\frac12 (x-\theta)^\top v^{-1} (x-\theta) \}$ and the IM's contour is 
\begin{align*}
\pi_x(\theta) & = \prob_\theta\{ R(X,\theta) \leq R(x,\theta) \} \\
& = 1-F_D\bigl( (x-\theta)^\top v^{-1} (x-\theta) \bigr), \quad \theta \in \RR^D.
\end{align*}
This is exactly the Gaussian possibility contour with mean $x$ and covariance matrix $v$.  More generally, by the famous theorem of \citet{wilks1938}, the relative likelihood $R(X^n,\theta)$, as a function of iid samples $X^n=(X_1,\ldots,X_n)$ from $\prob_\theta$, will be asymptotically distributed as $\exp\{-\frac12 \chisq(D)\}$.  Then the following approximation is expected when $n$ is large,
\[ \pi_{x^n}(\theta) \approx 1 - F_D\bigl( (\hat\theta_{x^n} - \theta)^\top J_{x^n} (\hat\theta_{x^n} - \theta) \bigr), \]
where $\hat\theta_{x^n}$ is the maximum likelihood estimator and $J_{x^n}$ is the observed Fisher information matrix.  The reader will notice that the right-hand side above is exactly the Gaussian possibility contour with mean vector $\hat\theta_{x^n}$ and covariance matrix $J_{x^n}^{-1}$, hence the connection between the possibilistic IM and the Gaussian possibility is very natural and holds quite generally, at least approximately.  Making this connection rigorous requires care, both in formulating the proper regularity conditions and in suitably bounding all the error terms.  Along these lines, under regularity conditions comparable to those advanced in \citet{lecam1970}, 
\citet{imbvm.ext} proved the following Bernstein--von Mises theorem for the possibilistic IM.  

\begin{thm}[\citealt{imbvm.ext}]
\label{thm:bvm}
If $X^n=(X_1,\ldots,X_n)$ consists of iid samples from $\prob_\Theta$, and if the model satisfies the regularity conditions presented in Appendix~D in the supplementary material,
as $n \to \infty$, the possibilistic IM's contour $\pi_{X^n}$ satisfies 
\[ \sup_{\theta \in \mathcal{T}} \bigl| \pi_{X^n}(\theta) - \gamma_{X^n}(\theta) \bigr| \to 0 \quad \text{in $\prob_\Theta$-probability}, \]
where $\mathcal{T}$ is an arbitrary compact subset of $\TT$, 
\[ \gamma_{X^n}(\theta) := \gamma_{\Theta + n^{-1/2}\Delta_\Theta(X^n), (n I_\Theta)^{-1}}(\theta) \]
is the centered and scaled Gaussian possibility contour, $I_\Theta$ the Fisher information matrix, and $\Delta_\Theta(X^n)$ a sequence that converges in distribution to $\nm_D(0, I_\Theta^{-1})$.  In addition, if the likelihood is twice differentiable, then, with $\prob_\Theta$-probability converging to 1 as $n \to \infty$, 
\[ \sup_{u \in \mathcal{U}} \bigl| \pi_{X^n}(\hat\theta_{X^n} + J_{X^n}^{-1/2} u) - \gamma(u) \bigr| \to 0, \]
where $\mathcal{U}$ is a fixed but arbitrary compact subset of $\RR^D$, and $\gamma$ is the standard Gaussian possibility with zero mean and identity covariance matrix.
\end{thm}

Among other things, Theorem~\ref{thm:bvm} implies consistency: $\uPi_{X^n}(H) \to 0$ in $\prob_\Theta$-probability for any $H$ with $\Theta \not\in H$.  Moreover, since the covariance matrix in the Gaussian limit agrees with the Cram\'er--Rao lower bound, Theorem~\ref{thm:bvm} implies that the possibilistic IM is asymptotically efficient and, hence, there's no loss of efficiency associated with the IM's exact validity (and imprecision).  \citet{imbvm.ext} give numerical illustrations that highlight the accuracy of the Gaussian approximation.

A consequence of 
Theorem~\ref{thm:bvm} is that uniform-kernel inner probabilistic approximations have Gaussian limits just like in the classical Bernstein--von Mises theorem.  

\begin{thm}
\label{thm:new.bvm}
Under the conditions of Theorem~\ref{thm:bvm}, if $\prior_{X^n}^\star$ is the uniform-kernel inner probabilistic approximation of the possibilistic IM $\uPi_{X^n}$ then, as $n \to \infty$, 
\[ 
\sup_H \bigl| \prior_{X^n}^\star(H) - \Gauss_{X^n}(H) \bigr| \to 0, \quad \text{in $\prob_\Theta$-probability}, \]
where $\Gauss_{X^n} = \nm_D(\Theta + n^{-1/2} \Delta_\Theta(X^n), (n I_\Theta)^{-1})$ is the approximating Gaussian distribution and the supremum is over all Borel subsets $H$ of $\TT$.  
\end{thm}

\begin{proof}
See Appendix~D
in the supplement.
\end{proof}

I'll give a quick proof sketch for scalar $\Theta$.  As in Section~\ref{SS:gamma}, 
using the uniform kernel with weight $w=\frac12$, the distribution function of $\prior_{x^n}^\star$ is  
\begin{align*}
\prior_{x^n}^\star\bigl( (-\infty, \theta] \bigr) 
& = \begin{cases} \tfrac12 \, \pi_{x^n}(\theta) & \text{if $\theta \leq \hat\theta_{x^n}$} \\ 1 - \tfrac12 \, \pi_{x^n}(\theta) & \text{if $\theta > \hat\theta_{x^n}$}. \end{cases} 
\end{align*}
Switching to local coordinates, i.e., $\theta = \hat\theta_{x^n} + J_{x^n}^{-1/2} z$ for $z$ fixed, the case $\theta \leq \hat\theta_{x^n}$ corresponds to $z \leq 0$ and Theorem~\ref{thm:bvm} implies that, with $Z \sim \nm(0,1)$,  
\begin{align*}
\tfrac12 \, \pi_{X^n}(\hat\theta_{X^n} + J_{X^n}^{-1/2} z) & \to \tfrac12 \, \gamma(z) \\
& = \tfrac12 \, \prob(Z^2 \geq z^2) \\
& = \prob(Z \leq z). 
\end{align*}
Similarly, if $z > 0$, then $1-\tfrac12 \, \pi_{X^n}(\hat\theta_{X^n} + J_{X^n}^{-1/2} z) \to \prob(Z \leq z)$.  Therefore, with probability tending to 1, the distribution of $J_{X^n}^{1/2}(\Theta - \hat\theta_{X^n})$, under $\Theta \sim \prior_{X^n}^\star$, is converging to a standard normal, just like in the classical Bernstein--von Mises theorem.  

A point that follows immediately from the above sketch is that assigning probability $w=\frac12$ to the boundary points in the inner probabilistic approximation is crucial; of course, any choice that assigns weights that converges to $\frac12$ works fine too.  More generally, since $\Gauss_{X^n}$ is an inner probabilistic approximation to the possibility measure with contour $\gamma_{X^n}$, and corresponds to taking a uniform kernel on the level sets of $\gamma_{X^n}$, it's necessary that the choice of kernel defining $\prior_{X^n}^\star$ at least be asymptotically uniform.  
This adds further support for my general, uniform-kernel recommendation in the inner probabilistic approximation.  But non-uniform kernels have a potential roll to play too, as shown in Section~\ref{SS:caveats} next.


\subsection{Marginalization risks and remedies}
\label{SS:caveats}




Except for the impossibly rare cases where a genuine prior distribution is available, reliable statistical inference is inherently imprecise---those familiar test and confidence procedures that control error rates all have an imprecise probabilistic characterization \citep{imchar}.  The point is that there's no single probability distribution that can reliably quantify uncertainty about the unknowns in a statistical model.  So, insisting that uncertainty quantification be probabilistic has risks:
\begin{quote}
[\citet{xie.singh.2012}] 
are thus recommending that we ignore the restriction to confidence sets or equivalent, and free confidence to allow the production of parameter distributions. Certainly distributions are easier to think about, are largely in accord with Fisher’s original proposal, and are more in the freedom of the Bayes approach, but they do overlook inherent risks... 
\citep{fraser.cd.discuss}
\end{quote}


Fraser's ``risks'' primarily concern marginalization \citep[e.g.,][]{dawid.stone.zidek.1973, fraser2011, balch.martin.ferson.2017}.  Roughly, whatever desirable properties a no-prior Bayes solution for $\Theta$ has, the corresponding no-prior Bayes solution for $\Phi=m(\Theta)$ obtained by probabilistic marginalization generally doesn't share those desirable properties.  
When \citet{schweder.hjort.cd.discuss} warn that ``joint [confidence distributions] should not be sought, we think, since they might easily lead the statistician astray,'' they're concerned about users being tempted to do these familiar probabilistic operations, thus creating a risk of unreliability.  
\citet{fraser2011} and \citet{fraser.etal.2016} highlight non-linearity in $m$ as a major contributor to this risk; see, also, \citet{martin.belief2024} and the discussion below.  

It's important to emphasize that {\em no probabilistic uncertainty quantification is safe from all such risks}, including my proposed inner probabilistic approximations.  So, the goal of the present section is simply to understand how marginalization with these approximations works and to identify which marginal inferences are safe.  To be safe from all such risks, one must break from the familiar probabilistic uncertainty quantification in one way or another: \citet{grunwald.safe} suggests explicitly restricting one's probabilistic inferences to those that are safe, whereas I've suggested relaxing from probabilities to possibilities so that all inferences are safe \citep{imreview}.

A quantification of uncertainty about $\Theta$ can be transformed to a quantification of uncertainty about $\Phi = m(\Theta)$.
First, the image $C_\alpha^m(x) = m\{ C_\alpha(x) \}$ of a confidence set $C_\alpha(x)$ for $\Theta$ is a confidence set for $\Phi$ because 
\begin{equation}
\label{eq:f.ci}
\Theta \in C_\alpha(x) \implies m(\Theta) \in C_\alpha^m(x). 
\end{equation}
The opposite implication generally doesn't hold, however.  So, while the transformed confidence set for $\Phi$ maintains the nominal coverage probability enjoyed by the original confidence for $\Theta$, it can be conservative; see Section~\ref{S:bf}.  

Next, take a no-prior Bayes posterior $\prior_x$ for $\Theta$ and map it to the corresponding marginal posterior $\prior_x^m$ for $\Phi=m(\Theta)$ via the probability calculus.  By \eqref{eq:f.ci}, however:
\begin{equation}
\label{eq:ocd}
\prior_x^m\{ C_\alpha^m(x) \} \geq \prior_x\{ C_\alpha(x) \}, \quad \alpha \in [0,1]. 
\end{equation}
That is, $\prior_x^m$ is inclined to assign more of its mass to the marginal confidence sets than $\prior_x$ assigned to the original confidence sets.  So, even if $\prior_x$ is a genuine confidence distribution, i.e., it assigns probability $1-\alpha$ to $C_\alpha(x)$ for each $\alpha \in [0,1]$, the marginal $\prior_x^m$ has a risk of being over-confident and, hence, anti-conservative. Illustrations of this over-confidence phenomenon are presented in Appendix~E of the supplement.
A relevant question is: if and how the above-defined inner probabilistic approximation can mitigate this risk?

Let $\uPi_x$ be the possibilistic IM as described above, quantifying uncertainty about $\Theta$, given $X=x$.  One way---but not the only way, see Section~\ref{S:bf}---to carry out possibilistic marginalization is by applying the general {\em extension} operation \citep[e.g.,][]{zadeh1975d, zadeh1978}.  Extension yields a marginal IM $\uPi_x^m$ for $\Phi=m(\Theta)$ with possibility contour 
\begin{equation}
\label{eq:mpi.ex}
\pi_x^m(\phi) = \sup_{\theta: m(\theta) = \phi} \pi_x(\theta), \quad \phi \in m(\TT). 
\end{equation}
For $C_\alpha(x)$ determined by the contour $\pi_x$ as in \eqref{eq:region}, its image $C_\alpha^m(x) = m\{C_\alpha(x)\}$ can be expressed as 
\[ C_\alpha^m(x) = \{m(\theta): \pi_x(\theta) \geq \alpha\} = \{ \phi: \pi_x^m(\phi) \geq \alpha \}. \]
That the possibilistic marginalization strategy appears organically in the common-sense treatment of confidence sets is no coincidence; see \citet{imchar}.  By Theorem~\ref{thm:char},  $\prior_x \in \cred(\uPi_x)$ is a mixture of kernels supported on the the confidence sets, so the derived marginal distribution $\prior_x^m$ of $\Phi=m(\Theta)$ has a similar mixture form; see the proof of Theorem~\ref{thm:cd} below.  But it's generally not true that the marginal distribution $\prior_x^{\star m}$ derived from an inner probabilistic approximation $\prior_x^\star$ of $\uPi_x$ via probability calculus is an inner probabilistic approximation of the extension-based marginal IM $\uPi_x^m$.  A relevant question is: when do the above two marginal distributions for $\Phi$ agree? 

Define $m$ to be {\em boundary-surjective} (with respect to the given $C_\alpha$'s) if every point in $\partial C_\alpha^m(x)$ corresponds to a point in $\partial C_\alpha(x)$, i.e., if $m^{-1}\{\partial C_\alpha^m(x)\} \subseteq \partial C_\alpha(x)$.  Linear functions are boundary-surjective, and an example to keep in mind is a two-dimensional case with $m(\theta_1,\theta_2) = \theta_1$ and, say, ellipsoidal confidence sets.  Then $m$ projects the two-dimensional confidence set to an interval the $\theta_1$ axis, and its endpoints correspond to the extreme tips of the joint confidence set in the $\theta_1$ direction.  But note that linearity isn't necessary, since $m(\theta_1, \theta_2) = \theta_1^2$ is also boundary-surjective in this case.  

\begin{thm}
\label{thm:cd}
If $m$ is boundary-surjective, then there exists an inner probabilistic approximation $\prior_x^\star$ of the possibilistic IM $\uPi_x$ for $\Theta$ such that the corresponding marginal distribution $\prior_x^{\star m}$ is an inner probabilistic approximation of the extension-based marginal IM for $\Phi=m(\Theta)$.  Furthermore, the aforementioned $\prior_x^\star$ has kernel $\kernel_x^\alpha$ fully supported on $m^{-1}\{ \partial C_\alpha^m(x) \} \subseteq \partial C_\alpha(x)$. 
\end{thm}

\begin{proof}
See Appendix~E of the supplement.
\end{proof}

\begin{figure}[t]
\begin{center}
\subfigure[Marginal IM contour]{\scalebox{0.4}{\includegraphics{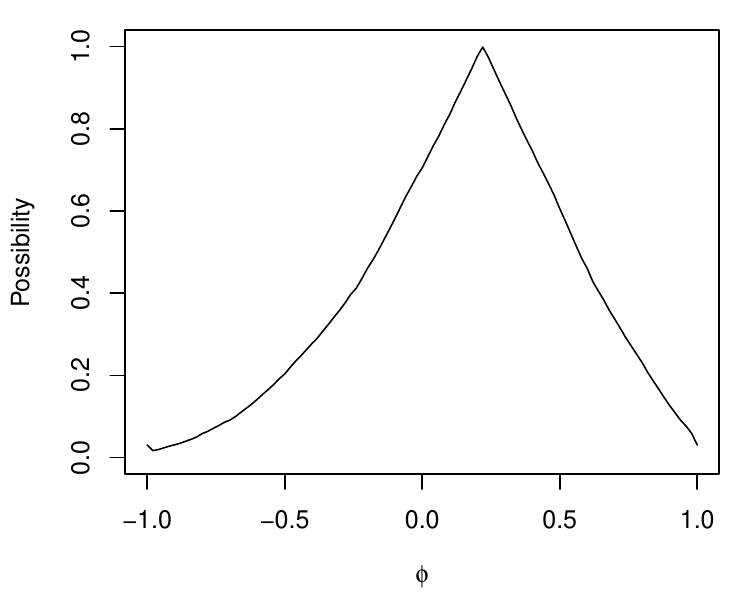}}}
\subfigure[Bayes marginal]{\scalebox{0.4}{\includegraphics{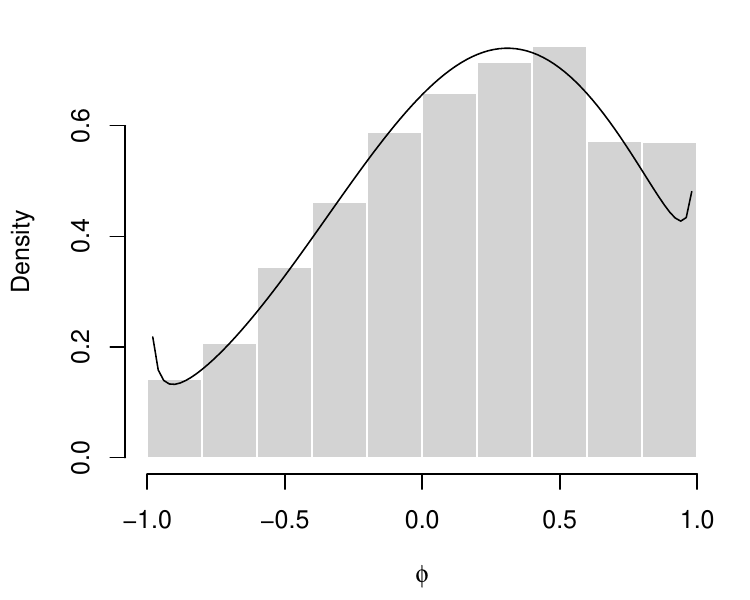}}}
\subfigure[Non-credibility]{\scalebox{0.4}{\includegraphics{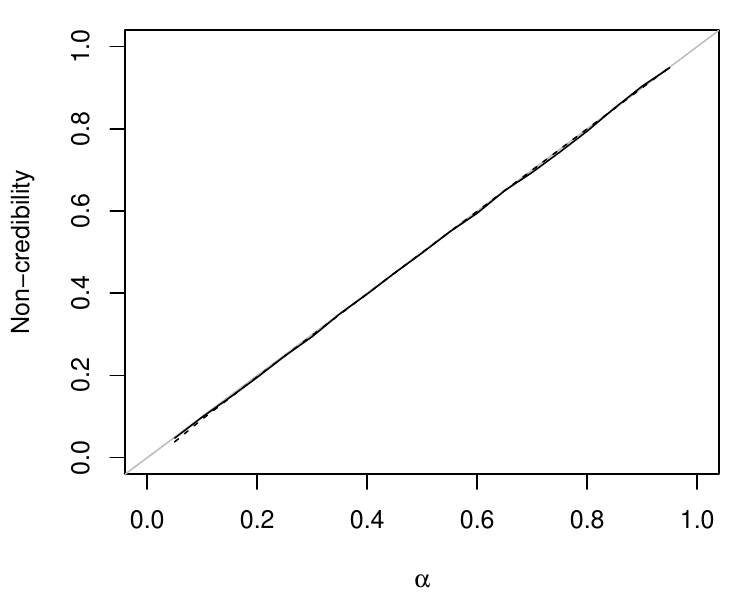}}}
\end{center}
\caption{Marginal inference results for $\Phi=\cos(a\Theta)$, with $a=1$, in the directional data example.  Panel~(a) shows the extension-based marginal IM contour for $\Phi$; Panel~(b) shows the Bayesian marginal posterior distribution for $\Phi$; and Panel~(c) shows the non-credibility function for the marginal Bayesian posterior (dashed) and the marginal IM's inner probabilistic approximation (solid).}
\label{fig:marg1}
\end{figure}

As an illustration, I'll revisit the directional data example presented in Section~\ref{SS:group} above, where the quantity of interest is $\Phi = \cos(a \Theta)$, for two different values of $a$.  Start with the case of $a=1$.  Since the Bayes/fiducial distribution---which is also an inner probabilistic approximation of the possibilistic IM by Theorem~\ref{thm:group}---is almost fully supported on the interval $(0,\pi)$, where $\theta \mapsto \cos(\theta)$ is one-to-one, this mapping is boundary-surjective.  In this one-dimensional case, Theorem~\ref{thm:cd} implies that the marginal Bayes/fiducial distribution for $\Phi$ is an inner probabilistic approximation of the marginal possibilistic IM for $\Phi$.  Figure~\ref{fig:marg1} shows, for $a=1$, the extension-based marginal IM contour for $\Phi$, samples of the marginal Bayes/fiducial with the corresponding density function overlaid, and the non-credibility function $\alpha \mapsto 1 - \prior_x^m\{ C_\alpha^m(x) \}$, for $\prior_x^m$ equal to the Bayes/fiducial marginal distribution and to the inner probabilistic approximation of the extension-based marginal IM.  The key observation is that the contour and the posterior distribution have similar shapes and the non-credibility function behaves exactly as expected of an inner probabilistic approximation, according to the definition \eqref{eq:inner}.  

\begin{figure}[t]
\begin{center}
\subfigure[Marginal IM contour]{\scalebox{0.4}{\includegraphics{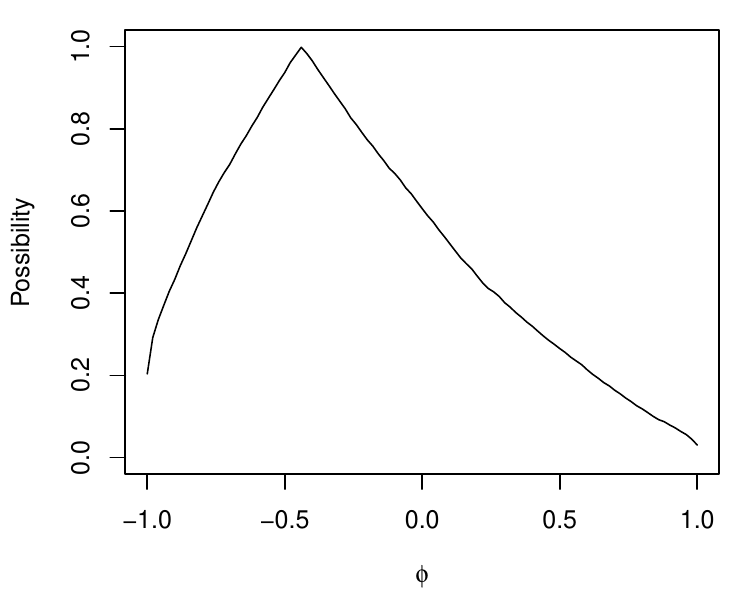}}}
\subfigure[Bayes marginal]{\scalebox{0.4}{\includegraphics{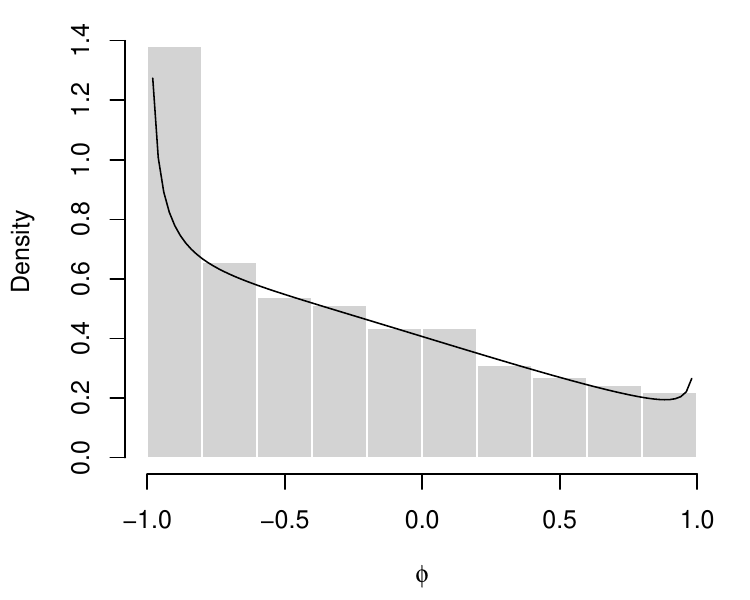}}}
\subfigure[Non-credibility]{\scalebox{0.4}{\includegraphics{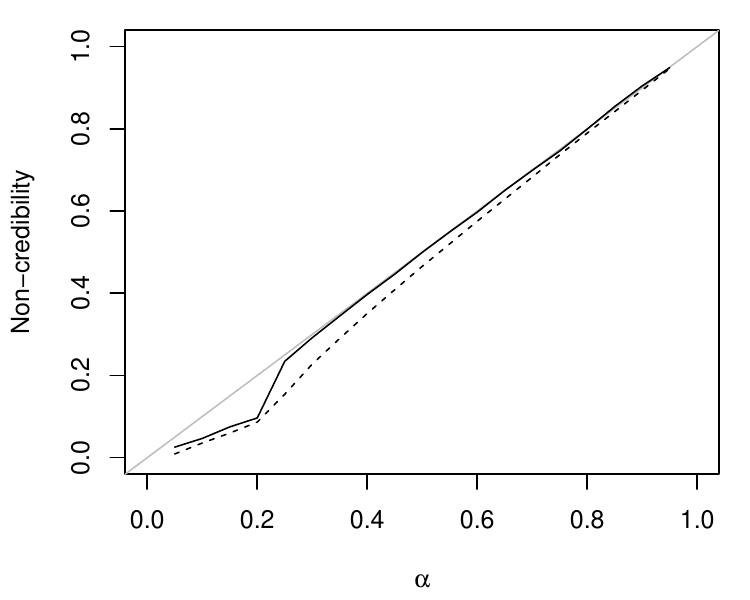}}}
\end{center}
\caption{Same as in Figure~\ref{fig:marg1}, just now for $a=1.5$.
}
\label{fig:marg2}
\end{figure}

Next, take $a=1.5$.  Then the boundary-surjective property no longer holds and, surprisingly, this small change in the quantity of interest has non-negligible impact on the results.  Figure~\ref{fig:marg2} shows the same plots as in Figure~\ref{fig:marg1}, this time for $a=1.5$.  The key observation is that, while the general shape of the marginal IM contour and the Bayes posterior distribution were similar in Figure~\ref{fig:marg1}, they are quite different in Figure~\ref{fig:marg2}; this is a consequence of the entirely different style of marginalization, from $\Theta$ to $\Phi$, with the IM's possibilistic marginalization preserving the validity properties enjoyed by both approaches' inference on $\Theta$ whereas the Bayesians' probabilistic marginalization doesn't.  The IM's inner probabilistic approximation (not shown) looks similar to the Bayesian posterior, but it's not the same.  Indeed, Figure~\ref{fig:marg2}(c) shows the non-credibility for the Bayes posterior distribution and the IM's inner probabilistic approximation, and there are clear signs of the Bayesians' over-confidence.  The IM's inner approximation also shows (milder) signs of over-confidence, but there's a perfectly reasonable explanation for this.  Recall from Section~\ref{S:inner} that exact inner probabilistic approximations need not exist, e.g., if the possibility contour is bounded away from 0.
In Figure~\ref{fig:marg2}(a), note that the contour is in fact bounded away from 0 and, consequently, the sets $C_\alpha^m(x)$ are equal to the full space $[-1,1]$ for sufficiently small $\alpha$.  Since any probability distribution $\prior_x^m$ supported on $[-1,1]$ must assign probability $1 > 1-\alpha$ to $C_\alpha^m(x) = [-1,1]$ for all sufficiently small $\alpha$, there is no $\prior_x^\star$ that satisfies \eqref{eq:inner} for all $\alpha$.  But the IM's {\em pseudo inner probabilistic approximation}---``pseudo'' because the inner probabilistic approximation strictly doesn't exist---quickly corrects for this over-confidence at small $\alpha$ and achieves the objective \eqref{eq:inner} for all $\alpha$ that are not too small.  

As an aside, recall that, in the one-dimensional case, the confidence sets $C_\alpha(x)$ in \eqref{eq:region} are typically two-sided intervals, as determined by the shape of the likelihood.  But what if only one-sided inference was desired, akin to the use of lower or upper confidence limits?  This amounts to imposing a constraint on the shape of the possibilsitic IM contour.  One way---but not the only or best way---is to treat this through the lens of marginalization.  Define a  shape-constrained ``marginal'' contour $\varpi_x(\phi) := \sup_{\theta: \theta \leq \phi} \pi_x(\theta)$, which is non-decreasing in $\phi$.  Then the level set $\{\phi: \varpi_x(\phi) \geq \alpha\}$ is a half-line of the form $[\varpi_x^{-1}(\alpha), \infty)$.  The corresponding inner probabilistic approximation for the lower bound $\Phi$ has distribution function $\phi \mapsto \varpi_x(\phi)$, which obtains by marginalizing a special inner probabilistic approximation for $\Theta$: interestingly, it's the one with an extreme choice of weight, $w=1$, analogous to those degenerate kernels supported on proper subsets of the boundary that appeared in Theorem~\ref{thm:cd}.  The corresponding ``marginal'' confidence limits are generally very conservative, so I would recommend a more efficient profiling strategy akin to that described in Section~\ref{S:bf} below.  The full details are beyond the scope of the present paper, but this is a focus of ongoing investigations and the results will be presented elsewhere.

I'll conclude this section by putting result in Theorem~\ref{thm:cd} in context.  It's a negative result in the sense that to achieve consistency between the marginals derived from an IM's inner probabilistic approximations and the corresponding inner approximations of the extension-based marginal IM requires the former to be based on an extreme choice of mixture kernel, one specifically tailored to the margin $\Phi = m(\Theta)$ and at odds with the general recommendation of a uniform kernel that was shown to be asymptotically efficient in Theorem~\ref{thm:new.bvm}.  That said, the marginal confidence sets $C_\alpha^m$ tend to be conservative, so the over-confident marginal $\prior_x^{\star m}$ derived from a more natural inner probabilistic approximation $\prior_x^\star$ might itself still be conservative; see Section~\ref{S:bf}.  Moreover, Theorem~\ref{thm:cd} can also be interpreted as a positive result for possibilistic marginalization.  That is, the marginalization risks associated with no-prior Bayes solutions can be avoided by first finding a marginal IM for $\Phi$, either by extension as above or (preferably) by profiling as in Section~\ref{S:bf}, and then finding a corresponding inner probabilistic approximation.  This incompatibility between confidence/calibration and probabilistic marginalization isn't new or unfamiliar, and the need for non-probabilistic resolutions is standard: for example, Bayesians choose default prior distributions for $\Theta$ tailored to the particular target $\Phi$ \citep[e.g.,][]{bernardo1979}.  What the IM framework offers is a principled way to manage the marginalization risks and, if desired, one can approximate the possibility by a probability after have the marginalization risks have been mitigated.


\section{Computation}
\label{S:computation}

Until recently, only naive/inefficient strategies for computing the IM contour were available.  In particular, the go-to strategy was to approximate $\pi_x$ via 
\begin{equation}
\label{eq:pi.naive}
\pi_{x}(\theta) \approx \frac1M \sum_{m=1}^M 1\{ R(X_{m,\theta}, \theta) \leq R(x, \theta) \},
\end{equation}
where $X_{m,\theta}$ are independent copies of the data $X$, drawn from $\prob_\theta$, for $m=1,\ldots,M$.  The above computation is feasible at a few different $\theta$ values, but all the practically relevant calculations---e.g., identifying the confidence set in \eqref{eq:region}---require evaluation over a sufficiently fine grid covering the relevant portions of $\TT$, and this is expensive.  Recently, \citet{immc} developed a new and more efficient strategy for doing the relevant IM computations that replaces most of the naive contour evaluations as in \eqref{eq:pi.naive} with Monte Carlo sampling from a sort of ``posterior distribution.''  That distribution is, in fact, the inner probabilistic approximation $\prior_x^\star$ of $\uPi_x$ defined in Section~\ref{S:inner} above.  Therefore, the same computational strategy for (approximately) sampling from $\prior_x^\star$ advanced in \citet{immc} can be reIMagined here as an implementation of the new no-prior Bayes solution.

The strategy in \citet{immc} is, roughly, to stitch together a suitable collection of Gaussian possibilistic IMs.  While the ``stitched'' collection of Gaussian possibilities isn't a Gaussian possibility, the use of a basic Gaussian building block can be justified by the asymptotic normality result in Section~\ref{SS:concentration}.  Interest here is solely in the inner probabilistic approximation, so it's not necessary for me to describe all the details involved in the ``stitched IM.'' For $\sigma$ a generic $D$-vector of positive numbers, consider a collection of Gaussian confidence/level sets 
\[ C_\alpha^\sigma(x) = \{ \theta: (\theta - \hat\theta_x)^\top \, J_x^\sigma \, (\theta - \hat\theta) \leq \chi_{D,1-\alpha}^2\}, \]
where $\chi_{D,1-\alpha}^2$ is the $1-\alpha$ quantile of the $\chisq(D)$ distribution, and $J_x^\sigma$ is an embellished version of the observed Fisher information matrix with spread at least partially controlled by the parameters in $\sigma$.  Specifically, if $J_x$ has spectral decomposition $J_x = E \Lambda E^\top$, then the embellished version $J_x^\sigma$ is defined as 
\begin{equation}
\label{eq:J.xi}
J_x^\sigma = E \, \text{diag}(\sigma^{-1}) \, \Lambda \, \text{diag}(\sigma^{-1}) \, E^\top, 
\end{equation}
where $\text{diag}(\sigma^{-1})$ is the diagonal matrix with elementwise reciprocal of the $\sigma$ vector on the diagonal.  Then $\sigma$ controls confidence set's size by manipulating the eigenvalues of the underlying observed Fisher information matrix, e.g., a large (resp.~small) entry in $\sigma$ stretches (resp.~contracts) the level set in the direction of the corresponding eigenvector.  Theorem~\ref{thm:bvm} says that the original $C_\alpha(x)$ will be roughly elliptical, so it's safe to assume that there exists a vector $\sigma$ such that the following inclusion holds snugly:
\begin{equation}
\label{eq:inclusion}
C_\alpha(x) \subseteq C_\alpha^\sigma(x). 
\end{equation}
Of course, this vector $\sigma$ generally depends on $(x,\alpha)$, so I'll denote this as $\sigma(x,\alpha)$.  A cartoon depiction of the proposed strategy is shown in Figure~\ref{fig:puff}.  The motivation for bounding $C_\alpha(x)$ inside $C_\alpha^\sigma(x)$ is that it's not possible to describe the former exactly but the latter is very simple, at least for fixed $\sigma$.  The data-driven algorithm recommended in \citet{immc}, relying on the foundations laid in \citet{imvar.ext}, for selecting $\sigma(x,\alpha)$ is presented in Appendix~F in the supplement.
Once this $\sigma(x,\alpha)$ has been determined, the inner probabilistic approximation uses the kernel 
\[ \kernel_x^\alpha = \unif\{ \partial C_\alpha^{\sigma(x,\alpha)}(x) \}, \quad \alpha \in (0,1). \]
Thanks to the direction of the inclusion in \eqref{eq:inclusion}, the proposed implementation of the inner probabilistic approximation is conservative.  So, the relevant properties enjoyed in theory by the inner probabilistic approximation of the IM should be at least approximately enjoyed by the practical proposal above; see Section~\ref{S:bf}.  

\begin{figure}[t]
\begin{center}
\scalebox{0.85}{
\begin{tikzpicture}
\node [cloud, cloud puffs=9, draw, color=black!20, fill=black!5, minimum width=6cm, minimum height=3.5cm] at (0, 0) {};
\draw (0, 0) ellipse (3.05cm and 1.8cm);
\filldraw[color=black, fill=black] (0,0) circle (1.5pt) node[anchor=west]{$\hat\theta_x$};
\draw (-0.75, 0) circle (0pt) node[anchor=east]{$C_\alpha(x)$}; 
\draw[dashed] (0,0) ellipse (4.5 and 1.05);
\draw (2.35, 1.4) circle (0pt) node[anchor=west]{$\partial C_\alpha^{\sigma(x,\alpha)}(x)$}; 
\end{tikzpicture}
}
\end{center}
\caption{Approximating the IM contour's $\alpha$-level set $C_\alpha(x)$ (gray) by an ellipse $C_\alpha^\sigma(x)$ with a ``good'' choice of $\sigma$ (solid) and with a ``bad'' choice of $\sigma$ (dashed).}
\label{fig:puff}
\end{figure}





\section{Example: Behrens--Fisher problem}
\label{S:bf}


Few examples have deeper practical and historical relevance than the {\em Behrens--Fisher problem} \citep{fisher1935a, fisher1939}.  The original problem is simple to state: independent samples of size $n_1$ and $n_2$, respectively, are available from two distinct normal populations, $\nm(\Theta_{11}, \Theta_{12}^2)$ and $\nm(\Theta_{21}, \Theta_{22}^2)$, with $\Theta=(\Theta_{11}, \Theta_{12}, \Theta_{21}, \Theta_{22})$ unknown, and the goal is marginal inference on the difference $\Phi = m(\Theta) = \Theta_{21}-\Theta_{11}$ of the two means.  The problem is straightforward if the two variances are known or if their ratio is known, but the fully unknown variance case has remained elusive---lots of candidate solutions are available but there's no consensus on which solution is ``right'' or ``best.''  See \citet{kimcohen1998} for a review. 

By far the most widely used solution to the Behrens--Fisher problem is the modification of the basic Student-t pivot with the degrees of freedom approximations due to \citet{welch1938}; this is implemented in R's {\tt t.test}.  Other standard approaches include the simple-but-conservative solution in 
\citet{scheffe1970} and the Bayesian solution proposed by \citet{jeffreys1940}, based on the right Haar prior for $\Theta$, which is  equivalent to Fisher's fiducial solution. Ironically, the solution proposed by Jeffreys is different from the Bayesian solution based on the Jeffreys prior.  

Start by considering inference on the full parameter $\Theta$.  This model has lots of structure so the relative likelihood-based, possibilistic IM construction for inference on $\Theta$ is conceptually and computationally straightforward.  Thanks to the model's underlying affine group invariance, it follows from Theorem~\ref{thm:group} that the Bayesian posterior distribution for $\Theta$ based on the right Haar prior, which is also Fisher's fiducial distribution, is also an inner probabilistic approximation of the IM for $\Theta$.  The corresponding marginal distribution for $\Phi$ is not an inner probabilistic approximation of the extension-based marginal IM, but the latter is sufficiently conservative that even this over-confident marginal itself is conservative.  


The approach I'm proposing here first carries out a different and more efficient IM marginalization strategy, based on validifying the profile relative likelihood.  This idea was first developed in \citet{martin.partial3} and its application to the Behrens--Fisher problem is considered in Example~5 there.  Roughly speaking, the difference between this profile likelihood-based marginal IM construction and the aforementioned extension-based construction is where the marginalization is carried out: the former first eliminates nuisance parameters in the relative likelihood and then constructs the IM contour directly for $\Phi$, whereas the latter first constructs the IM contour for $\Theta$ and then marginalizes to $\Phi$.  \citet{imbvm.ext} confirmed that, at least asymptotically, the profile likelihood-based marginal IM construction is more efficient than the extension-based construction.  My proposal here is to extract the inner probabilistic approximation of this profile likelihood-based marginal possibilistic IM for $\Phi$.  This requires no change in the computational approach outlined in Section~\ref{S:computation}, and the exact validity of the marginal IM implies, e.g., that credible intervals for $\Phi$ based on this ``posterior'' are exact confidence intervals.  The only wrinkle is that the relative profile likelihood doesn't have a closed-form expression, and its distribution depends on a nuisance parameter.  This increases the computational cost of the inner probabilistic approximation compared to other examples (see Appendix~G),
but this pays off in terms of efficiency gains.  

\begin{figure}[t]
\begin{center}
\scalebox{0.5}{\includegraphics{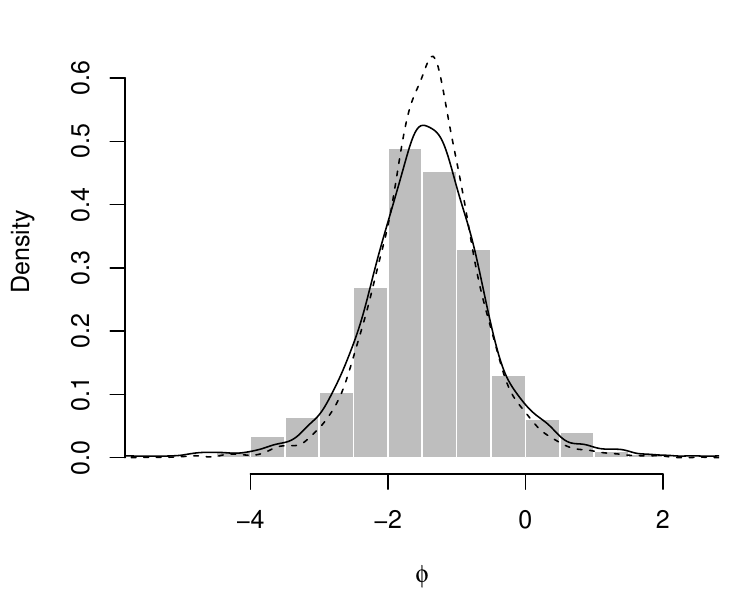}}
\end{center}
\caption{Histogram summarizing the distribution of $\Phi$ in the Behrens--Fisher problem based on the marginal IM's inner probabilistic approximation. The solid and dashed lines are (kernel estimates of) the Bayesian posterior distributions for $\Phi$ based on the right Haar and Jeffreys priors, respectively.}
\label{fig:bf.density}
\end{figure}

The go-to real-data illustration is the example on travel times to work via two different routes \citep[][p.~83]{lehmann1975}.  The relevant summary statistics---sample sizes, sample means, and sample standard deviations---are as follows: $n_1=5$, $\hat\theta_{11}=7.580$, and $\hat\theta_{12}= 2.237$; $n_2=11$, $\hat\theta_{21}= 6.136$, and $\hat\theta_{22}= 0.073$.  The wide discrepancy between the two standard deviations, $\hat\theta_{12}$ and $\hat\theta_{22}$, would make it difficult to justify assuming the two variances are equal.  Figure~\ref{fig:bf.density} shows a histogram of the samples of $\Phi$ from the inner probabilistic approximation of the marginal IM, with (kernel estimates) of the density functions from the right Haar and Jeffreys prior Bayes solutions overlaid; the former also agrees with Fisher's fiducial distribution.  The point is that all three distributions are similar, with the new inner probabilistic approximation and the right Haar prior posterior being a bit more diffuse than the Jeffreys prior posterior.  

For further comparison, I run a small simulation study.  I focus on a rather severely imbalanced case---with $n_1=3$ and $n_2=20$---to ensure that the differences in performance are apparent.  Otherwise, the simulation settings are standard: $\Theta_{11}=2$, $\Theta_{21}=0$, $\Theta_{12}^2=1$, and $\Theta_{22}^2 = 2$.  I generated 10000 samples under this setup, and Table~\ref{tab:bf} presents the coverage probability and expected length of the various 90\% confidence intervals for $\Phi$.  Notably, only the marginal possibilistic IM's inner probabilistic approximation is able to hit the target coverage probability nearly on the nose, and, as desired, it's more efficient than the valid-but-conservative Bayes solution based on the right Haar prior, which agrees with the marginal distribution for $\Phi$ derived from the inner probabilistic approximation of the IM for the full parameter $\Theta$.  

\begin{table}[t]
\begin{center}
\begin{tabular}{ccc}
\hline
Method & Coverage & Length \\
\hline 
Bayes with Jeffreys prior & 0.861 & 2.31 \\
Bayes with right Haar prior & 0.929 & 3.26 \\
{\tt t.test} with Welch df & 0.881 & 2.79 \\
Marginal IM's inner approximation & 0.904 & 2.71 \\
\hline 
\end{tabular}
\end{center}
\caption{Coverage probabilities and mean lengths for the 90\% confidence intervals returned by the various methods.  Standard errors for all of the coverage probabilities and mean lengths are roughly 0.003 and 0.01, respectively.}
\label{tab:bf}
\end{table}










\section{Conclusion}
\label{S:discuss}

This paper offers a new perspective on no-prior Bayes inference, one that has certain close ties to Bayesian inference but is not actually Bayesian at all.  My approach starts with an inferential model (IM) framework for data-driven uncertainty quantification that prioritizes reliability, insisting on calibration of its data-dependent degrees of belief.  The kind of calibration IMs require makes them incompatible with probabilistic Bayesian inference, and it's for this reason that the IM's output is possibilistic, i.e., takes the mathematical form of a possibility measure.  However, if one desires probabilistic uncertainty quantification, then the IM can accommodate this with an {\em inner probabilistic approximation} of its possibilistic output.  Aside from being exactly probability matching in certain ways, the proposed solution agrees with existing no-prior Bayes solutions in group invariant problems where there's an overall consensus, and is also asymptotically efficient by a version of the celebrated Bernstein--von Mises theorem.  And there's literally no choice of prior distribution required: for the given model and data, the possibilistic IM is what it is, so it's just a matter of extracting a suitable inner probabilistic approximation.  

The characterization of the contents of the IM's credal set in Theorem~\ref{thm:char} suggests a strategy for evaluating, at least approximately, the inner probabilistic approximation via Monte Carlo.  I explored this thoroughly in \citet{immc}, but mostly in the context of possibilistic IMs.  Here I use this same strategy for my proposed reIMagined no-prior Bayesian inference, and I show that this leads to a new and broadly reliable solution to the technically challenging and practically relevant Behrens--Fisher problem.  Nothing about this approach was tailored to the Behrens--Fisher problem specifically, so I fully expect similarly good performance in many other important applications.  Surely, the computational efficiency of my proposed solution can be improved, so I welcome advancements along these lines from computationally savvy readers.  

The focus here has been on {\em no-prior} solutions, but there are cases in which incomplete or partial prior information is available, including high-dimensional problems in which structural assumptions like sparsity are common.  Bayesian solutions can't work with incomplete prior information---a prior distribution, perhaps diffuse, must be specified for every aspect of the unknown $\Theta$.  Genuine partial-prior possibilistic IMs have recently been developed \citep[e.g.,][]{martin.partial2}, and an interesting idea would be to apply the same inner probabilistic approximation proposed here in the no-prior case to the partial-prior case in the above references.  Just as in this paper, these inner probabilistic approximations would enjoy some of the inherent reliability properties satisfied by the partial-prior IM, but the details remain to be investigated.

\section*{Acknowledgments}

The author thanks the Editor, Associate Editor, two anonymous reviewers, and Max Raner for their valuable comments on an earlier version of this paper. This work is partially supported by the U.S.~National Science Foundation, grants SES--2051225 and DMS--2412628.

\appendix

\section{Further IM insights, intuition, etc.}
\label{A:logic}

In the main text, I gave only the mathematical definition and brief explanation of the possibilistic inferential model (IM) and its basic properties.  But these developments are deeper and more insightful than can be communicated in such a brief explanation, so I'll share a bit more here about the IM intuition and logic.  I'll organize this presentation as a list of answers to some common/frequently-asked questions. 

\begin{ques}
{\em Why is imprecision needed?}
The false confidence theorem \citep{balch.martin.ferson.2017} says that for any precise data-dependent probability distribution $\prior_X$, e.g., a Bayesian or fiducial posterior, and for any thresholds $\rho$ and $\tau$, there exists a hypothesis $H$ such that 
\[ H \not\ni \Theta \quad \text{and} \quad \prob_\Theta\{ \prior_X(H) > \tau \} > \rho. \]
This is called {\em false confidence} because, with $\tau$ large and $\rho$ not small, the posterior tends (at non-negligible rate $\rho$) to be confident (high probability $\tau$) in the truthfulness of $H$ even though $H$ is actually false.  There's a sense in which this result is trivial: take $H$ equal to the complement of $\Theta$, so that $H$ is false and, at least in the absolutely continuous case, $\prior_X(H)=1$ almost surely.  But there's more nuance here because it's not just for these extreme/trivial $H$ that false confidence manifests; it's not even just for ``big'' sets that happen to be false.  The shape of $H$ and the effects of non-linearity, just like in Section~\ref{SS:caveats} in the present paper, play an important albeit not yet well understood role.  

In addition, there's no reason why the size of a hypothesis should matter at all---data are compatible with a hypothesis as soon as they're compatible with any single point in the hypothesis.  Size/measure only becomes relevant when integration enters the picture, but there's no justification for making this step except that statisticians can't help it, the theory of integration is burned into our minds.  Ideally, integration could be replaced by some other calculus that doesn't rely on notions of size, no differential element.  

Fortunately, both of the issues identified above can be resolved at once if ordinary probability can be replaced by a suitable imprecise probability.  The proof of the false confidence theorem leverages the additivity of the posterior, so, if the additivity assumption is dropped, then the proof breaks down.  But, also, certain data-dependent imprecise probabilities, namely, (possibilistic) IMs, have been shown to be immune to false confidence.  Moreover, possibility theory specifically operates under a different calculus than integration: the possibilistic calculus is optimization, so there is no differential element and the size of hypotheses is irrelevant when it comes to if and to what extent they are compatible with data; see \citet{martin.basu}.  
\end{ques}

\begin{ques}
{\em Why the possibilistic form specifically?}
It helps to start with being clear about what the objectives are when it comes to the formulation of a new and improved framework for statistical uncertainty quantification, so I list here three desiderata:
\begin{enumerate}
\item {\em Mathematically rigorous.}  Statistical reasoning shouldn't be based on a patchwork of disparate pieces but, rather, a coherent whole from which an answer to any question about the relevant unknowns can be obtained.
\vspace{-1mm} 
\item {\em As simple and ``familiar-looking'' as possible.}  Much of the theory that is currently being used for statistical inference (e.g., p-values for tests) generally works.  So, a new framework doesn't need to be completely different and, in fact, it shouldn't give altogether different solutions to classical inference problems.  
\vspace{-1mm}
\item {\em Reliable.}  Not only should procedures (tests, confidence sets, etc) derived from ones uncertainty quantification be reliable in the sense that they control frequentist error rates, but the (lower and upper) probabilities and expectations that it assigns to relevant hypotheses or assessments should be calibrated too.  In fact, it's this reliability property that gives the uncertainty quantification its meaning.  
\end{enumerate} 

To my knowledge, subject to Desiderata~3 above, the only options for a mathematically rigorous framework of uncertainty quantification are those that can be described by an imprecise probability of some sort.  Possibility theory is arguably one of the simplest imprecise probability theories and, since the possibilistic IM framework that I've been advancing has close connections to p-values for likelihood ratio tests, confidence distributions, and now no-prior Bayes solutions, it clearly satisfies the first two desiderata.  

The main text highlighted some of the possibilistic IM framework's reliability properties, but there is a more to say as it concerns the topic of this section: why the possibilistic form?  The property I referred to as {\em validity} in the main paper---Equation~\eqref{eq:valid}---is actually quite strong.  The frequentist error rate control guarantees it implies for test and confidence set procedures derived from the IM output are important, but doesn't do justice to the strength of this property.  As \citet{martin.partial} and \citet{cella.martin.probing} have shown, the validity property in Equation~\eqref{eq:valid} of the main paper is equivalent to 
\begin{equation}
\label{eq:valid.unif}
\sup_{\Theta \in \TT} \prob_\Theta\{ \uPi_X(H) \leq \alpha \text{ for some $H$ with $H \ni \Theta$} \} \leq \alpha, \quad \alpha \in [0,1]. 
\end{equation}
This is similar to the property stated in Equation~\eqref{eq:valid.alt} of the main text, but with one key difference: the property in \eqref{eq:valid.alt} establishes that the IM is calibrated ``pointwise'' in $H$, whereas \eqref{eq:valid.unif} shows that, in fact, the IM is calibrated ``uniformly'' in $H$.  This is relevant because modern statisticians and data scientists aren't quantifying uncertainty for themselves, they're developing general tools and software to assist other users in quantifying their statistical uncertainty.  Since I can't possibly know what a user might want to do with the uncertainty quantification framework that I offer them, if I want to guarantee that it's safe---which is important both for science/society and for me as the developer of such a tool---then I must ensure that calibration is maintained even if the user is, e.g., peeking at the data first to decide what hypotheses to consider.  The pointwise validity result in \eqref{eq:valid.alt} of the main paper doesn't guarantee this, but the uniform validity result in \eqref{eq:valid.unif} does.  

There are mathematical obstacles that would make it difficult for any non-possibilistic quantification of uncertainty to satisfy the strong validity property in Equation~\eqref{eq:valid} of the main paper, and hence the uniform calibration property in \eqref{eq:valid.unif} above.  Indeed, if $\uOmega_x$ was some other upper probability quantifying uncertainty about $\Theta$, given $X=x$, then one could define the analogue of its ``contour,'' i.e., $\theta \mapsto \uOmega_x(\{\theta\})$.  But this may not meet the requirement of a possibility contour, that is, generally, 
\[ \sup_{\theta \in \TT} \uOmega_x(\{\theta\}) < 1, \quad \text{and often $\ll 1$}. \]
Nothing in imprecise probability theory says that this ``contour'' must satisfy this supremum-equals-1 property and, in fact, in many (non-statistical) cases there are good reasons why such a property would be undesirable.  But, as \citet[][p.~226]{shafer1976} explains, statistical inference problems are special and well-suited for this supremum-equals-1 property.  What I can add to Shafer's justification is that it is virtually impossible for anything other than a possibilistic quantification of uncertainty, one for which the supremum-equals-1 property holds, to satisfy the validity property as stated in Equation~\eqref{eq:valid} of the main paper and, therefore, the equivalent uniform calibration property in \eqref{eq:valid.unif} above.  The point is that, if $\uOmega_X(\{\Theta\})$ is bounded away from 1 for almost all $X$, then it can't be stochastically no less than $\unif(0,1)$.  Moreover, if one was able to find a non-possibilistic IM $(\lOmega_x, \uOmega_x)$ that does satisfy the validity property in Equation~\eqref{eq:valid} in the main text, then there is a possibilistic IM that is valid and no less efficient.  So, there is at least no loss of generality in focusing on IMs that take on the possibilistic form.  
\end{ques}

\begin{ques}
{\em What else can be done beyond assigning beliefs to hypotheses about model parameters?}  
Classical inference on the parameters is a fully-specified statistical model is the right context to describe this paper's contribution, namely, a new perspective on no-prior Bayesian inference, but there are many other kinds of problems and statistical tasks that scientists are interested in.  Fortunately, IMs can contribute along these lines too, and they already have.  First, it's often the case that the quantity of interest is not best described as the parameter of a statistical model, but as some feature (or functional) of the underlying distribution $\prob$, which doesn't have a familiar parametric form.  The construction of approximately valid IMs for such cases is presented in \citet{cella.martin.imrisk}; work on removing the ``approximately'' adjective is ongoing.  Prediction is also an important problem and usually requires a nonparametric touch.  IM advances along these lines can be found in \citet{imconformal.supervised}, with a more possibility-centric perspective offered in \citet[][Sec.~6.3]{martin.partial3}.  Sometimes inference isn't the objective, the focus is on decision-making, i.e., choosing an appropriate (data-driven) action.  In \citet{imdec}, I propose the use of Choquet integration to use the IM output to evaluate upper (and lower) expected losses and recommend choosing actions that minimize the upper expected loss.  I also show that the IM's inherent reliability in inference carries over to the decision problem as well: it's a provably rare event that the IM's upper expected loss assessment is substantially more optimistic than an oracle's, uniformly over actions.  Finally, novel and provably reliable IM solutions for model assessment and selection, information fusion (e.g., meta-analysis), etc.~are currently in development.  
\end{ques}

\section{Details from Section~\ref{SS:characterization}}
\label{A:weights}

A point emphasized in Section~\ref{SS:characterization} is that inner probabilistic approximations of a given IM $\uPi_x$ aren't unique.  The reason is that the characterization result and the ``maximally diffuse'' objective don't determine exactly how the kernel $\kernel_x^\alpha$ allocates its mass to the boundary $\partial C_\alpha(x)$.  The goal of the present section is to expand a bit on this ambiguity to better understand the role played by the kernel.  

To start, I'll focus on the case of a scalar parameter, like in the illustrations presented in Section~\ref{S:inner}.  In such a case, the kernel can be expressed generally as 
\[ \kernel_x^\alpha = w_{x,\alpha} \, \delta_{a(x,\alpha)} + (1-w_{x,\alpha}) \, \delta_{b(x, \alpha)}, \]
where $w_{x,\alpha}$ is a data and level-dependent weight in $[0,1]$, $a(x,\alpha)$ and $b(x,\alpha)$ are the lower and upper endpoints of the confidence interval $C_\alpha(x)$, and $\delta_z$ is a probability distribution concentrating all of its mass on $z$.  In this case, the kernel is fully determined up to the weight $w_{x,\alpha}$.  My general recommendation was to take $w_{x,\alpha}$ to be constant equal to $\frac12$, which has a two-fold motivation: it's consistent with a version of the indifference principle and weights converging to $\frac12$ is necessary in order for the inner probabilistic approximation to satisfy a Bernstein--von Mises theorem.  Nevertheless, there's still a choice of weight that can be made, so it's worth having some understanding of what this entails.  

For the gamma model $X \sim \prob_\Theta = \gam(n,\Theta)$ in Section~\ref{SS:gamma} of the main text, Figure~\ref{fig:cdf.toy}(a) shows the distribution functions associated with three different inner probabilistic approximations, each determined by a choice of the weight $w_{x,\alpha}$.  Two of the three were based on constant-weight kernels: one with $w=0.5$ and one with $w \approx 0.45$; my rationale behind the choice $w \approx 0.45$ is described below.  Since these weights are numerically very similar, it's not surprising that the corresponding distribution functions are likewise similar.  The third of the three distribution functions in Figure~\ref{fig:cdf.toy}(a) is that based on the Bayes solution that employs the default right Haar prior with density $\theta \mapsto \theta^{-1}$ with respect to Lebesgue measure.  I claimed in the main text that this Bayesian solution can't be expressed in terms of the kernel above with constant weight; a non-constant weight $w_{x,\alpha}$ is needed.  This suggests the following question: what non-constant weight corresponds to the right Haar prior Bayes solution?

\begin{figure}[t]
\begin{center}
\scalebox{0.6}{\includegraphics{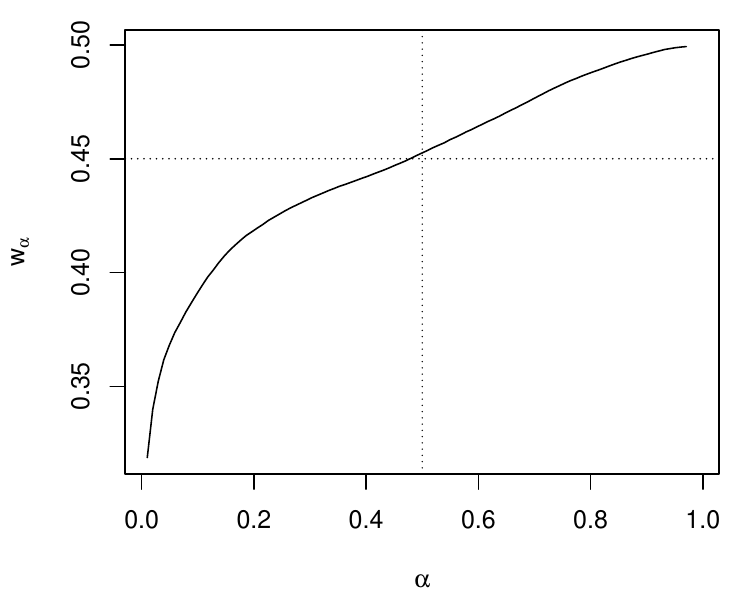}}
\end{center}
\caption{Plot of $\alpha \mapsto w_\alpha$ as defined in \eqref{eq:w.alpha} for the gamma model with $n=7$ and $x=14$.}
\label{fig:w.alpha}
\end{figure}

In general, suppose that $Q_x(\theta)$ is the distribution function corresponding to an inner probabilistic approximation $\prior_x$; let $q_x$ be the corresponding density with respect to Lebesgue measure.  Let $\hat\theta_x$ be the maximum likelihood estimator.  Following the derivation in Section~\ref{SS:gamma}, for $\theta \leq \hat\theta_x$, the distribution function $Q_x(\theta)$ can be expressed as 
\[ Q_x(\theta) = \int_0^1 w_\alpha \, 1\{a(x,\alpha) \geq \theta\} \, d\alpha = \int_0^{\pi_x(\theta)} w_\alpha \, d\alpha, \]
where I've dropped the subscript ``$x$'' in $w_{x,\alpha}$ for simplicity.  Applying Leibniz's integral rule gives the identity 
\[ q_x(\theta) = w_{\pi_x(\theta)} \, \dot\pi_x(\theta), \quad \theta \leq \hat\theta_x, \]
where $\dot\pi_x(\theta)$ is the partial derivative of $\pi_x(\theta)$ with respect to $\theta$.  This implies that 
\begin{equation}
\label{eq:w.alpha}
w_\alpha = \frac{q_x(\theta_{x,\alpha})}{\dot\pi_x(\theta_{x,\alpha})}, \quad \alpha \in [0,1], 
\end{equation}
where $\theta_{x,\alpha}$ is such that $\theta_{x,\alpha} \leq \hat\theta_x$ and $\pi_x(\theta_{x,\alpha}) = \alpha$.  This calculation can't be done analytically for the gamma illustration in Section~\ref{SS:gamma}, but it can be done numerically, and the plot of $\alpha \mapsto w_\alpha$ is shown in Figure~\ref{fig:w.alpha}.  Notice that this curve ranges from below to above the $w=0.45$ line, with $w_{0.5} \approx 0.45$; this explains my choice of weight $w \approx 0.45$ as the best constant-weight approximation of the right Haar prior Bayes posterior.  

The formula \eqref{eq:w.alpha} for the non-constant weight is a bit abstract and can only be evaluated numerically in the gamma example above.  So it may help to provide a simpler, proof-of-concept illustration---in particular, the Gaussian location model.  In that case, the posterior density is $q_x(\theta) = (2\pi)^{-1/2} e^{-(\theta-x)^2/2}$, and the contour derivative is 
\[ \dot\pi_x(\theta) = 2(\theta - x) \times f_1((\theta-x)^2), \]
where $f_1(z) = (2\pi)^{-1/2} z^{-1/2} e^{-z/2}$ is the $\chisq(1)$ density function. Then the ratio is 
\[ \frac{q_x(\theta)}{\dot\pi_x(\theta)} = \frac12, \quad \theta < \hat\theta_x = x. \]
This gives further justification for the choice of a constant weight $w=\frac12$ in the IM's inner probabilistic approximation's kernel for the Gaussian location model.  

More extreme choices of weights can be considered.  For example, in the scalar parameter case, an extreme choice is taking $w=0$ or $w=1$.  The large-sample results in Section~\ref{SS:concentration} strongly suggest that such a choice isn't warranted, so I wouldn't recommend this.  The point is that $w=0$ and $w=1$ both introduce bias that can't be beneficial.  When the parameter $\Theta$ is a vector or something more complex, then there are more choices of ``extreme'' kernels, some of which might not be entirely unreasonable.  For instance, if $\Phi = m(\Theta)$ is a scalar feature of interest, then one might consider an ``extreme'' kernel that puts all of its weight on the extremes of the function $m$ on each confidence set boundary.  That is, one might consider a kernel 
\[ \kernel_x^\alpha = \tfrac12 \, \delta_{\underline m_\alpha} + \tfrac12 \, \delta_{\overline m_\alpha}, \]
where 
\[ \underline m_\alpha = \min_{\theta \in \partial C_\alpha(x)} m(\theta) \quad \text{and} \quad \overline m_\alpha = \max_{\theta \in \partial C_\alpha(x)} m(\theta). \]
With this choice, the marginal distribution $\prior_x^m$ of $\Phi = m(\Theta)$ derived from the inner probabilistic approximation with the above kernel is my recommended inner probabilistic approximation of the marginal IM $\uPi_x^m$ derived from $\Pi_x$ via extension.  The point is that ``extreme choices'' of kernels might not be unreasonable, and the justification of such a choice would be based on marginalization considerations, when there's a particular feature $\Phi=m(\Theta)$ of $\Theta$ of interest.  When there's no single feature that's of high priority, it makes sense to make a neutral choice of kernel, and, aside from the large-sample properties, it's precisely the uniform kernel's neutrality that makes it a good choice.

\section{Details from Section~\ref{SS:group}}
\label{A:group}

\subsection{Theory}



This section provides some additional background and details about the invariant model setting and the result in Theorem~\ref{thm:group}.  

Recall that $\{\prob_\theta: \theta \in \TT\}$ is a family of probability distributions supported on $\XX$ that is also invariant with respect to a group of transformations $\G$ from $\XX$ to itself in the sense of Equation~\eqref{eq:invariant} in the main text.  If the distributions $\prob_{\theta}$ have densities with respect to some underlying dominating measure, which I'll assume here, then invariance means 
\begin{equation}
\label{eq:invariant.density}
p_\theta(x) = p_{\bar g \theta}(gx) \, \chi(g), \quad x \in \XX, \; \theta \in \TT, \; g \in \G, 
\end{equation}
where $\chi(g)$ is the ``multiplier,'' a change-of-variables Jacobian term; see \citet[][Ch.~6]{schervish1995} and \citet{eaton1989} for details.  A further simplification that I'll adopt here, common in the literature \citep[e.g.,][p.~371]{schervish1995}, is to assume $\TT = \G = \Gbar$.
Recall that I'm also assuming here that the group $\G$ with respect to which the model is invariant is unique.  There are cases, however, where the model is invariant with respect to more than one group, and I'll elaborate on this case at the end of this subsection.  

The set $\G x = \{gx: g \in \G\} \subseteq \XX$ is called the {\em orbit} of $\G$ corresponding to $x$.  The orbits partition $\XX$ into equivalence classes, so every point $x \in \XX$ falls on exactly one orbit.  This partition can be used to construct a new coordinate system on $\XX$ which will be useful for us in what follows.  Identify $x \in \XX$ with $(g_x, u_x)$, where $u_x \in \UU$ denotes the label of orbit $\G x$ and $g_x \in \G$ denotes the position 
of $x$ on the orbit $\G x$. 

\begin{asmp}
Suppose that the model $\{\prob_\theta: \theta \in \TT\}$ is invariant with respect to a (unique) group $\G$, and that the corresponding density functions $\{p_\theta: \theta \in \TT\}$ satisfy \eqref{eq:invariant.density}.  For simplicity, assume $\TT = \G = \Gbar$.  
In addition, assume that the following hold:
\begin{itemize}
\item[A1.] The left Haar measure $\lambda$ and the corresponding right Haar measure $\rho$ on (the Borel $\sigma$-algebra of) $\G$ exist and are unique up to scalar multiples.
\vspace{-2mm}
\item[A2.] There exists a bijection $t: \XX \to \G \times \UU$, with both $t$ and $t^{-1}$ measurable, that maps $x \in \XX$ to its position--orbit coordinates $(g_x, u_x) \in \G \times \UU$.  
\vspace{-2mm}
\item[A3.] The distribution of $t(X) = (G, U) \in \G \times \UU$ induced by the distribution of $X \sim \prob_{\theta}$ has a density with respect to $\lambda \times \mu$ for some measure $\mu$ on $\UU$.
\end{itemize} 
\end{asmp}

A couple quick remarks are in order.  First, for A1, existence and uniqueness of the left and associated right Haar measures is a classical result, which holds under the mild condition that $\G$ be a locally compact topological group; see \citet{halmos.measure} and \citet{nachbin1965}.  
For A3, existence of a joint density with respect to a product measure ensures that there's no difficulty in defining a conditional distribution for $G$, given $U=u$, which is needed below.  Finally, $U=U_X$ is an ancillary statistic as a function of $X \sim \prob_\theta$, meaning that its marginal distribution doesn't depend on $\theta$.


The simplest example is a location model where $\G = \Gbar = (\RR, +)$.  This is an abelian group, so the left and right Haar measures are the same and both equal to Lebesgue measure.  The function $x \mapsto t(x)$ in A2 consists of two components: in its ``$g_x$'' coordinate an equivariant function of $x$ that estimates the location and, in its ``$u_x$'' component, an invariant function of $x$, such as residuals.  For example, $g_x = \bar x$ the arithmetic mean of $x=(x_1,\ldots,x_n)$ and $u_x = \{x_i - \bar x: i=1,\ldots,n\}$.  Note that the $u_x$ coordinate satisfies a constraint, so, after it's represented in a suitable lower-dimensional space $\UU$, $\mu$ can be taken as Lebesgue measure there.  There are many other problems that fit this general form; see, e.g., Chapters~1--2 of \citet{fraser1968}, including the exercises. 

Since I'm assuming $\TT = \G = \Gbar$, generic $\theta \in \TT$ can be identified as transformations in $\G$; the same goes for the uncertain $\Theta$.  That means $\theta$ can be inverted to $\theta^{-1}$ and can be composed with other transformations in $\G$.  A key result \citep[e.g.,][Cor.~6.64]{schervish1995} is that the density of $X$ or, equivalently, of $t(X) = (G,U)$, under $\prob_{\theta}$, is given by 
\begin{equation}
\label{eq:GU.joint}
p_\theta(g, u) = f(\theta^{-1} \circ g, u), 
\end{equation}
where the function $f: \G \times \UU \to \RR$ doesn't directly depend on $\theta$.  The particular form of $f$ isn't important for what follows.  
\begin{itemize}
\item Lemma~6.65 in \citet{schervish1995} gives a Bayesian posterior distribution $\prior_x^\rho$ for $\Theta$ that has the following density with respect to the right Haar prior measure $\rho$: 
\begin{equation}
\label{eq:schervish}
q_x^\rho(\theta) = \frac{d \prior_x^\rho}{d\rho}(\theta) \propto f(\theta^{-1} \circ g, u), \quad \theta \in \TT, \quad x \equiv (g,u).
\end{equation}
This also agrees with the accepted fiducial distribution for $\Theta$, given $x$.
\vspace{-2mm}
\item Proposition~2 in \citet{martin.isipta2023} shows that the relative likelihood is given by 
\[ R(x,\theta) = d_u \, f(\theta^{-1} \circ g, u), \quad \theta \in \TT, \quad x \equiv (g,u), \]
where $d_u$ depends only on the $u$-component of $x$. 
\end{itemize} 
The above points are crucial to the following calculation.  Indeed, since the $U$ component of $X$ is ancillary, the {\em minimum complexity principle} in \citet{martin.partial2} says the IM construction should condition on the observation $U=u$.  Write $\pi_{g|u}(\theta)$ for the possibilistic IM's contour, where the subscript is meant to emphasize that the conditional distribution of $X \equiv (G,U)$, given the observed value of $u$, is used in the validification step \eqref{eq:contour}.  Then the IM contour is defined as 
\begin{equation}
\label{eq:contour.conditional}
\pi_{g|u}(\theta) = \prob_\theta\{ R(X,\theta) \leq R(x,\theta) \mid U=u \}, \quad \theta \in \TT. 
\end{equation}
Then the following identities are consequences of the two results above that link the relative likelihood and the posterior density:
\begin{align*}
\pi_{g|u}(\theta) & = \prob_\theta\{ f(\theta^{-1} \circ G, u) \leq f(\theta^{-1} \circ g, u) \mid U=u \} \\
& = \prob\{ f(H, u) \leq f(\theta^{-1} \circ g, u) \mid U=u \} \\
& = \prior_x^\rho\{ f(H, u) \leq f(\theta^{-1} \circ g, u) \} \\
& = \prior_x^\rho\{ q_x^\rho(\Theta) \leq q_x^\rho(\theta) \},
\end{align*}
where the first line is by simplification of the definition of $\pi_{g|u}$; the second line is by the fact that $H := \theta^{-1} \circ G$ is a pivot with respect to the conditional distribution of $G$, given $U=u$, under $\prob_\theta$ \citep[e.g.,][Cor.~6.67]{schervish1995}; the third line is by the fact that $H := \Theta^{-1} \circ g$ is also a pivot with respect to the Bayes posterior $\Theta \sim \prior_x^\rho$ and has the same distribution as $H$ in previous line; and the last line is by simplification via \eqref{eq:schervish}. 

The last line in the above display can be recognized as the probability-to-possibility transform applied to the Bayesian/fiducial posterior distribution.  This identifies the posterior distribution $\prior_x^\rho$ as an inner probabilistic approximation of the possibilistic IM $\uPi_x = \uPi_{g|u}$.  This provides a new/different proof of the main result in \citet{martin.isipta2023}, of which Theorem~\ref{thm:group} above is a generalization.  

As mentioned above, there are cases in which the group $\G$ with respect to which the model is invariant is not unique.  A key example of this is in the case of Gaussian models, which are invariant with respect to linear (or affine) transformations $gx = Ax + b$, indexed by pairs $(A,b)$ consisting of a matrix $A$ and a vector $b$.  Since the group indexed by all such matrices $A$ is ``too large'' (in the sense that the underlying group is not amenable), it's natural to restrict the transformations to those indexed by certain sub-classes of matrices, corresponding to one or another lower-dimensional matrix factorization.  It turns out that the right Haar prior depends on which class of matrix factorizations is used.  For example, in the bivariate normal model indexed by means $\mu_1$ and $\mu_2$, standard deviations $\sigma_1$ and $\sigma_2$, and correlation $\rho$, there are two right Haar priors \citep[][Sec.~5]{sun.berger.2007}, with respective density functions 
\[ (\mu_1, \mu_2, \sigma_1, \sigma_2, \rho) \mapsto \frac{1}{\sigma_2^2(1-\rho^2)} \quad \text{and} \quad (\mu_1, \mu_2, \sigma_1, \sigma_2, \rho) \mapsto \frac{1}{\sigma_1^2(1-\rho^2)}. \]
Of course, these two distinct priors lead to (slightly) different Bayes solutions with different statistical properties.  Some might find this troubling: since the model is what determines the statistical inference problem, not the choice of group representation, it's problematic if the Bayesian solution depends on a somewhat ad hoc choice between one group/right Haar prior or another.  

In the construction of the IM solution described above, the dependence on the group structure is implicit: it's in the identification of the $(G,U)$ components, where $G$ is an equivariant function of $X$ and $U$ is a (maximal) invariant function of $X$ which, of course, depend on $\G$.  So, in general, different groups with respect to which the model is invariant would suggest a different ``$(G,U)$'' representation.  Since this representation determines how the conditioning is carried out in \eqref{eq:contour.conditional}, this IM solution also depends on the particular choice of group $\G$.  But I want to emphasize that conditioning on an ancillary statistic isn't {\em required} for the IM construction; this is only {\em recommended} when an appropriate ancillary statistic is available.  If the group isn't unique, then there are multiple ancillary statistics available, yielding distinct conditional IMs.  There are some natural ways one might remove the group-specific conditional IM's undesirable dependence on the choice of group, but this is not really relevant to the present paper, so I'll save this investigation for future work; see, also, the next paragraph.  Alternatively, when there are multiple ancillary statistics, arguably none of them are fully ``appropriate.''  In that case, one might prefer to ignore conditioning and proceed with the basic likelihood-based IM construction as in Equation~\eqref{eq:contour} of the main text.  A corresponding inner probabilistic approximation retains the desirable probability matching and Bernstein--von Mises properties, but whether it corresponds to any Bayesian posterior distribution is an interesting open question.  

To conclude this discussion about non-unique groups, let me consider the aforementioned Gaussian case specifically.  There, the maximum likelihood estimator is a complete and sufficient statistic, so, by Basu's theorem, whether one conditions on an ancillary statistic or not has no effect on the IM solution.  Consequently, the same argument used above to prove Theorem~\ref{thm:group} applied to the present case gives 
\[ \pi_x(\theta) = \prob_\theta\{ R(X,\theta) \leq R(x,\theta) \} = \prior_x^\rho\{ q_x^\rho(\Theta) \leq q_x^\rho(\theta)\}, \quad \theta \in \TT, \]
for any right Haar prior $\rho$.  Therefore, in the multivariate Gaussian case, it follows that {\em all} right Haar prior Bayes solutions are inner probabilistic approximations of the IM with contour as defined in Equation~\eqref{eq:contour} of the main text.  Compare this with the results in \citet{eaton.sudderth.2012} which imply, for example, that highest posterior density regions corresponding to different right Haar priors in these multivariate normal models are exactly probability matching.  Outside the Gaussian family, completeness typically doesn't hold, so I currently don't know if the above claim concerning ``all right Haar prior Bayes solutions'' applies more broadly.

\subsection{Directional data example}
\label{AA:wheel}

Let $X=(X_1,\ldots,X_n)$ be the observable angles for $n$ independent spins of a roulette wheel.  The model assumed here is the {\em von Mises distribution}, with density 
\[ p_\theta(x) = \{2\pi I_0(\kappa)\}^{-1} \exp\{\kappa \cos(x - \theta)\}, \quad x,\theta \in [0, 2\pi). \]
Here $\kappa > 0$ is a known concentration parameter, and $I_0$ denotes the Bessel function of order 0.  The unknown mean angle $\Theta$ is the parameter to be inferred from the data.  

Digging into this model further, using the cosine formula 
\[ \cos(x - \theta) = \cos x \cos\theta + \sin x \sin \theta, \]
it's easy to see that the minimal sufficient statistic for this model is the pair 
\[ Y = (\bar C, \bar S) = \Bigl( \frac1n \sum_{i=1}^n \cos X_i, \, \frac1n\sum_{i=1}^n \sin X_i \Bigr). \]
Relative to the posited von Mises model, $Y$ can be treated as ``the data.'' Since $(C_i, S_i) = (\cos X_i, \sin X_i)$ denotes the Cartesian coordinates of the point on the unit circle corresponding to the angle $X_i$, it's clear that $Y=(\bar C, \bar S)$ is generally inside the unit circle.  The more ``concentrated'' the original data points are around a particular angle, the closer $Y$ will be to the boundary of the circle.  Convert this average position to polar coordinates:
\[ G = \arctan(\bar S / \bar C) \quad \text{and} \quad U = ( \bar C^2 + \bar S^2)^{1/2}. \]
Here $G$ is the angle the vector $(\bar C, \bar S)$ makes with the horizontal axis and $U$ is the Euclidean length of $(\bar C, \bar S)$.  It's clear that $(G,U)$ is a bijection of the minimal sufficient statistic $Y$, so write $t(X) = (G,U)$.  I'm using the notation $G$ and $U$ to align with that introduced in the previous subsection.  That is, there's a group of transformations $\G$ that corresponds to rotations of points in the plane; technically, $\G$ is the special orthogonal group $\text{SO}(2)$ consisting of orthogonal matrices with unit determinant.  Then $G$ can be interpreted as both an angle or as a rotation by that angle.  Similarly, the interior of the unit circle can be partitioned into distinct concentric circles, which correspond to orbits, and $U$ determines which of these orbits $Y$ sits on.  

The details here align with the general setup above, so I can proceed to carry out the fiducial or default-prior Bayes and IM analyses.  It's a standard result \citep[e.g.,][Eq.~4.5.5]{mardia.jupp.book} that the joint density for $t(X)=(G,U)$ factors as 
\begin{align*}
p_\theta(g, u) & = p_\theta(g \mid u) \, p(u) \\
& = \{2\pi I_0(\kappa u)\}^{-1} \exp\{\kappa u \cos(g - \theta)\} \, p(u),
\end{align*}
where $p(u)$ denotes the marginal density for $U$, which doesn't depend on $\theta$ and isn't relevant here.  The expression for the conditional density of $G$, given $U=u$, is easily seen to be that of a von Mises distribution but with concentration parameter $\kappa u$.  The influence of conditioning on $U$ makes intuitive sense because, as indicated above, the statistic $U$ acts like a measure of how ``concentrated'' the observed angles are around a common angle.  It's the conditional density above that drives both the fiducial and IM analyses.  Indeed, the no-prior Bayes (and fiducial) density is 
\[ q_x(\theta) \propto \exp\{\kappa u \cos(g - \theta)\}, \quad \theta \in (0,2\pi], \]
and, after simplification, the IM's possibility contour is 
\[ \pi_{g|u}(\theta) = \prob\{ \cos H \leq \cos(g - \theta) \mid U=u\},\quad \theta \in [0,2\pi), \]
where the probability is with respect to the conditional distribution of $H := G-\Theta$, given $U=u$, which is simply a von Mises distribution with mean angle 0 and concentration parameter $\kappa u$.  This contour function is easy to evaluate numerically and to plot as in the main text.  By setting a sufficiently fine grid on the interval $(0,2\pi)$ and then evaluating the contour on that grid, it's easy to obtain the level set $C_\alpha(x)$ for any $\alpha$.  From here it's straightforward to sample from the inner probabilistic approximation, etc.

\section{Details from Section~\ref{SS:concentration}}
\label{A:concentration}

\subsection{Setup and regularity conditions}

\newcommand{\sL}{\mathcal{L}}

Certain regularity conditions are required in order to establish asymptotic concentration properties of estimators, posterior distributions, etc., and the same is true for IMs and the inner probabilistic approximations under consideration here.  Roughly, these conditions ensure that the log-likelihood function is smooth enough that it can be well-approximated by a quadratic function.  One common set of regularity conditions are the classical {\em Cram\'er conditions} \citep{cramer.book}, versions of which can be found in the standard texts, including \citet[][Theorem~3.10]{lehmann.casella.1998} and \citet[][Theorem~7.63]{schervish1995}.  Here I'll adopt the more modern set of sufficient conditions originating in \citet{lecam1956, lecam1960, lecam1970} and \citet{hajek1972}; see, also, \citet{bickel1998} and \citet{vaart1998}. 

The posited model specifies a class $\{\prob_\theta: \theta \in \TT\}$ of probability distributions, supported on $\XX$, with $\prob_\theta$ having a density function $p_\theta(x)$ relative to a $\sigma$-finite measure $\nu$ on $\XX$.  Then the data $X^n=(X_1,\ldots,X_n)$ is assumed to be iid of size $n$ with common distribution $\prob_\Theta$, where $\Theta \in \TT$ is the uncertain ``true value'' of the parameter.  Following \citet[][Ch.~2]{bickel1998}, define the (natural) logarithm and square-root density functions:
\[ \ell_\theta(x) = \log p_\theta(x) \quad \text{and} \quad s_\theta(x) = p_\theta(x)^{1/2}. \]
The ``dot'' notation, e.g., $\dot g_\theta(x)$, represents a function that behaves like the derivative of $g_\theta(x)$ with respect to $\theta$ for pointwise in $x$.  If the usual partial derivative of $g_\theta(x)$ with respect to $\theta$ exists, then $\dot g_\theta(x)$ is that derivative; but suitable functions $\dot g_\theta(x)$ may exist even when the ordinary derivative fails to exist.  In particular, below I'll assume existence of a suitable ``derivative'' $\theta \mapsto \dot s_\theta(x)$ of the square-root density.  Finally, let $\sL_2(\nu)$ denote the set of measurable functions on $\XX$ that are square $\nu$-integrable. 

\begin{asmp}
The parameter space $\TT$ is open and there exists a vector $\dot s_\theta(x) = \{ \dot s_{\theta,d}(x): d=1,\ldots,D\}$, with $\dot s_{\theta,d}$ an element of $\sL_2(\nu)$, such that the following conditions hold: 
\begin{itemize}
\item[A1.] the maps $\theta \mapsto \dot s_{\theta,d}$ from $\TT$ to $\sL_2(\nu)$ are continuous for each $d=1,\ldots,D$; 
\item[A2.] at each $\theta \in \TT$, 
\begin{equation}
\label{eq:dqm}
\int \bigl| s_{\theta + u}(x) - s_\theta(x) - u^\top \dot s_\theta(x) \bigr|^2 \, \nu(dx) = o(\|u\|^2), \quad u \to 0 \in \RR^D; 
\end{equation}
\item[A3.] and the $D \times D$ matrix $\int \dot s_\theta(x) \, \dot s_\theta(x)^\top \, \nu(dx)$ is non-singular for each $\theta \in \TT$. 
\end{itemize} 
\end{asmp}

The condition in \eqref{eq:dqm} is often described as $\theta \mapsto s_\theta$ being {\em differentiable in quadratic mean}.  Note that this condition does not require the square-root density to actually be differentiable at $\theta$, only that it be locally ``linearizable'' in an average sense, like a differentiable-at-$\theta$ function would be.  The classical Cram\'er conditions assume more than two continuous derivatives, so \eqref{eq:dqm}, which does not even require existence of a first derivative, is significantly weaker; sufficient conditions for \eqref{eq:dqm} are given in \citet[][Lemma~7.6]{vaart1998}.  Then the {\em score function} $\dot\ell_\theta(x)$ is defined in terms of $\dot s_\theta(x)$ as 
\[ \dot\ell_\theta(x) = \frac{2\dot s_\theta(x)}{s_\theta(x)} \, 1\{s_\theta(x) > 0\}, \]
and it can be shown that $\int \dot\ell_\theta(x) \, \prob_\theta(dx) = 0$ for each $\theta$.  Moreover, Condition~A3 above implies non-singularity of the Fisher information matrix $I_\theta = \int \dot\ell_\theta(x) \ell_\theta(x)^\top \, \prob_\theta(dx)$ for each $\theta \in \TT$.  \citet[][Prop.~2.1.1]{bickel1998} provide sufficient conditions for A1--A3.   

The above conditions are rather mild, and hold in a broad range of problems, including exponential families.  One further condition is required for the possibilistic Bernstein--von Mises theorem in Section~\ref{SS:concentration}, namely, that the maximum likelihood estimator is consistent, i.e., $\hat\theta_{X^n} \to \Theta$ in $\prob_\Theta$-probability as $n \to \infty$.  This, is not automatic, but holds quite broadly when the parameter space dimension is fixed, as I'm assuming here.

\subsection{Proof of Theorem~\ref{thm:new.bvm}}

The intuition behind the Bernstein--von Mises result itself, and the proof that follows, is that both the inner probabilistic approximation and the Gaussian that it's merging with are fully determined by corresponding possibility contours.  So, any difference in these distributions will imply a difference in the possibility contours, but the contours themselves are merging by Theorem~\ref{thm:bvm} and, therefore, any difference in the distributions must be vanishing asymptotically.  The following details make this argument rigorous.

I'll start by justifying the above claim that any difference between $\prior_{X^n}^\star$ and the Gaussian approximation $\Gauss_{X^n}$ is due to differences in the corresponding possibility contours.  The first observation is that $\prior_{X^n}^\star$ has the characterization as a uniform mixture of uniform kernels by Theorem~\ref{thm:char}.  The second observation is that, similarly, $\Gauss_{X^n}$ is the inner probabilistic approximation of the Gaussian possibility measure with contour $\gamma_{X^n}$ and, therefore, the characterization result in Theorem~\ref{thm:char} applies to $\Gauss_{X^n}$ too.  That is, if $\kernel_{X^n, \pi}^\alpha$ and $\kernel_{X^n, \gamma}^\alpha$ are the uniform kernels supported on the level sets $\{ \theta: \pi_{X^n}(\theta)=\alpha\}$ and $\{\theta: \gamma_{X^n}(\theta)=\alpha\}$, then 
\[ \prior_{X^n}^\star(\cdot) = \int_0^1 \kernel_{X^n,\pi}^\alpha(\cdot) \, d\alpha \quad \text{and} \quad \Gauss_{X^n}(\cdot) = \int_0^1 \kernel_{X^n,\gamma}^\alpha(\cdot) \, d\alpha. \]
Then, for any Borel measurable subset $H$ of $\TT$, 
\[ \bigl| \prior_{X^n}^\star(H) - \Gauss_{X^n}(H) \bigr| \leq \int_0^1 \bigl| \kernel_{X^n,\pi}^\alpha(H) - \kernel_{X^n,\gamma}^\alpha(H) \bigr| \, d\alpha, \]
so $d_\text{\sc tv}(\prior_{X^n}^\star, \Gauss_{X^n}) > \eps$ implies that there exists an $H$ such that $| \kernel_{X^n,\pi}^\alpha(H) - \kernel_{X^n,\gamma}^\alpha(H) |$ exceeds $\eps$ on a set of $\alpha$'s having positive Lebesgue measure.  This implies that there exists an open subset of $H$ and a non-empty interval $(\alpha_1, \alpha_2)$ such that 
\[ \pi_{X^n}(\theta) \in (\alpha_1, \alpha_2) \quad \text{and} \quad \gamma_{X^n}(\theta) > \alpha_2 \quad \text{for all $\theta$ in that subset;} \]
of course, the last inequality could be replaced by ``$\gamma_{X^n}(\theta) < \alpha_1$ and the positions of $\pi_{X^n}$ and $\gamma_{X^n}$ in the above display could be reversed, but these changes don't affect the argument that follows.  What matters is that there exists a point $\theta$---depending implicitly on data $X^n$, on $H$, and on $\eps$---such that 
\[ |\pi_{X^n}(\theta) - \gamma_{X^n}(\theta)| > \alpha_2 - \alpha_1 > 0. \]
That Theorem~\ref{thm:bvm} implies the merging of $\pi_{X^n}$ and $\gamma_{X^n}$ {\em uniformly} means that the event in the above display has vanishing $\prob_\Theta$-probability as $n \to \infty$.  

What's not fully satisfactory about the above conclusion is that the positive bound ``$\alpha_2 - \alpha_1$'' is only implicit.  An important point that the above argument reveals is that any differences in how $\prior_{X^n}^\star$ and $\Gauss_{X^n}$ assign probability will manifest through the contour level sets.  Since these are also the sets that determine the distributions themselves, it's without loss of generality to consider hypotheses $H$ that take the form of lower level sets of either of the defining contours.  For concreteness, consider 
\[ H = \{\theta: \pi_{X^n}(\theta) \leq \tau\}, \quad \tau \in (0,1). \]
By definition of the inner probabilistic approximation, $\prior_{X^n}^\star(H) = \sup_{\theta \in H} \pi_{X^n}(\theta) = \tau$.  Moreover, since $H$ is bounded and excludes the mode of $\gamma_{X^n}$, it's easy to check that $\Gauss_{X^n}(H)$ satisfies 
\[ \inf_{\theta \in \partial H} \gamma_{X^n}(\theta) \leq \Gauss_{X^n}(H) \leq \sup_{\theta \in \partial H} \gamma_{X^n}(\theta). \]
Consequently, 
\begin{align*}
\prior_{X^n}^\star(H) - \Gauss_{X^n}(H) & \leq \tau - \inf_{\theta \in \partial H} \gamma_{X^n}(\theta) \\
& = \sup_{\theta \in \partial H} \{ \tau - \gamma_{X^n}(\theta) \} \\
& = \sup_{\theta \in \partial H} \{ \pi_{X^n}(\theta) - \gamma_{X^n}(\theta) \} \\
& \leq \sup_{\theta \in \partial H} |\pi_{X^n}(\theta) - \gamma_{X^n}(\theta)|,  
\end{align*}
where the second-to-last line follows because $\pi_{X^n}(\theta) \equiv \tau$ on $\partial H$.  Similarly, 
\begin{align*}
\prior_{X^n}^\star(H) - \Gauss_{X^n}(H) & \geq \tau - \sup_{\theta \in \partial H} \gamma_{X^n}(\theta) \\
& = \inf_{\theta \in \partial H} \{ \tau - \gamma_{X^n}(\theta) \} \\
& = \inf_{\theta \in \partial H} \{ \pi_{X^n}(\theta) - \gamma_{X^n}(\theta) \} \\
& \geq -\sup_{\theta \in \partial H} |\pi_{X^n}(\theta) - \gamma_{X^n}(\theta) |. 
\end{align*}
Putting these two inequalities together gives 
\[ | \prior_{X^n}^\star(H) - \Gauss_{X^n}(H) | \leq \sup_{\theta \in \partial H} | \pi_{X^n}(\theta) - \gamma_{X^n}(\theta) |. \]
Therefore, 
\[ | \prior_{X^n}^\star(H) - \Gauss_{X^n}(H) | > \eps \implies \sup_{\theta \in \partial H} | \pi_{X^n}(\theta) - \gamma_{X^n}(\theta) | > \eps, \]
and, by Theorem~\ref{thm:bvm}, the latter event has vanishing $\prob_\Theta$-probability as $n \to \infty$.  

Putting everything together, $d_\text{\sc tv}(\prior_{X^n}^\star, \Gauss_{X^n}) > \eps$ implies existence of a set $H$ for which $| \prior_{X^n}^\star(H) - \Gauss_{X^n}(H) | > \eps$.  This implies a non-negligible difference in the contours so, without loss of generality, the problematic $H$---whose existence is guaranteed---can be taken as a lower level set corresponding to one of the contours.  When $H$ is such a lower level set, $| \prior_{X^n}^\star(H) - \Gauss_{X^n}(H) | > \eps$ implies that $|\pi_{X^n}(\theta) - \gamma_{X^n}(\theta)| > \eps$ for all $\theta$ on the boundary of $H$.  Since the latter event has vanishing $\prob_\Theta$-probability by Theorem~\ref{thm:bvm}, so must the event $d_\text{\sc tv}(\prior_{X^n}^\star, \Gauss_{X^n}) > \eps$, thus proving Theorem~\ref{thm:new.bvm}.

\section{Details from Section~\ref{SS:caveats}}
\label{A:caveats}

\subsection{Examples of over-confidence}

A claim was made in Section~\ref{SS:caveats} that confidence distributions (CDs)---or no-prior Bayes solutions in general---often transform into {\em over-confidence distributions} (oCDs) when (non-linear) marginalization is carried out.  Here I show just a couple simple illustrations to justify this claim.  While my details are different, the insights were inspired by the analysis in \citet{fraser2011}.  

\begin{figure}[t]
\begin{center}
\subfigure[$x=0$]{\scalebox{0.55}{\includegraphics{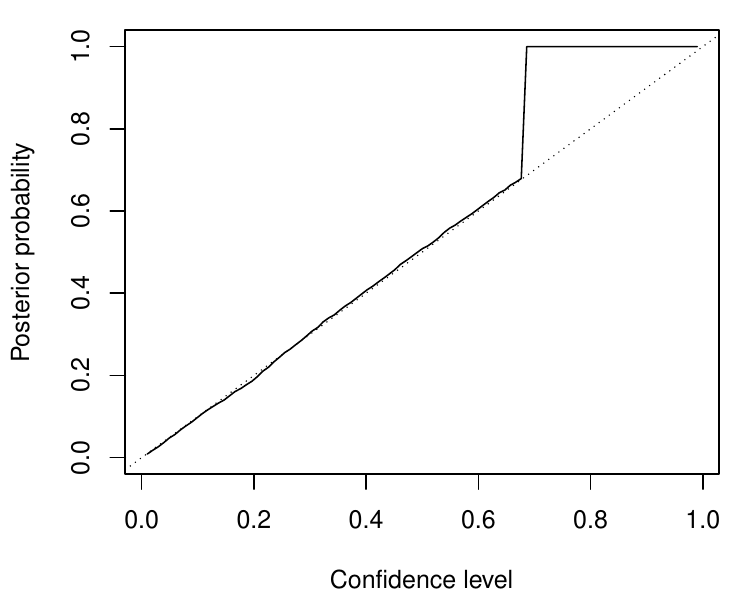}}}
\subfigure[$x=0.5$]{\scalebox{0.55}{\includegraphics{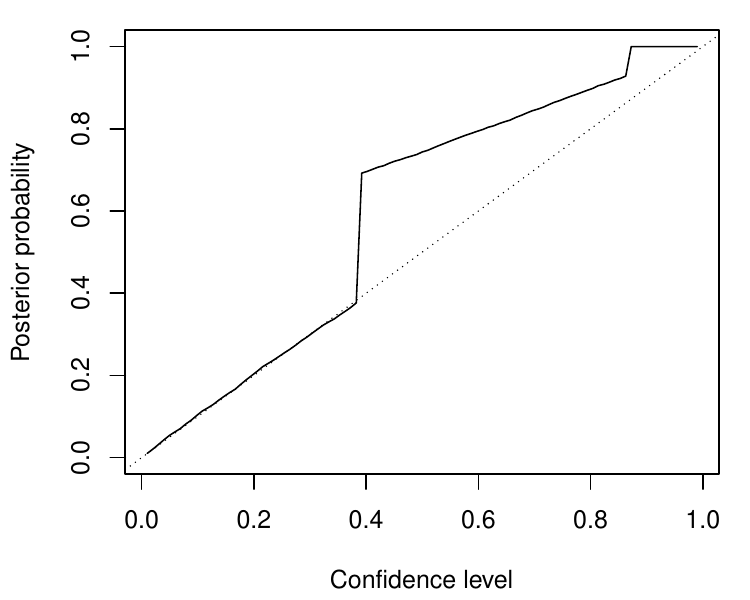}}}
\subfigure[$x=1$]{\scalebox{0.55}{\includegraphics{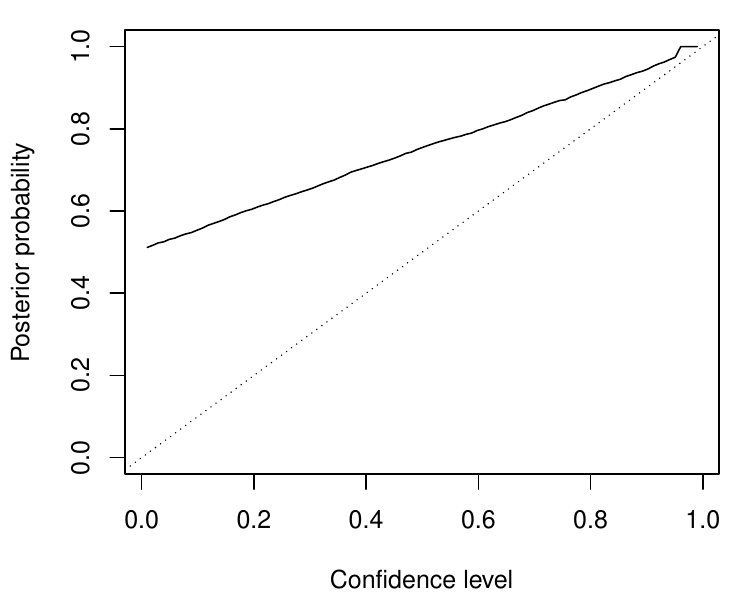}}}
\subfigure[$x=2$]{\scalebox{0.55}{\includegraphics{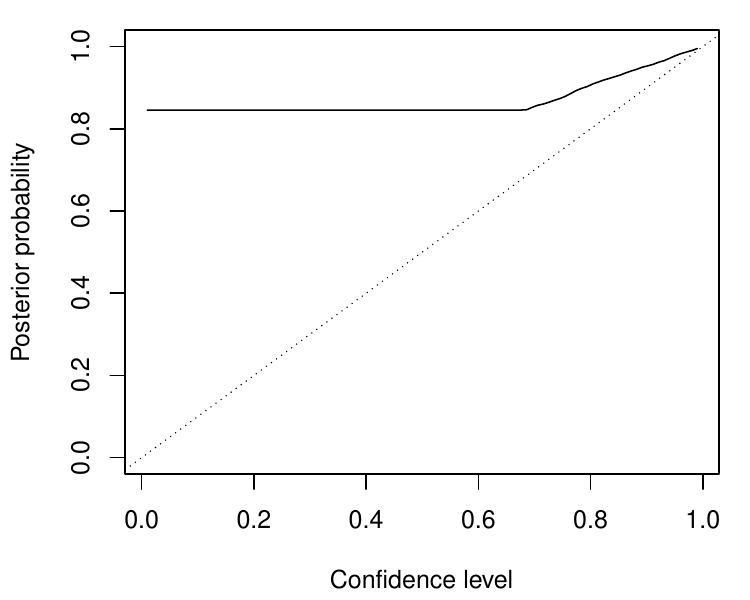}}}
\end{center}
\caption{Plots of the marginal posterior probability assigned $m\{ C_\alpha(x) \}$ as a function of the level $\alpha$ for several different $x$, where $m$ is the bounded function \eqref{eq:m.bounded}.}
\label{fig:bounded}
\end{figure}

The first concerns a single sample $X$ from a simple Gaussian model with known variance, say $\sigma^2=1$, but unknown mean $\Theta$.  Of course, there is an exact confidence interval $C_\alpha(x) = x \pm {\tt qnorm}(1-\frac{\alpha}{2})$ and the standard CD, $\prior_x^\star = \nm(x,1)$, assigns exactly probability $1-\alpha$ to $C_\alpha(x)$.  But now suppose the goal is marginal inference on $\Phi = m(\Theta)$ where $m$ is {\em bounded} and, hence, non-linear, e.g., 
\begin{equation}
\label{eq:m.bounded}
m(\theta) = \max\{-1, \min(1, \theta)\}, \quad \theta \in \RR. 
\end{equation}
The claim is that the marginal CD of $m(\Theta)$ will assign more than probability $1-\alpha$ to the image $m\{C_\alpha(x)\}$ of the original confidence interval under $m$.  Figure~\ref{fig:bounded} shows the (marginal) posterior probability assigned to $m\{ C_\alpha(x) \}$ as a function of the level $\alpha$ for several different realizations $x$.  That the curve generally falls above the diagonal line is a demonstration of the aforementioned over-confidence.  

\begin{figure}[t]
\begin{center}
\subfigure[$x=(0.1, 0.3)$]{\scalebox{0.55}{\includegraphics{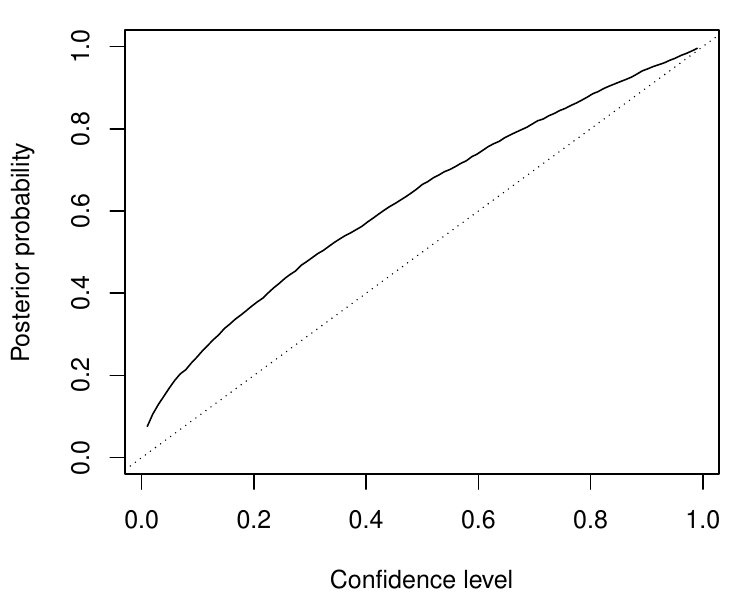}}}
\subfigure[$x=(0.1, 3)$]{\scalebox{0.55}{\includegraphics{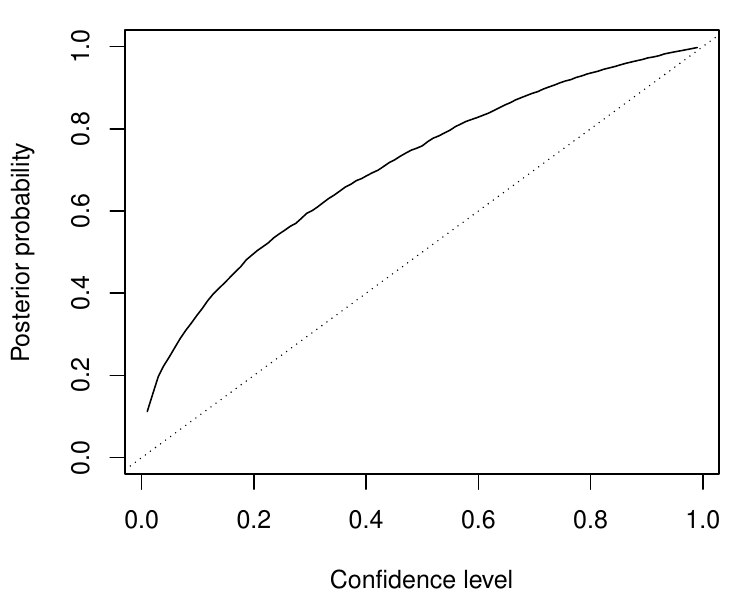}}}
\end{center}
\caption{Plots of the marginal posterior probability assigned $m\{ C_\alpha(x) \}$ as a function of the level $\alpha$ for several different $x$, where $m(\theta) = \|\theta\|^2$ is the squared norm.}
\label{fig:norm2}
\end{figure}

As a second example, suppose that $X=(X_1,X_2)$ is a single realization from a bivariate normal with mean vector $\Theta=(\Theta_1,\Theta_2)$ and identity covariance matrix.  Again, it's straightforward to construct a confidence disc $C_\alpha(x)$ and a confidence distribution $\prior_x^\star = \nm_2(x, I)$.  Consider marginalization via the familiar non-linear function $m(\theta) = \|\theta\|^2$.  I say ``familiar'' because this is an infamous example going to back at least to \citet{stein1959} and more recently in \citet{fraser2011}, \citet{balch.martin.ferson.2017}, and elsewhere.  Figure~\ref{fig:norm2} shows the same plots of marginal posterior probability assigned to $m\{ C_\alpha(x) \}$ as a function of the level $\alpha$ for different realizations $x$.  The over-confidence appears for any $x$ that's away from the origin; the discrepancy between the curve and the diagonal line is increasing in $\|x\|^2$ to a point, and then remains roughly constant as $\|x\|^2 \to \infty$.  

\subsection{Proof of Theorem~\ref{thm:cd}}

Take a generic $\prior_x \in \cred(\uPi_x)$, which has the mixture form in \eqref{eq:char} with kernel $\kernel_x^\alpha$ supported on $C_\alpha(x)$.  If $\prior_x^m$ denote the derived marginal distribution for $\Phi = m(\Theta)$, then, for any $\beta \in [0,1]$, 
\begin{align*}
\prior_x^m\{ C_\beta^m(x)^c \} & = \int_0^1 \kernel_x^\alpha[ m^{-1}\{ C_\beta^m(x)^c \} ] \, d\alpha \\
& = \int_0^\beta \kernel_x^\alpha[ m^{-1}\{ C_\beta^m(x)^c \} ] \, d\alpha, 
\end{align*}
where the second equality follows because 
\[ m^{-1}\{ C_\beta^m(x)^c \} \cap C_\alpha(x) = \{\theta: \pi_x(\theta) < \beta \text{ , } \pi_x(\theta) \geq \alpha\}, \]
and the right-hand side is clearly empty when $\alpha > \beta$.  The above integrand is bounded by 1, so 
\[ \prior_x^m\{ C_\beta^m(x)^c\} \leq \beta, \quad \text{for all $\beta \in [0,1]$}, \]
which implies $\prior_x^m \in \cred(\uPi_x^m)$ and confirms \eqref{eq:ocd}.  $\prior_x^m$ is an inner probabilistic approximation of $\uPi_x^m$ if and only equality in the above display holds for all $\beta$.  This holds if and only if $\kernel_x^\alpha$ is fully supported on $m^{-1}\{ C_\beta^m(x)^c \} \cap C_\alpha(x)$ for all $\beta \geq \alpha$ or, equivalently, if and only if $\kernel_x^\alpha$ is fully supported on $m^{-1}\{ \partial C_\alpha^m(x) \}$.  Since $\prior_x$ is an inner probabilistic approximation of $\uPi_x$ if and only if $\kernel_x^\alpha$ is fully supported on $\partial C_\alpha(x)$, the inner probabilistic approximation of $\uPi_x^m$ can be the marginal derived from an inner probabilistic approximation $\prior_x$ only if $m$ is boundary-surjective.

\section{Computational details from Section~\ref{S:computation}}
\label{A:computation}

Here I review some of the relevant details presented in \citet{imvar.ext} on the choice of $\sigma=\sigma(x,\alpha)$ in Section~\ref{S:computation}.  In this review I'll focus on use of a Gaussian variational family, which is motivated by the asymptotic results in Section~\ref{SS:concentration}, but similar developments can surely be made with other distributional families.  The key features of the Gaussian are that its credible sets can be described in closed-form (ellipsoids) and that it's straightforward to sample on the boundary of these credible sets.  

Let $\mathsf{G}_x^\sigma$ denote the Gaussian distribution $\nm_D(\hat\theta_x, (J_x^\sigma)^{-1})$ with mean equal to the maximum likelihood estimator and covariance matrix the inverse of the $\sigma$-adjusted observed Fisher information matrix in \eqref{eq:J.xi}.  Recall that $\sigma$ scales the eigenvalues of $J_x^\sigma$, in particular, the eigenvalues are decreasing elementwise in $\sigma$.  The corresponding Gaussian credible/confidence sets are 
\[ C_\alpha^\sigma(x) = \{\theta: (\theta - \hat\theta_x)^\top J_x^\sigma (\theta - \hat\theta_x) \leq \chi_{D,1-\alpha}^2\}, \quad \alpha \in [0,1]. \]
The intuition is that $C_\alpha^\sigma(x)$ can be appropriately inflated or deflated by varying $\sigma$, and this is relevant because the Gaussian approximation $\mathsf{G}_x^\sigma$ assigns probability at least $1-\alpha$ to the original IM's $\alpha$-cut $C_\alpha(x)$ if $C_\alpha^\sigma(x) \supseteq C_\alpha(x)$ or, equivalently, if 
\[ \sup_{\theta \not\in C_\alpha^\sigma(x)} \pi_x(\theta) \leq \alpha. \]
Since the contour $\pi_x$ is itself approximately Gaussian \citep{imbvm.ext} and the maximum likelihood estimator $\hat\theta_x$, also the mode of $\pi_x$, is in $C_\alpha^\sigma(x)$, the above supremum should be attained on the boundary $\partial C_\alpha^\sigma(x)$.  Moreover, since equality in the above display implies a near-perfect match between the IM's and the posited Gaussian $\alpha$-cuts, a reasonable goal is to find a root to the function 
\[ g_\alpha(\sigma) := \max_{\theta \in \partial C_\alpha^\sigma(x)} \pi_x(\theta) - \alpha. \]
Design of an iterative algorithm to find this root requires care, primarily because evaluating $\pi_x$ is expensive; so the goal is to evaluate $g_\alpha(\sigma)$ with as few $\pi_x$ evaluations as possible.  \citet{imvar.ext} propose to represent the boundary of $C_\alpha^\sigma(x)$ by $2D$-many vectors 
\begin{equation}
\label{eq:posts}
\vartheta_s^{\sigma, \pm} := \hat\theta_x \pm \bigl\{ \chi_{D,1-\alpha}^2 \, \sigma_s \, / \, \lambda_s \}^{1/2} \, e_s, \quad s=1,\ldots,d, 
\end{equation}
where $(\lambda_s, e_s)$ is the eigenvalue--eigenvector pair corresponding to the $s^\text{th}$ largest eigenvalue in the spectral decomposition of $J_x$.  Define the vector function $\hat g_\alpha$ with components 
\begin{equation}
\label{eq:ghat.xi}
\hat g_{\alpha,s}(\sigma) = \max\{ \pi_x(\vartheta_s^{\sigma, +}), \pi_x(\vartheta_s^{\sigma,-})\} - \alpha, \quad s=1,\ldots,D.
\end{equation}
At least intuitively, negative and positive $\hat g_{\alpha,s}(\sigma)$ indicate that the $\alpha$-cut $C_\alpha^\sigma(x)$ is too large and too small, respectively in the $e_s$-direction.  From here, \citet{imvar.ext} suggest applying a stochastic approximation algorithm---see \citet{robbinsmonro} and \citet{kushner}---to construct a sequence $(\sigma^{(t)}: t \geq 1)$ of $D$-vectors that converges to a root of $\hat g_\alpha$ and, hence, an approximate root of $g_\alpha$.  For an initial guess $\sigma^{(0)}$, the specific sequence is defined as 
\[ \sigma_s^{(t+1)} = \sigma_s^{(t)} + w_{t+1} \, \hat g_{\alpha,s}(\sigma^{(t)}), \quad s=1,\ldots,D, \quad t \geq 0, \]
where $(w_t)$ is a deterministic sequence that satisfies 
\[ \sum_{t=1}^\infty w_t = \infty \quad \text{and} \quad \sum_{t=1}^\infty w_t^2 < \infty. \]
These steps are iterated until (practical) convergence is achieved, and the limit is what I called $\sigma(x,\alpha)$ in Section~\ref{S:computation}.  The basic steps are outlined in Algorithm~\ref{algo:varim}, but I refer to \citet{imvar.ext} for further details and discussion.  

\begin{algorithm}[t]
\SetAlgoLined
requires: data $x$, eigen-pairs $(\lambda_s, e_s)$, and ability to evaluate $\pi_x$\; 
initialize: $\alpha$-level, guess $\sigma^{(0)}$, step size sequence $(w_t)$, and threshold $\eps > 0$\; 
set: {\tt stop = FALSE}, $t=0$\; 
\While{{\tt !stop}}{
construct the representative points $\{\vartheta_s^{\sigma^{(t)}, \pm}: s=1,\ldots,D\}$ as in \eqref{eq:posts}\; 
evaluate $\hat g_{\alpha,s}(\sigma_s^{(t)})$ for $s=1,\ldots,D$ as in \eqref{eq:ghat.xi}\;
update $\sigma_s^{(t+1)} = \sigma_s^{(t)} \pm w_{t+1} \, \hat{g}_{\alpha,s}(\sigma_s^{(t)})$ for $s=1,\ldots,D$\;
\eIf{$\max_s |\sigma_s^{(t+1)} - \sigma_s^{(t)}| < \eps$}{
  $\sigma(z,\alpha) = \sigma^{(t+1)}$\;
  {\tt stop = TRUE}\;
}{
  $t \gets t+1$\;
}
}
return $\sigma(x,\alpha)$\; 
\caption{Determining $\sigma(x,\alpha)$---from \citet{imvar.ext}}
\label{algo:varim}
\end{algorithm}

\section{Additional examples}
\label{A:examples}

\subsection{Bivariate normal correlation}

A challenging one-parameter inference problem is that where $X=(X_1,\ldots,X_n)$ are iid, with $X_i=(X_{1i},X_{2i})$ a bivariate normal random vector with zero means, unit standard deviations, and correlation $\Theta \in \TT = (-1,1)$ to be inferred.  Ironically, the known mean and variance case is more difficult---it's a curved exponential family so the minimal sufficient statistic is not complete and there are various ancillary statistics available to condition on \citep[e.g.,][]{basu1964}.  How this curvature affects asymptotic inference is detailed in \citet{reid2003}.  A likelihood-based possibilistic IM, which is exactly valid for all sample sizes and asymptotically efficient, is easy to construct in theory \citep[][Example~2]{martin.basu}, but naive computation is relatively expensive because there's no pivot structure and no closed-form expression for the maximum likelihood estimator.  This motivates the faster, approximate solution suggested in \citet{immc} and Section~\ref{S:computation} above, based on the variational approximation strategy developed in \citet{imvar.ext}.  

A relevant feature of the exact IM solution is that, when $\Theta$ is relatively close the extremes, i.e., $\Theta \approx \pm 1$, the IM's possibility contour tends to be very asymmetric.  Consequently, approximation suggested in \citet{immc} and Section~\ref{S:computation} above, which are based on stitching together symmetric, Gaussian credible intervals, tend to be a bit conservative.  Other asymmetric distributions, e.g., skewed normal \citep[e.g.,][]{azzalini}, could potentially replace the Gaussian recommendation made in \citet{immc}, but that's beyond the scope of the present paper.  I'm mentioning this point here because the same conservative behavior will be observed below with the inner probabilistic approximation; this isn't a shortcoming of the inner probabilistic approximation itself, but of the choice to bound the IM's asymmetric level sets using symmetric Gaussian credible limits.  

\begin{figure}[t]
\begin{center}
\subfigure[Posterior results in Efron's example]{\scalebox{0.55}{\includegraphics{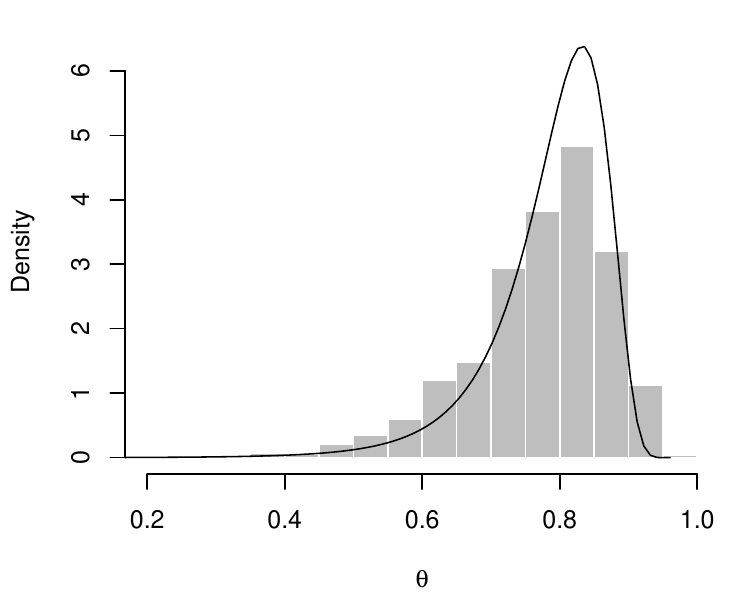}}}
\subfigure[Cumulative distribution function of $U_X$]{\scalebox{0.55}{\includegraphics{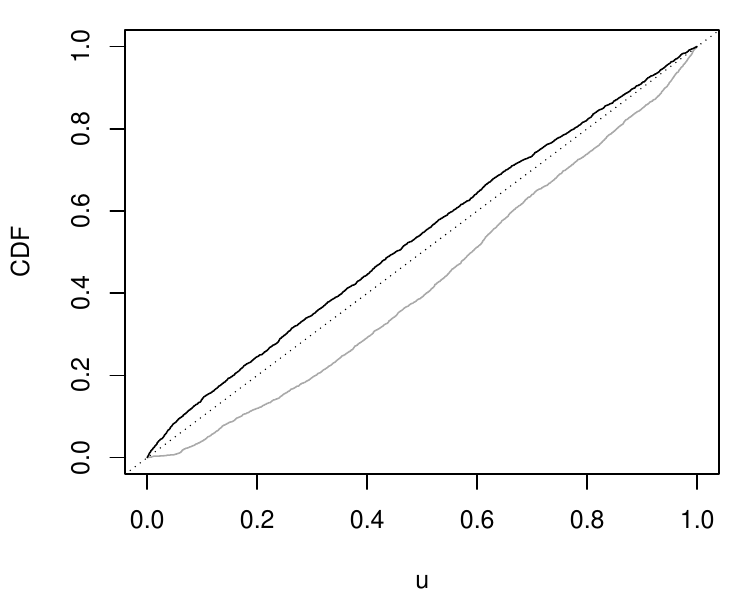}}}
\end{center}
\caption{Comparison of Jeffreys-prior Bayes and the possibilistic IM's inner probabilistic approximation.  In Panel~(a), the solid line is the Bayes posterior density and the bars are based on samples from the inner probabilistic approximation.  In Panel~(b), the black and gray lines correspond to the Bayes and inner probabilistic approximation solutions, respectively.  Definition and explanation of $U_X$ is given in the text.}
\label{fig:bvn}
\end{figure}

For illustration, let $X=(X_1,\ldots,X_n)$ denote (a centered and scaled version of) the law school admissions data analyzed in \citet{efron1982}, where $n=15$ and the maximum likelihood estimator of $\Theta$ is $\hat\theta_x = 0.789$.  Figure~\ref{fig:bvn}(a) shows a plot of the Bayesian posterior density, based on Jeffreys prior, along with samples from the inner probabilistic approximation.  (The latter is based on first reparametrizing in terms of Fisher's z-transformation, i.e., $\Psi = \tanh^{-1}\Theta$, obtaining the inner probabilistic approximation of $\Psi$, and then transforming back to $\Theta = \tanh\Psi$.)  Both distributions are concentrated around the maximum likelihood estimator and have similar shapes, but the Bayes posterior is a little tighter than inner probabilistic approximation.  

Is the Bayes posterior distribution too concentrated in the above illustration?  If so, then is this aggressive behavior systematic or just a fluke in this one example?  To answer these questions, I carried out a brief simulation study, involving 3000 data sets of size $n=10$ drawn from a standard bivariate normal distribution with $\Theta=0.7$.  For both methods, and each data set $X$, I evaluate 
\[ U_X = 1 - \bigl| 2 Q_X(\Theta) - 1 \bigr|, \]
where $Q_X(\Theta)$ is the cumulative distribution function of the derived posterior distribution, based on the Bayes posterior or the inner probabilistic approximation, evaluated at the true $\Theta=0.7$.  If the random variable $U_X$ is stochastically no smaller than $\unif(0,1)$, then the corresponding equal-tailed posterior credible intervals have frequentist coverage probability at or above the nominal level; if $U_X$ is stochastically smaller than $\unif(0,1)$, the credible intervals undercover.  Figure~\ref{fig:bvn}(b) plots the cumulative distribution function of $U_X$ for the two methods and we find that the Jeffreys-prior Bayes solution and the inner probabilistic approximation solution yield $U_X$ that are stochastically smaller and larger than $\unif(0,1)$, respectively.  This means that the former's credible intervals generally undercover while the latter's generally overcover; therefore, the Bayes posterior density displayed in Figure~\ref{fig:bvn}(a) might indeed be ``too concentrated.'' 

\subsection{Gamma shape and scale}

Inference on the parameters of a gamma model is a classical problem.  When the shape parameter is unknown, this model lacks the group transformation structure that makes specification of default priors having suitable reliability properties non-trivial.  The go-to no-prior Bayes solution invokes Jeffreys prior on the shape--scale parameter pair.  Alternatively, a possibilistic IM solution is readily available and doesn't require any prior specification and has reliability guarantees.  This IM solution is somewhat expensive to compute, however, at least if a naive approach is taken.  \citet{immc} showed how this naive approach can be drastically improved and it's on this more sophisticated/efficient approach that my solution is based.  Indeed, here I'm proposing the inner probabilistic approximation of that possibilistic IM, and the sampling from that ``posterior distribution'' is based on the strategy put forward in \citet{immc}.  

\begin{figure}[t]
\begin{center}
\subfigure[shape--scale parametrization]{\scalebox{0.55}{\includegraphics{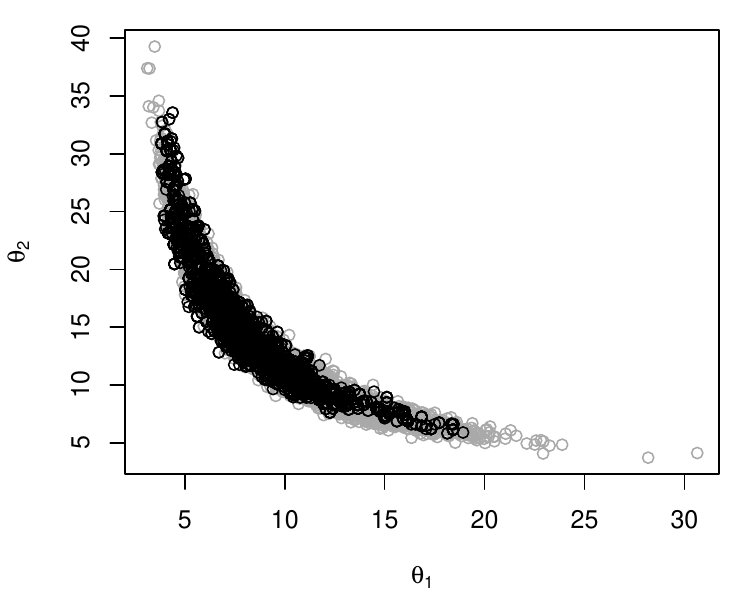}}}
\subfigure[log(shape)--log(scale) parametrization]{\scalebox{0.55}{\includegraphics{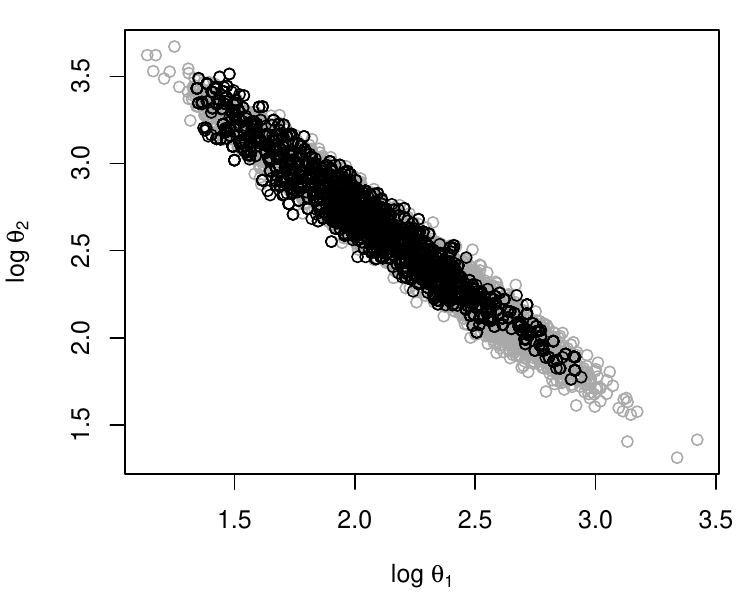}}}
\end{center}
\caption{Black points are samples from the Jeffreys-prior Bayes posterior distribution in the gamma model; gray points are samples from the IM's inner probabilistic approximation.}
\label{fig:gamma.rat}
\end{figure}

As a first illustration, reconsider the data analyzed in \citet{fraser.reid.wong.1997} on the survival time (in weeks) for $n=20$ rats exposed to a certain amount of radiation.  Assuming a gamma model with unknown shape and scale parameters, the posterior distributions for the two no-prior Bayes solutions---Jeffreys prior Bayes and the IM's inner probabilistic approximation---are shown in Figure~\ref{fig:gamma.rat}, both in the original and log-transformed parametrization.  The two posteriors look generally similar and, after log-transformation, they both have Gaussian-like elliptical shapes.  The only notable difference is that the Bayes solution is a bit more tightly concentrated than the IM-based solution.

\begin{figure}[t]
\begin{center}
\scalebox{0.6}{\includegraphics{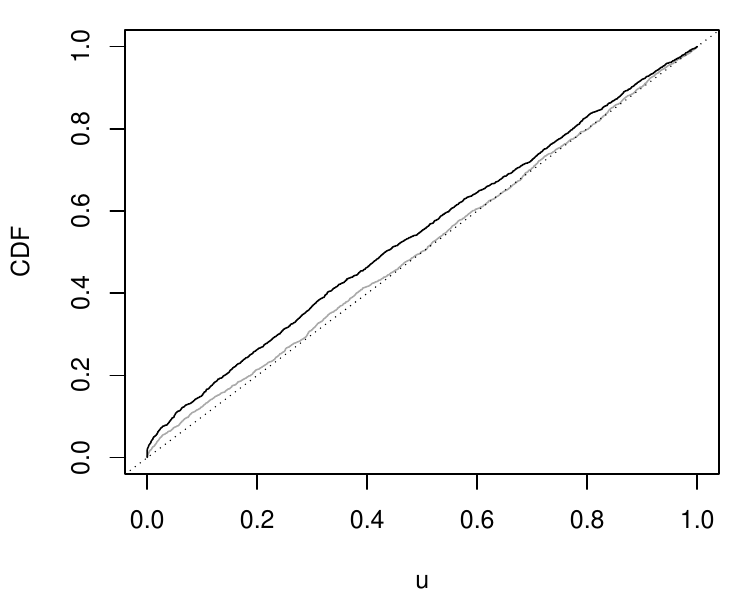}}
\end{center}
\caption{Comparison of Jeffreys-prior Bayes and the possibilistic IM's inner probabilistic approximation, in particular, distribution functions of $U_X$ corresponding to the Bayes (black) and IM inner probabilistic approximation (gray). }
\label{fig:gamma.sim}
\end{figure}

Like in the bivariate normal correlation example above, it's natural to ask if the Bayes solution is actually too tightly concentrated.  To assess this, I'll do a brief simulation study.  Here I draw 2500 samples of size $n=20$, with true shape and scale parameters set to be $\Theta_1=7$ and $\Theta_2=3$, respectively.  Similar to the bivariate normal example in the previous section, for both no-prior Bayes solutions, define the quadratic form of the form 
\[ \kappa_x(\theta) = (\log\theta - \hat\mu_x)^\top \hat\sigma_x^{-1} (\log\theta - \hat\mu_x), \]
where $\hat\mu_x$ and $\hat\sigma_x$ are estimates of the posterior mean and posterior covariance matrix of the log-transformed shape and scale parameters.  Then I let 
\[ Q_x(\theta) = \prior_x\{ \kappa_x(\widetilde\Theta) \leq \kappa_x(\theta) \} \]
denote the distribution function of the quadratic form relative to the posterior distribution and, finally, I define 
\[ U_X = 1 - \bigl| 2 Q_X(\Theta) - 1 \bigr|, \]
where $\Theta=(7,3)$ is the true shape and scale parameters.  Figure~\ref{fig:gamma.sim} plots the distribution functions of $U_X$ for each of the two methods.  Being above the diagonal line indicates that the elliptical credible sets derived from the method's posterior samples have frequentist coverage probability below the nominal level; similarly, being on or below the line indicates that the credible sets have frequentist coverage probability at or above the nominal level.  In this case, the Jeffreys-prior Bayes solution shows severe undercoverage while the IM's inner probabilistic approximation shows signs of very slight undercoverage for 90 and 95\% credible regions.  This suggests that the Jeffreys-prior Bayes solutions displayed in Figure~\ref{fig:gamma.rat} might indeed be too tightly concentrated. 

\bibliographystyle{apalike}
\bibliography{/Users/rgmarti3/Dropbox/Research/mybib.bib}

\begin{thebibliography}{}

\bibitem[Azzalini, 1985]{azzalini}
Azzalini, A. (1985).
\newblock A class of distributions which includes the normal ones.
\newblock {\em Scand. J. Statist.}, 12(2):171--178.

\bibitem[Balch et~al., 2019]{balch.martin.ferson.2017}
Balch, M.~S., Martin, R., and Ferson, S. (2019).
\newblock Satellite conjunction analysis and the false confidence theorem.
\newblock {\em Proc. Royal Soc. A}, 475(2227):2018.0565.

\bibitem[Basu, 1964]{basu1964}
Basu, D. (1964).
\newblock Recovery of ancillary information.
\newblock {\em Sankhy\={a} Ser. A}, 26:3--16.

\bibitem[Berger, 2006]{berger2006}
Berger, J. (2006).
\newblock The case for objective {B}ayesian analysis.
\newblock {\em Bayesian Anal.}, 1(3):385--402.

\bibitem[Berger et~al., 2024]{berger.objective.book}
Berger, J., Bernardo, J., and Sun, D. (2024).
\newblock {\em Objective {B}ayesian {I}nference}.
\newblock World Scientific Publishing Co.

\bibitem[Bernardo, 1979]{bernardo1979}
Bernardo, J.-M. (1979).
\newblock Reference posterior distributions for {B}ayesian inference.
\newblock {\em J. Roy. Statist. Soc. Ser. B}, 41:113--147.

\bibitem[Bernardo and Smith, 1994]{bernardo.smith.book}
Bernardo, J.-M. and Smith, A. F.~M. (1994).
\newblock {\em Bayesian {T}heory}.
\newblock Wiley Series in Probability and Mathematical Statistics: Probability
  and Mathematical Statistics. John Wiley \& Sons, Ltd., Chichester.

\bibitem[Bickel et~al., 1998]{bickel1998}
Bickel, P.~J., Klaassen, C. A.~J., Ritov, Y., and Wellner, J.~A. (1998).
\newblock {\em Efficient and {A}daptive {E}stimation for {S}emiparametric
  {M}odels}.
\newblock Springer-Verlag, New York.

\bibitem[Birnbaum, 1961]{birnbaum1961}
Birnbaum, A. (1961).
\newblock Confidence curves: an omnibus technique for estimation and testing
  statistical hypotheses.
\newblock {\em J. Amer. Statist. Assoc.}, 56:246--249.

\bibitem[Blaker and Spj{\o}tvoll, 2000]{blaker.spjotvoll.2000}
Blaker, H. and Spj{\o}tvoll, E. (2000).
\newblock Paradoxes and improvements in interval estimation.
\newblock {\em Amer. Statist.}, 54(4):242--247.

\bibitem[Cella and Martin, 2022a]{cella.martin.imrisk}
Cella, L. and Martin, R. (2022a).
\newblock Direct and approximately valid probabilistic inference on a class of
  statistical functionals.
\newblock {\em Internat. J. Approx. Reason.}, 151:205--224.

\bibitem[Cella and Martin, 2022b]{imconformal.supervised}
Cella, L. and Martin, R. (2022b).
\newblock Valid inferential models for prediction in supervised learning
  problems.
\newblock {\em Internat. J. Approx. Reason.}, 150:1--18.

\bibitem[Cella and Martin, 2023]{cella.martin.probing}
Cella, L. and Martin, R. (2023).
\newblock Possibility-theoretic statistical inference offers performance and
  probativeness assurances.
\newblock {\em Internat. J. Approx. Reason.}, 163:109060.

\bibitem[Cella and Martin, 2025]{imvar.ext}
Cella, L. and Martin, R. (2025).
\newblock Computationally efficient variational-like approximations of
  possibilistic inferential models.
\newblock {\em Internat. J. Approx. Reason.}, 186:Paper No. 109506.

\bibitem[Chang and Pollard, 1997]{chang.pollard.1997}
Chang, J.~T. and Pollard, D. (1997).
\newblock Conditioning as disintegration.
\newblock {\em Statist. Neerlandica}, 51(3):287--317.

\bibitem[Cortinovis and Caron, 2024]{cortinovis.caron.2024}
Cortinovis, S. and Caron, F. (2024).
\newblock {B}ayes-assisted confidence regions: {F}ocal point estimator and
  bounded-influence priors.
\newblock {\tt arXiv:2410.20169}.

\bibitem[Couso et~al., 2001]{cuoso.etal.2001}
Couso, I., Montes, S., and Gil, P. (2001).
\newblock The necessity of the strong {$\alpha$}-cuts of a fuzzy set.
\newblock {\em Internat. J. Uncertain. Fuzziness Knowledge-Based Systems},
  9(2):249--262.

\bibitem[Cover and Thomas, 2006]{cover.thomas.book}
Cover, T.~M. and Thomas, J.~A. (2006).
\newblock {\em Elements of {I}nformation {T}heory}.
\newblock Wiley-Interscience [John Wiley \& Sons], Hoboken, NJ, second edition.

\bibitem[Cox, 2006]{cox2006}
Cox, D.~R. (2006).
\newblock {\em Principles of Statistical Inference}.
\newblock Cambridge University Press, Cambridge.

\bibitem[Cram\'er, 1946]{cramer.book}
Cram\'er, H. (1946).
\newblock {\em Mathematical {M}ethods of {S}tatistics}, volume vol. 9 of {\em
  Princeton Mathematical Series}.
\newblock Princeton University Press, Princeton, NJ.

\bibitem[Cunen et~al., 2020]{prsa.conf}
Cunen, C., Hjort, N.~L., and Schweder, T. (2020).
\newblock Confidence in confidence distributions!
\newblock {\em Proc. Roy. Soc. A}, 476:20190781.

\bibitem[Datta and Ghosh, 1995]{datta.ghosh.1995}
Datta, G.~S. and Ghosh, J.~K. (1995).
\newblock On priors providing frequentist validity for {B}ayesian inference.
\newblock {\em Biometrika}, 82(1):37--45.

\bibitem[Dawid, 2024]{dawid2020}
Dawid, A.~P. (2024).
\newblock Fiducial inference then and now.
\newblock In Berger, J., Meng, X.-L., Reid, N., and Xie, M.-g., editors, {\em
  Handbook of {B}ayesian, {F}iducial, and {F}requentist {I}nference}, pages
  83--105. Chapman \& Hall/CRC Press.

\bibitem[Dawid et~al., 1973]{dawid.stone.zidek.1973}
Dawid, A.~P., Stone, M., and Zidek, J.~V. (1973).
\newblock Marginalization paradoxes in {B}ayesian and structural inference.
\newblock {\em J. Roy. Statist. Soc. Ser. B}, 35:189--233.
\newblock With discussion and reply by the authors.

\bibitem[De~Cooman, 1997]{cooman.poss1}
De~Cooman, G. (1997).
\newblock Possibility theory. {I}. {T}he measure- and integral-theoretic
  groundwork.
\newblock {\em Internat. J. Gen. Systems}, 25(4):291--323.

\bibitem[De~Cooman and Aeyels, 1999]{cooman.aeyels.1999}
De~Cooman, G. and Aeyels, D. (1999).
\newblock Supremum preserving upper probabilities.
\newblock {\em Inform. Sci.}, 118(1-4):173--212.

\bibitem[Dempster, 1966]{dempster1966}
Dempster, A.~P. (1966).
\newblock New methods for reasoning towards posterior distributions based on
  sample data.
\newblock {\em Ann. Math. Statist.}, 37:355--374.

\bibitem[Den{\oe}ux, 2006]{denoeux2006}
Den{\oe}ux, T. (2006).
\newblock Constructing belief functions from sample data using multinomial
  confidence regions.
\newblock {\em Internat. J. of Approx. Reason.}, 42(3):228--252.

\bibitem[Den{\oe}ux, 2014]{denoeux2014}
Den{\oe}ux, T. (2014).
\newblock Likelihood-based belief function: justification and some extensions
  to low-quality data.
\newblock {\em Internat. J. Approx. Reason.}, 55(7):1535--1547.

\bibitem[Den{\oe}ux, 2023a]{denoeux.fuzzy.2023}
Den{\oe}ux, T. (2023a).
\newblock Parametric families of continuous belief functions based on
  generalized {G}aussian random fuzzy numbers.
\newblock {\em Fuzzy Sets and Systems}, 471:Paper No. 108679, 33.

\bibitem[Den{\oe}ux, 2023b]{denoeux.fuzzy.2022}
Den{\oe}ux, T. (2023b).
\newblock Reasoning with fuzzy and uncertain evidence using epistemic random
  fuzzy sets: general framework and practical models.
\newblock {\em Fuzzy Sets and Systems}, 453:1--36.

\bibitem[Destercke and Dubois, 2014]{destercke.dubois.2014}
Destercke, S. and Dubois, D. (2014).
\newblock Special cases.
\newblock In {\em Introduction to {I}mprecise {P}robabilities}, Wiley Ser.
  Probab. Stat., pages 79--92. Wiley, Chichester.

\bibitem[Dubois, 2006]{dubois2006}
Dubois, D. (2006).
\newblock Possibility theory and statistical reasoning.
\newblock {\em Comput. Statist. Data Anal.}, 51(1):47--69.

\bibitem[Dubois and Den{\oe}ux, 2010]{dubois.denoeux.2010}
Dubois, D. and Den{\oe}ux, T. (2010).
\newblock Statistical inference with belief functions and possibility measures:
  a discussion of basic assumptions.
\newblock In {\em International Conference on Soft Methods in Probability and
  Statistics (SMPS 2010)}, volume~77 of {\em Advances in {I}ntelligent and
  {S}oft {C}omputing}, pages 217--225. Springer.

\bibitem[Dubois et~al., 2004]{dubois.etal.2004}
Dubois, D., Foulloy, L., Mauris, G., and Prade, H. (2004).
\newblock Probability-possibility transformations, triangular fuzzy sets, and
  probabilistic inequalities.
\newblock {\em Reliab. Comput.}, 10(4):273--297.

\bibitem[Dubois and Prade, 1988]{dubois.prade.book}
Dubois, D. and Prade, H. (1988).
\newblock {\em Possibility {T}heory}.
\newblock Plenum Press, New York.

\bibitem[Dubois and Prade, 1990]{dubois.prade.1990}
Dubois, D. and Prade, H. (1990).
\newblock Consonant approximations of belief functions.
\newblock {\em Internat. J. Approx. Reason.}, 4(5-6):419--449.

\bibitem[Dubois et~al., 2008]{dubois.prade.smets.2008}
Dubois, D., Prade, H., and Smets, P. (2008).
\newblock A definition of subjective possibility.
\newblock {\em Internat. J. Approx. Reason.}, 48(2):352--364.

\bibitem[Eaton, 1989]{eaton1989}
Eaton, M.~L. (1989).
\newblock {\em Group {I}nvariance {A}pplications in {S}tatistics}.
\newblock Institute of Mathematical Statistics, Hayward, CA.

\bibitem[Eaton and Sudderth, 2012]{eaton.sudderth.2012}
Eaton, M.~L. and Sudderth, W.~D. (2012).
\newblock Invariance, model matching and probability matching.
\newblock {\em Sankhya A}, 74(2):170--193.

\bibitem[Efron, 1982]{efron1982}
Efron, B. (1982).
\newblock {\em The {J}ackknife, the {B}ootstrap and other {R}esampling
  {P}lans}, volume~38 of {\em CBMS-NSF Regional Conference Series in Applied
  Mathematics}.
\newblock Society for Industrial and Applied Mathematics (SIAM), Philadelphia,
  Pa.

\bibitem[Efron, 1998]{efron1998}
Efron, B. (1998).
\newblock R. {A}. {F}isher in the 21st century.
\newblock {\em Statist. Sci.}, 13(2):95--122.

\bibitem[Efron, 2013]{efron.cd.discuss}
Efron, B. (2013).
\newblock Discussion: ``{C}onfidence distribution, the frequentist distribution
  estimator of a parameter: a review'' [mr3047496].
\newblock {\em Int. Stat. Rev.}, 81(1):41--42.

\bibitem[Fisher, 1933]{fisher1933}
Fisher, R.~A. (1933).
\newblock The concepts of inverse probability and fiducial probability
  referring to unknown parameters.
\newblock {\em Proc. R. Soc. Lond. A.}, 139:343--348.

\bibitem[Fisher, 1935a]{fisher1935a}
Fisher, R.~A. (1935a).
\newblock The fiducial argument in statistical inference.
\newblock {\em Ann. Eugenics}, 6:391--398.

\bibitem[Fisher, 1935b]{fisher1935b}
Fisher, R.~A. (1935b).
\newblock The logic of inductive inference.
\newblock {\em J. Roy. Statist. Soc.}, 98:39--82.

\bibitem[Fisher, 1939]{fisher1939}
Fisher, R.~A. (1939).
\newblock The comparison of samples with possibly unequal variances.
\newblock {\em Ann. Eugenics}, 9:174--180.

\bibitem[Fraser, 1968]{fraser1968}
Fraser, D. A.~S. (1968).
\newblock {\em The {S}tructure of {I}nference}.
\newblock John Wiley \& Sons Inc., New York.

\bibitem[Fraser, 1991]{fraser1991}
Fraser, D. A.~S. (1991).
\newblock Statistical inference: likelihood to significance.
\newblock {\em J. Amer. Statist. Assoc.}, 86(414):258--265.

\bibitem[Fraser, 2011a]{fraser2011}
Fraser, D. A.~S. (2011a).
\newblock Is {B}ayes posterior just quick and dirty confidence?
\newblock {\em Statist. Sci.}, 26(3):299--316.

\bibitem[Fraser, 2011b]{fraser2011.rejoinder}
Fraser, D. A.~S. (2011b).
\newblock Rejoinder: ``{I}s {B}ayes posterior just quick and dirty
  confidence?''.
\newblock {\em Statist. Sci.}, 26(3):329--331.

\bibitem[Fraser, 2013]{fraser.cd.discuss}
Fraser, D. A.~S. (2013).
\newblock Discussion: ``{C}onfidence distribution, the frequentist distribution
  estimator of a parameter: a review'' [mr3047496].
\newblock {\em Int. Stat. Rev.}, 81(1):42--48.

\bibitem[Fraser, 2014]{fraser.copss}
Fraser, D. A.~S. (2014).
\newblock Why does statistics have two theories?
\newblock In Lin, X., Genest, C., Banks, D.~L., Molenberghs, G., Scott, D.~W.,
  and Wang, J.-L., editors, {\em Past, Present, and Future of Statistical
  Science}, chapter~22. Chapman \& Hall/CRC Press.

\bibitem[Fraser et~al., 2016]{fraser.etal.2016}
Fraser, D. A.~S., B\'edard, M., Wong, A., Lin, W., and Fraser, A.~M. (2016).
\newblock Bayes, reproducibility and the quest for truth.
\newblock {\em Statist. Sci.}, 31(4):578--590.

\bibitem[Fraser et~al., 1997]{fraser.reid.wong.1997}
Fraser, D. A.~S., Reid, N., and Wong, A. (1997).
\newblock Simple and accurate inference for the mean of a gamma model.
\newblock {\em Canad. J. Statist.}, 25(1):91--99.

\bibitem[Gr\"unwald, 2018]{grunwald.safe}
Gr\"unwald, P. (2018).
\newblock Safe probability.
\newblock {\em J. Statist. Plann. Inference}, 195:47--63.

\bibitem[Gr\"{u}nwald, 2023]{grunwald.epost}
Gr\"{u}nwald, P.~D. (2023).
\newblock The e-posterior.
\newblock {\em Philos. Trans. Roy. Soc. A}, 381(2247):Paper No. 20220146, 21.

\bibitem[H\'ajek, 1972]{hajek1972}
H\'ajek, J. (1972).
\newblock Local asymptotic minimax and admissibility in estimation.
\newblock In {\em Proceedings of the {S}ixth {B}erkeley {S}ymposium on
  {M}athematical {S}tatistics and {P}robability ({U}niv. {C}alifornia,
  {B}erkeley, {C}alif., 1970/1971), {V}ol. {I}: {T}heory of statistics}, pages
  175--194. Univ. California Press, Berkeley, CA.

\bibitem[Halmos, 1950]{halmos.measure}
Halmos, P.~R. (1950).
\newblock {\em Measure {T}heory}.
\newblock D. Van Nostrand Co., Inc., New York, N. Y.

\bibitem[Hannig et~al., 2016]{hannig.review}
Hannig, J., Iyer, H., Lai, R. C.~S., and Lee, T. C.~M. (2016).
\newblock Generalized fiducial inference: a review and new results.
\newblock {\em J. Amer. Statist. Assoc.}, 111(515):1346--1361.

\bibitem[Hose, 2022]{hose2022thesis}
Hose, D. (2022).
\newblock {\em Possibilistic {R}easoning with {I}mprecise {P}robabilities:
  {S}tatistical {I}nference and {D}ynamic {F}iltering}.
\newblock PhD thesis, University of Stuttgart.
\newblock \url{https://dominikhose.github.io/dissertation/diss_dhose.pdf}.

\bibitem[Jaynes, 2003]{jaynes2003}
Jaynes, E.~T. (2003).
\newblock {\em Probability {T}heory: {T}he {L}ogic of {S}cience}.
\newblock Cambridge University Press, Cambridge.

\bibitem[Jeffreys, 1940]{jeffreys1940}
Jeffreys, H. (1940).
\newblock Note on the {B}ehrens-{F}isher formula.
\newblock {\em Ann. Eugenics}, 10:48--51.

\bibitem[Jeffreys, 1946]{jeffreys1946}
Jeffreys, H. (1946).
\newblock An invariant form for the prior probability in estimation problems.
\newblock {\em Proc. Roy. Soc. London Ser. A}, 186:453--461.

\bibitem[Kass and Wasserman, 1996]{kass.wasserman.1996}
Kass, R.~E. and Wasserman, L. (1996).
\newblock The selection of prior distributions by formal rules.
\newblock {\em J. Amer. Statist. Assoc.}, 91(435):1343--1370.

\bibitem[Keynes, 1921]{keynes.probability}
Keynes, J.~M. (1921).
\newblock {\em A {T}reatise on {P}robability}.
\newblock Macmillan and Co.

\bibitem[Kim and Cohen, 1998]{kimcohen1998}
Kim, S.-H. and Cohen, A.~S. (1998).
\newblock On the {B}ehrens-{F}isher problem: A review.
\newblock {\em J. Educ. Behav. Stat.}, 23(4):356--377.

\bibitem[Kushner and Yin, 2003]{kushner}
Kushner, H.~J. and Yin, G.~G. (2003).
\newblock {\em Stochastic {A}pproximation and {R}ecursive {A}lgorithms and
  {A}pplications}.
\newblock Springer-Verlag, New York, second edition.

\bibitem[Le~Cam, 1956]{lecam1956}
Le~Cam, L. (1956).
\newblock On the asymptotic theory of estimation and testing hypotheses.
\newblock In {\em Proceedings of the {T}hird {B}erkeley {S}ymposium on
  {M}athematical {S}tatistics and {P}robability, 1954--1955, vol. {I}}, pages
  129--156. Univ. California Press, Berkeley-Los Angeles, Calif.

\bibitem[Le~Cam, 1960]{lecam1960}
Le~Cam, L. (1960).
\newblock Locally asymptotically normal families of distributions. {C}ertain
  approximations to families of distributions and their use in the theory of
  estimation and testing hypotheses.
\newblock {\em Univ. California Publ. Statist.}, 3:37--98.

\bibitem[Le~Cam, 1970]{lecam1970}
Le~Cam, L. (1970).
\newblock On the assumptions used to prove asymptotic normality of maximum
  likelihood estimates.
\newblock {\em Ann. Math. Statist.}, 41:802--828.

\bibitem[Lehmann, 1975]{lehmann1975}
Lehmann, E.~L. (1975).
\newblock {\em Nonparametrics: Statistical Methods Based on Ranks}.
\newblock Holden-Day Inc., San Francisco, Calif.

\bibitem[Lehmann and Casella, 1998]{lehmann.casella.1998}
Lehmann, E.~L. and Casella, G. (1998).
\newblock {\em Theory of {P}oint {E}stimation}.
\newblock Springer Texts in Statistics. Springer-Verlag, New York, second
  edition.

\bibitem[Lindley, 1958]{lindley1958}
Lindley, D.~V. (1958).
\newblock Fiducial distributions and {B}ayes' theorem.
\newblock {\em J. Roy. Statist. Soc. Ser. B}, 20:102--107.

\bibitem[Liu et~al., 2022]{liu.liu.xie.2022}
Liu, D., Liu, R.~Y., and Xie, M.-g. (2022).
\newblock Nonparametric fusion learning for multiparameters: synthesize
  inferences from diverse sources using data depth and confidence distribution.
\newblock {\em J. Amer. Statist. Assoc.}, 117(540):2086--2104.

\bibitem[Mardia and Jupp, 2000]{mardia.jupp.book}
Mardia, K.~V. and Jupp, P.~E. (2000).
\newblock {\em Directional {S}tatistics}.
\newblock Wiley Series in Probability and Statistics. John Wiley \& Sons, Ltd.,
  Chichester.

\bibitem[Martin, 2019]{martin.nonadditive}
Martin, R. (2019).
\newblock False confidence, non-additive beliefs, and valid statistical
  inference.
\newblock {\em Internat. J. Approx. Reason.}, 113:39--73.

\bibitem[Martin, 2021a]{imchar}
Martin, R. (2021a).
\newblock An imprecise-probabilistic characterization of frequentist
  statistical inference.
\newblock {\tt arXiv:2112.10904}.

\bibitem[Martin, 2021b]{imdec}
Martin, R. (2021b).
\newblock Inferential models and the decision-theoretic implications of the
  validity property.
\newblock {\tt arXiv:2112.13247}.

\bibitem[Martin, 2022a]{martin.partial}
Martin, R. (2022a).
\newblock Valid and efficient imprecise-probabilistic inference with partial
  priors, {I}. {F}irst results.
\newblock {\tt arXiv:2203.06703}.

\bibitem[Martin, 2022b]{martin.partial2}
Martin, R. (2022b).
\newblock Valid and efficient imprecise-probabilistic inference with partial
  priors, {II}. {G}eneral framework.
\newblock {\tt arXiv:2211.14567}.

\bibitem[Martin, 2023a]{martin.isipta2023}
Martin, R. (2023a).
\newblock Fiducial inference viewed through a possibility-theoretic inferential
  model lens.
\newblock In Miranda, E., Montes, I., Quaeghebeur, E., and Vantaggi, B.,
  editors, {\em Proceedings of the Thirteenth International Symposium on
  Imprecise Probability: Theories and Applications}, volume 215 of {\em
  Proceedings of Machine Learning Research}, pages 299--310. PMLR.

\bibitem[Martin, 2023b]{martin.partial3}
Martin, R. (2023b).
\newblock Valid and efficient imprecise-probabilistic inference with partial
  priors, {III}. {M}arginalization.
\newblock {\tt arXiv:2309.13454}.

\bibitem[Martin, 2024a]{martin.basu}
Martin, R. (2024a).
\newblock A possibility-theoretic solution to {B}asu's {B}ayesian--frequentist
  via media.
\newblock {\em Sankhya A}, 86:43--70.

\bibitem[Martin, 2024b]{martin.belief2024}
Martin, R. (2024b).
\newblock Which statistical hypotheses are afflicted by false confidence?
\newblock In Bi, Y., Jousselme, A.-L., and Denoeux, T., editors, {\em BELIEF
  2024}, volume 14909 of {\em Lecture Notes in Artificial Intelligence}, pages
  140--149, Switzerland. Springer Nature.

\bibitem[Martin, 2026a]{immc}
Martin, R. (2026a).
\newblock An efficient {M}onte {C}arlo method for valid prior-free
  possibilistic statistical inference.
\newblock {\em J. Amer. Statist. Assoc.}, to appear; {\tt arXiv:2501.10585}.

\bibitem[Martin, 2026b]{imreview}
Martin, R. (2026b).
\newblock Possibilistic inferential models: a review.
\newblock {\em J. Amer. Statist. Assoc.}, 121(553):807--826.

\bibitem[Martin and Liu, 2013]{imbasics}
Martin, R. and Liu, C. (2013).
\newblock Inferential models: a framework for prior-free posterior
  probabilistic inference.
\newblock {\em J. Amer. Statist. Assoc.}, 108(501):301--313.

\bibitem[Martin and Liu, 2015]{imbook}
Martin, R. and Liu, C. (2015).
\newblock {\em Inferential {M}odels}, volume 147 of {\em Monographs on
  Statistics and Applied Probability}.
\newblock CRC Press, Boca Raton, FL.

\bibitem[Martin et~al., 2026]{imdec.ext}
Martin, R., Prim, S.-N., and Williams, J. (2026).
\newblock Decision-making with possibilistic inferential models.
\newblock {\em Internat. J. Approx. Reason.}, 196:Paper No. 109720.

\bibitem[Martin and Williams, 2025]{imbvm.ext}
Martin, R. and Williams, J.~P. (2025).
\newblock Asymptotic efficiency of inferential models and a possibilistic
  {B}ernstein--von {M}ises theorem.
\newblock {\em Internat. J. Approx. Reason.}, 180:Paper No. 109389.

\bibitem[Murph et~al., 2024]{murph.etal.fiducial}
Murph, A., Hannig, J., and Williams, J.~P. (2024).
\newblock Introduction to generalized fiducial inference.
\newblock In Berger, J., Meng, X.-L., Reid, N., and Xie, M.-g., editors, {\em
  Handbook of {B}ayesian, {F}iducial, and {F}requentist {I}nference}, pages
  276--299. Chapman \& Hall/CRC Press.

\bibitem[Nachbin, 1965]{nachbin1965}
Nachbin, L. (1965).
\newblock {\em The {H}aar {I}ntegral}.
\newblock D. Van Nostrand Co., Inc., Princeton, N.J.-Toronto-London.

\bibitem[Neal, 2003]{neal2003}
Neal, R.~M. (2003).
\newblock Slice sampling.
\newblock {\em Ann. Statist.}, 31(3):705--767.
\newblock With discussions and a rejoinder by the author.

\bibitem[Neyman, 1941]{neyman1941}
Neyman, J. (1941).
\newblock Fiducial argument and the theory of confidence intervals.
\newblock {\em Biometrika}, 32:128--150.

\bibitem[Pereira and Stern, 2022]{MR4426408}
Pereira, C. A.~B. and Stern, J.~M. (2022).
\newblock The {$e$}-value: a fully {B}ayesian significance measure for precise
  statistical hypotheses and its research program.
\newblock {\em S\~{a}o Paulo J. Math. Sci.}, 16(1):566--584.

\bibitem[Reid, 2003]{reid2003}
Reid, N. (2003).
\newblock Asymptotics and the theory of inference.
\newblock {\em Ann. Statist.}, 31(6):1695--1731.

\bibitem[Robbins and Monro, 1951]{robbinsmonro}
Robbins, H. and Monro, S. (1951).
\newblock A stochastic approximation method.
\newblock {\em Ann. Math. Statistics}, 22:400--407.

\bibitem[Savage, 1961]{savage1961}
Savage, L.~J. (1961).
\newblock The foundations of statistics reconsidered.
\newblock In {\em Proc. 4th {B}erkeley {S}ympos. {M}ath. {S}tatist. and
  {P}rob., {V}ol. {I}}, pages 575--586. Univ. California Press, Berkeley,
  Calif.

\bibitem[Scheff{\'e}, 1970]{scheffe1970}
Scheff{\'e}, H. (1970).
\newblock Practical solutions of the {B}ehrens--{F}isher problem.
\newblock {\em J. Amer. Statist. Assoc.}, 65:1501--1508.

\bibitem[Schervish, 1995]{schervish1995}
Schervish, M.~J. (1995).
\newblock {\em Theory of {S}tatistics}.
\newblock Springer-Verlag, New York.

\bibitem[Schweder and Hjort, 2002]{schweder.hjort.2002}
Schweder, T. and Hjort, N.~L. (2002).
\newblock Confidence and likelihood.
\newblock {\em Scand. J. Statist.}, 29(2):309--332.

\bibitem[Schweder and Hjort, 2013]{schweder.hjort.cd.discuss}
Schweder, T. and Hjort, N.~L. (2013).
\newblock Discussion: ``{C}onfidence distribution, the frequentist distribution
  estimator of a parameter: a review''.
\newblock {\em Int. Stat. Rev.}, 81(1):56--68.

\bibitem[Schweder and Hjort, 2016]{schweder.hjort.book}
Schweder, T. and Hjort, N.~L. (2016).
\newblock {\em Confidence, {L}ikelihood, {P}robability}, volume~41 of {\em
  Cambridge Series in Statistical and Probabilistic Mathematics}.
\newblock Cambridge University Press, New York.

\bibitem[Shafer, 1976]{shafer1976}
Shafer, G. (1976).
\newblock {\em A {M}athematical {T}heory of {E}vidence}.
\newblock Princeton University Press, Princeton, N.J.

\bibitem[Shafer, 1982]{shafer1982}
Shafer, G. (1982).
\newblock Belief functions and parametric models.
\newblock {\em J. Roy. Statist. Soc. Ser. B}, 44(3):322--352.
\newblock With discussion.

\bibitem[Shafer, 1987]{shafer1987}
Shafer, G. (1987).
\newblock Belief functions and possibility measures.
\newblock In Bezdek, J.~C., editor, {\em The Analysis of Fuzzy Information,
  Vol. 1: Mathematics and Logic}, pages 51--84. CRC.

\bibitem[Shapley, 1953]{shapley1953}
Shapley, L.~S. (1953).
\newblock A value for {$n$}-person games.
\newblock In {\em Contributions to the {T}heory of {G}ames, vol. 2}, volume no.
  28 of {\em Ann. of Math. Stud.}, pages 307--317. Princeton Univ. Press,
  Princeton, NJ.

\bibitem[Smets and Kennes, 1994]{smets.kennes.1994}
Smets, P. and Kennes, R. (1994).
\newblock The transferable belief model.
\newblock {\em Artificial Intelligence}, 66(2):191--234.

\bibitem[Spj{\o}tvoll, 1983]{spjotvoll1983}
Spj{\o}tvoll, E. (1983).
\newblock Preference functions.
\newblock In {\em A {F}estschrift for {E}rich {L}. {L}ehmann}, Wadsworth
  Statist./Probab. Ser., pages 409--432. Wadsworth, Belmont, Calif.

\bibitem[Stein, 1959]{stein1959}
Stein, C. (1959).
\newblock An example of wide discrepancy between fiducial and confidence
  intervals.
\newblock {\em Ann. Math. Statist.}, 30:877--880.

\bibitem[Sun and Berger, 2007]{sun.berger.2007}
Sun, D. and Berger, J.~O. (2007).
\newblock Objective {B}ayesian analysis for the multivariate normal model.
\newblock In {\em Bayesian {S}tatistics 8}, Oxford Sci. Publ., pages 525--562.
  Oxford Univ. Press, Oxford.

\bibitem[Taraldsen, 2021]{taraldsen.2021.cd}
Taraldsen, G. (2021).
\newblock Joint confidence distributions.
\newblock \url{https://doi.org/10.13140/RG.2.2.33079.85920}.

\bibitem[Thornton and Xie, 2024]{thornton.xie.cd}
Thornton, S. and Xie, M.-g. (2024).
\newblock Bridging {B}ayesian, frequentist and fiducial ({BFF}) inferences
  using confidence distributions.
\newblock In Berger, J., Meng, X.-L., Reid, N., and Xie, M.-g., editors, {\em
  Handbook of {B}ayesian, {F}iducial, and {F}requentist {I}nference}, pages
  106--132. Chapman \& Hall/CRC Press.

\bibitem[Troffaes and de~Cooman, 2014]{lower.previsions.book}
Troffaes, M. C.~M. and de~Cooman, G. (2014).
\newblock {\em Lower {P}revisions}.
\newblock Wiley Series in Probability and Statistics. John Wiley \& Sons, Ltd.,
  Chichester.

\bibitem[van~der Vaart, 1998]{vaart1998}
van~der Vaart, A.~W. (1998).
\newblock {\em Asymptotic {S}tatistics}.
\newblock Cambridge University Press, Cambridge.

\bibitem[Walley, 1991]{walley1991}
Walley, P. (1991).
\newblock {\em Statistical {R}easoning with {I}mprecise {P}robabilities},
  volume~42 of {\em Monographs on Statistics and Applied Probability}.
\newblock Chapman \& Hall Ltd., London.

\bibitem[Wasserman, 1990a]{wasserman1990b}
Wasserman, L.~A. (1990a).
\newblock Belief functions and statistical inference.
\newblock {\em Canad. J. Statist.}, 18(3):183--196.

\bibitem[Wasserman, 1990b]{wasserman1990}
Wasserman, L.~A. (1990b).
\newblock Prior envelopes based on belief functions.
\newblock {\em Ann. Statist.}, 18(1):454--464.

\bibitem[Welch, 1938]{welch1938}
Welch, B.~L. (1938).
\newblock The significance of the difference between two means when the
  population variances are unequal.
\newblock {\em Biometrika}, 29:350--362.

\bibitem[Wilks, 1938]{wilks1938}
Wilks, S.~S. (1938).
\newblock The large-sample distribution of the likelihood ratio for testing
  composite hypotheses.
\newblock {\em Ann. Math. Statist}, 9:60--62.

\bibitem[Xie and Singh, 2013]{xie.singh.2012}
Xie, M. and Singh, K. (2013).
\newblock Confidence distribution, the frequentist distribution estimator of a
  parameter: a review.
\newblock {\em Int. Stat. Rev.}, 81(1):3--39.

\bibitem[Zabell, 1992]{zabell1992}
Zabell, S.~L. (1992).
\newblock R. {A}. {F}isher and the fiducial argument.
\newblock {\em Statist. Sci.}, 7(3):369--387.

\bibitem[Zadeh, 1975]{zadeh1975d}
Zadeh, L.~A. (1975).
\newblock Fuzzy logic and approximate reasoning.
\newblock {\em Synthese}, 30(3--4):407--428.

\bibitem[Zadeh, 1978]{zadeh1978}
Zadeh, L.~A. (1978).
\newblock Fuzzy sets as a basis for a theory of possibility.
\newblock {\em Fuzzy Sets and Systems}, 1(1):3--28.

\end{thebibliography}

\end{document}